\documentclass[11pt,draftclsnofoot,onecolumn]{IEEEtran}

\usepackage{graphicx,figsize,subfigure}
\usepackage{epsfig}
\usepackage{amsmath}
\usepackage{bbm}
\usepackage{amssymb}
\usepackage{wrapfig}	
\usepackage{times,color}
\usepackage{cite,footnote,xspace,syntonly,algorithm,algorithmic,bm}
\usepackage[caption=false,font=footnotesize]{subfig}
\usepackage{ifpdf}
\usepackage{psfrag}
\usepackage{tikz,chemarrow} %for drawing figures
\usepackage[framemethod=tikz]{mdframed}
\usepackage{amsfonts}
\usepackage{pifont}% http://ctan.org/pkg/pifont
\input{mysymbol.sty}
\newtheorem{definition}{\hspace*{-11pt}\bf Definition}

\newcommand{\xmark}{\text{\ding{55}}}

\long\def\symbolfootnote[#1]#2{\begingroup
\def\thefootnote{\fnsymbol{footnote}}
\footnote[#1]{#2}\endgroup} \psfull

\begin{document}
%--------------------------------------------The First Page---------------------------------------------------------------
% paper title
\title{\huge Connecting the Dots: Identifying Network Structure\\ via Graph Signal Processing$^\dag$}

\author{{\it Gonzalo~Mateos, Santiago Segarra, Antonio G. Marques, and Alejandro Ribeiro$^\ast$}}

%\markboth{IEEE SIGNAL PROCESSING MAGAZINE (REVISED)}
\maketitle \maketitle \symbolfootnote[0]{$\dag$ Work in this paper
was supported by the USA NSF award CCF-1750428, the Spanish MINECO grants OMICRON (TEC2013-41604-R) and KLINILYCS (TEC2016-75361-R), and an MIT IDSS seed grant.} 
\symbolfootnote[0]{$\ast$ G. Mateos is with the Dept. of Electrical and Computer Engineering, University of Rochester. S. Segarra is with the Dept. of Electrical and Computer Engineering, Rice University. A. G. Marques is with the Dept. of Signal Theory and Communications, King Juan Carlos University. A. Ribeiro is with the Dept. of Electrical and Systems Engineering,
	University of Pennsylvania.   Emails: \texttt{gmateosb@ece.rochester.edu}, \texttt{segarra@rice.edu}, \texttt{antonio.garcia.marques@urjc.es}, and \texttt{aribeiro@seas.upenn.edu}.}

\vspace*{-45pt}
%\begin{center}
%\small{\bf Submitted: }\today\\
%\end{center}
%\vspace*{10pt}

\begin{abstract}
Network topology inference is a prominent problem in Network Science. Most graph signal processing (GSP) efforts to date assume that the underlying network is known, and then analyze how the graph's algebraic and spectral characteristics impact the properties of the graph signals of interest.  {Such an assumption is often untenable beyond applications dealing with e.g., directly observable social and infrastructure networks;} and typically adopted graph construction schemes are largely informal, distinctly lacking an element of validation. This tutorial offers an overview of graph learning methods developed to bridge the aforementioned gap, by using information available from graph signals to infer the underlying graph topology. Fairly mature statistical approaches are surveyed first, where correlation analysis takes center stage along with its connections to covariance selection and high-dimensional regression for learning Gaussian graphical models. Recent GSP-based network inference frameworks are also described, which postulate that the network exists as a latent underlying structure, and that observations are generated as a result of a network process defined in such a graph. A number of arguably more nascent topics are also briefly outlined, including inference of dynamic networks, nonlinear models of pairwise interaction, as well as extensions to directed graphs and their relation to causal inference. All in all, this paper introduces readers to challenges and opportunities for signal processing research in emerging topic areas at the crossroads of modeling, prediction, and control of complex behavior arising in  networked systems that evolve over time.
\end{abstract}

% % % % % % % % % % % % % % % % % % % % % % % % % % % % % % % % % % % % % % % %
%                         Section I                                           %
% % % % % % % % % % % % % % % % % % % % % % % % % % % % % % % % % % % % % % % %

\section{Introduction}
\label{sec:intro}

Coping with the challenges found at the intersection of Network Science and Big Data necessitates 
fundamental breakthroughs in modeling, identification, and controllability of distributed network processes -- often conceptualized as signals defined on graphs~\cite{gsp2018tutorial}. For instance, graph-supported signals can model vehicle congestion levels over road networks, economic activity observed over a network of production flows between industrial sectors, infectious states of individuals susceptible to an epidemic disease spreading on a social network, gene expression levels defined on top of gene regulatory networks, brain activity signals supported on  brain connectivity networks, and fake news that diffuse in online social networks, just to name a few. There is an evident mismatch between our scientific understanding of signals defined over regular domains (time or space) and graph-supported signals. Knowledge about time series was developed over the course of decades and boosted by technology-driven needs in areas such as communications, speech processing, and control. On the contrary, the prevalence of network-related signal processing (SP) problems and the access to quality network data are recent events. Making sense of large-scale datasets from a network-centric perspective will constitute a crucial step to obtain new insights in various areas in science and engineering; and SP can play a key role to that end.

Under the assumption that the signals are related to the topology of the graph where they are supported, the goal of graph signal processing (GSP) is to develop algorithms that fruitfully leverage this relational structure, and can make inferences about these relationships even when they are only partially observed. Most GSP efforts to date \emph{assume that the underlying network topology is known}, and then analyze how the graph's algebraic and spectral characteristics impact the properties of the graph signals of interest.  {This is feasible in applications involving physical networks, or, when the relevant links are tangible and can be directly observed (e.g., when studying flows in transportation networks, monitoring cascading failures in power grids, maximizing influence on social networks, and tracking the dynamic structure of the World Wide Web). However, in many other settings the network may represent a conceptual model of pairwise relationships among entities.}
%not be observable or such an assumption can be untenable} in practice 
%and  
In exploratory studies of e.g., functional brain connectivity or regulation among genes,  inference of nontrivial pairwise interactions between signal elements (i.e., blood-oxygen-level dependent time series per voxel or gene expression levels, respectively) is often a goal per se.  {In these settings and beyond, arguably most graph construction schemes are largely informal, distinctly lacking an element of validation. Even for infrastructure networks, their sheer size, (un)intentional reconfiguration, in addition to privacy or security constraints enforced by administrators may render the acquisition of updated topology information a challenging endeavor.} Accordingly, a fundamental question is how to use  {observations of} graph signals to learn the underlying network structure, or, a judicious network model of said data facilitating efficient signal representation, visualization, prediction, (nonlinear) dimensionality reduction, and (spectral) clustering. 

In this tutorial we offer an overview of graph learning methods developed to bridge the aforementioned gap, by using information available from graph signals to infer the underlying graph topology (see Section \ref{sec:prelim} for a general, yet formal problem statement).  {For the topology inference problem to be well posed, it is clear that one must assume a data model linking the observations to the unknown graph. One can thus formulate the graph learning task as an inverse problem, with a criterion adapted to the properties of the observations (e.g., via a probabilistic model, smoothness, or graph stationarity) and regularized to encourage desirable characteristics of the sought network.} Fairly mature statistical approaches are surveyed first in Section \ref{sec:stat}, where linear correlation analysis takes center stage along with its connections to covariance selection and high-dimensional regression for learning Gaussian graphical models. In Section \ref{sec:GSP_foundations} we shift gears and review recent GSP advances including the graph Fourier transform (GFT) and related notions of signal variation on graphs, as well as graph filter models of network diffusion and an innovative characterization of stationarity for random network processes. These concepts are at the heart of recent GSP-based topology inference frameworks that postulate that the network exists as a latent underlying structure, and that observations are generated as a result of a network process defined on such a graph. Within this class, focus is laid first on methods that infer graph structure from observed signals assumed to be smooth over the graph (Section \ref{sec:smooth}).  {Historically, the success of signal and information processing algorithms has hinged upon exploiting signal structure. It is thus only natural that one would opt for graphs under which data admit certain regularity.} Next, Section \ref{sec:topoid_diffusion} deals with a family of approaches that postulate a graph filter-based generative model for the observations that diffused on the sought network.  {Such linear models of network diffusion arise with, e.g., distributed control, multi-agent coordination, opinion formation, brain disease progression, and molecular communications. The central part of this tutorial is summarized in Section \ref{sec:discussion}, where we take a step back and compare the methods surveyed so far. In doing so we emphasize the new GSP perspective to the topology identification problem, and offer insights on the choice of the `right' graph learning algorithm for a given network-analytic application.} 

\begin{figure*}[t]
	\centering
	\includegraphics[width=\linewidth]{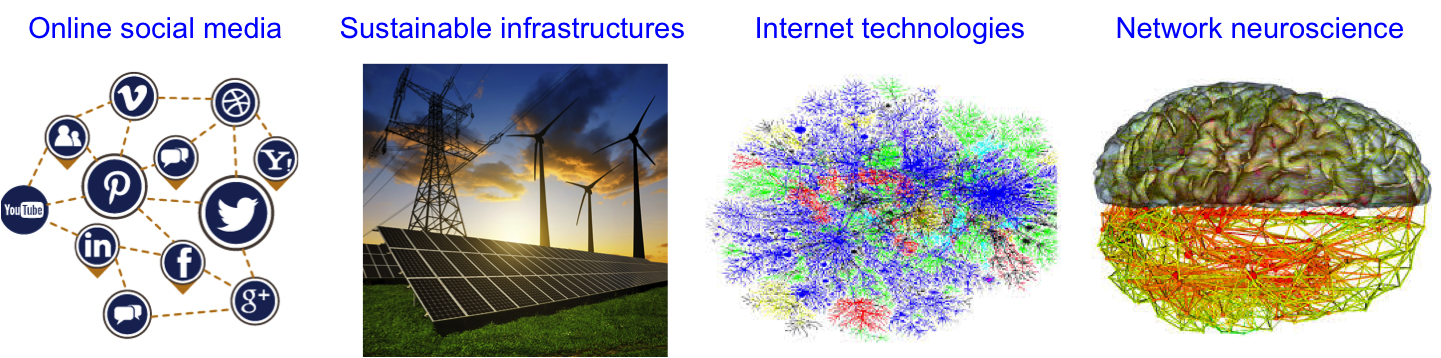}
	\vspace*{-0.5cm}
	\caption{Advancing our scientific understanding of fundamental issues arising with networked systems can have sizable impact in neuroscience, social networks, smart cities, Internet-of-Things technologies, networking
and sensing integration for cyber-physical systems, as well as green grid analytics for seamless integration of
renewable sources of energy. To that end, the signal processing body of knowledge is one of the core required components at the intersection of data and network sciences.}
	\label{F:network_examples}
\end{figure*}

A number of arguably more nascent topics are also briefly outlined in Section \ref{sec:emerging}, including inference of dynamic and multi-networks, nonlinear models of pairwise interaction, as well as extensions to directed (di)graphs and their relation to causal inference. Through rigorous problem formulations, intuitive reasoning, and illustration of applications (Section \ref{sec:applications}), this tutorial introduces readers to challenges and opportunities for SP research in emerging topic areas at the crossroads of modeling, prediction, and control of complex behavior arising in networked systems; see Figure~\ref{F:network_examples} and the research outlook in Section \ref{sec:conc}.

%%%%%%%%%%%%%%%   F   I   G   U   R    E   %%%%%%%%%%%%%%%%%%%%%%%%%%%%%%%%%%%%%%%%
%
%%%%%%%%%%%%%%%%%%%%%%%%%%%%%%%%%%%%%%%%%%%%%%%%%%%%%%%%%%%%%%%%%%%

\noindent \textbf{Notation.} The entries of a matrix $\mathbf{X}$ and a (column) vector $\mathbf{x}$ are denoted by $X_{ij}$ and $x_i$, respectively. Sets are represented by calligraphic capital letters, and $|\cdot|$ stands for the cardinality of a set or the
magnitude of a scalar.  % and $\bbX_{\ccalI}$ denotes a submatrix of $\bbX$ formed by selecting the rows of $\bbX$ indexed by $\ccalI$. 
The notation $(\cdot)^T$, $(\cdot)^H$ and $(\cdot)^\dag$ stands for transpose, conjugate transpose and matrix pseudo-inverse, respectively; $\mathbf{0}$ and $\mathbf{1}$ refer to the all-zero and all-one vectors; while $\bbI$ denotes the identity matrix. For a vector $\bbx$, $\diag(\mathbf{x})$ is a diagonal matrix whose $i$-th diagonal entry is $x_i$; when applied to a matrix, $\diag(\bbX)$ is a vector collecting the diagonal elements of $\bbX$. 
The operators $\otimes$, $\odot$, $\circ$, $\text{vec}(\cdot)$, $\text{trace}(\cdot)$ and $\E{\cdot}$ stand for Kronecker product, Khatri-Rao (columnwise Kronecker) product, Hadamard (entry-wise) product,  matrix vectorization, matrix trace, and expectation, respectively. The indicator function $\ind{\cdot}$ takes the value $1$ if the logical condition in the argument holds true, and $0$ otherwise. For matrix $\bbX$, $\| \bbX \|_p$ denotes the $\ell_p$-norm of $\textrm{vec}(\bbX)$ ($\|\bbX\|_F=\|\bbX\|_2$ stands for the Frobenius norm), whereas $\| \bbX \|_{M(p)}$ is the matrix norm induced by the vector $\ell_p$-norm. Lastly, $\mathrm{range}(\bbX)$ refers to the column space of $\bbX$.

% % % % % % % % % % % % % % % % % % % % % % % % % % % % % % % % % % % % % % % %
%                         Section II                                          %
% % % % % % % % % % % % % % % % % % % % % % % % % % % % % % % % % % % % % % % %

\section{Graph-theoretic preliminaries and problem statement}
\label{sec:prelim}

As the Data Science revolution keeps gaining momentum, it is only natural that complex signals with irregular structure
become increasingly of interest. While there are many possible sources and models of added complexity, a general proximity relationship between signal elements is not only a plausible but a ubiquitous model across science and engineering. 

To develop such a model, consider signals whose values are associated with nodes of a weighted, undirected, and connected graph. Formally, we consider the signal $\bbx = [x_1, \ldots, x_{N}]^T \in \reals^N$ and the generally unknown weighted graph  $\ccalG(\ccalV,\ccalE, \bbW)$, where $\ccalV=\{1,\ldots,N\}$ is a set of $N$ vertices or nodes and $\ccalE\subseteq \ccalV\times\ccalV$ is the set of edges. Scalar $x_i$ denotes the signal value at node $i\in\ccalV$. The map $\bbW: \ccalV\times \ccalV \to \reals_{+}$ from the set of unordered pairs of vertices to the nonnegative reals
associates a weight $W_{ij}\geq 0$ with the edge $(i,j)\in \ccalE$, while $W_{ij}\equiv 0$ for $(i,j)\neq \ccalE$. The symmetric coefficients $W_{ij}=W_{ji}$ represent the strength of the connection (i.e., the similarity or influence) between nodes $i$ and $j$. In terms of the signal $\bbx$, this means that when the weight $W_{ij}$ is large, the signal values $x_i$ and $x_j$ tend to be similar. Conversely, when the weight $W_{ij}$ is small or, in the extremum, when we have $W_{ij}=0$, the signal values $x_i$ and $x_j$ are not directly related except for what is implied by their separate connections to other nodes. Such an interpretation of the edge weights establishes a link between the signal values and the graph topology, which at a high level supports the feasibility of inferring $\ccalG$ from signal observations.

As a more general algebraic descriptor of network structure (i.e., topology), associated with the graph $\ccalG$ one can introduce the so-called \emph{graph-shift operator} $\mathbf{S}$~\cite{SandryMouraSPG_TSP13}. The shift $\mathbf{S}\in\mathbb{R}^{N\times N}$ is a matrix whose entry $S_{ij}$ can be nonzero only if $i=j$ or if $(i,j)\in\mathcal{E}$. Thus, the sparsity pattern of the matrix $\bbS$ captures the local structure of $\ccalG$, but we make no specific assumptions on the values of its nonzero entries. Widely-adopted choices for $\mathbf{S}$ are the adjacency matrix $\bbW$~\cite{SandryMouraSPG_TSP13,DSP_freq_analysis}, the combinatorial graph Laplacian $\bbL:=\textrm{diag}(\bbW\mathbf{1})-\bbW$~\cite{EmergingFieldGSP}, or, their various degree-normalized counterparts. For probabilistic graphical models of random $\bbx$, one could adopt covariance or precision matrices as graph shifts encoding conditional (in)dependence relationships among nodal random variables; see e.g.,~\cite[Ch. 7.3.3]{kolaczyk2009book} and Section \ref{sec:stat}. Other application-specific alternatives have been proposed as well; see~\cite{gsp2018tutorial} and references therein. In any case, parameterizing graph topology via a graph-shift operator of choice can offer additional flexibility when it comes to formulating constrained optimization problems to estimate graph structure. As it will become clear in the sequel, such a generality can have a major impact on the performance and computational complexity of the ensuing algorithms.

\begin{mdframed}[hidealllines=true,backgroundcolor=gray!20]
\textbf{Why graph \textit{shift}?} To justify the adopted graph \emph{shift} terminology, consider the directed cycle graph whose circulant adjacency matrix $\bbW_{dc}$ is zero, except for entries $W_{ij}=1$ whenever $i=\mathrm{mod}_N(j)+1$, where $\mathrm{mod}_N(x)$ denotes the modulus (remainder) obtained after dividing $x$ by $N$. Such a graph can be used to represent the domain of discrete-time periodic signals with period $N$. If $\bbS=\bbW_{dc}$, then $\bby=\bbS\bbx$ implements a circular shift of the entries in $\bbx$, which corresponds to a one-unit time delay under the aforementioned interpretation~\cite{gsp2018tutorial}.  {Notice though that in general $\bbS$ need not be neither invertible nor isometric, an important departure from the shift in discrete-time signal processing.} The intuition behind $\mathbf{S}$ for general graphs beyond the directed cycle, is to capture a linear transformation that can be computed locally at the nodes of the graph. Formally, if signal $\mathbf{y}$ is given by $\mathbf{y}=\mathbf{S}\mathbf{x}$, then node $i$ can compute $y_i$ as a linear combination of the signal values $x_j$ at node $i$'s neighbors $\ccalN_i:=\{j: (i,j)\in\ccalE\}$. For example, one can think of an individual's opinion formation process as one of weighing in the views of close friends regarding the subject matter. As elaborated in Section \ref{ssec:diffusion}, $\bbS$ also serves as the main building block to define more general linear shift-invariant operators for graph signals, namely graph filters~\cite{SandryMouraSPG_TSP13}. The shift operator is also at the heart of graph stationarity notions for random network processes (Section \ref{ssec:stationarity}).
\end{mdframed}

All elements are now in place to state a general network topology identification problem. 

\vspace{0.2cm}
 {\noindent\fbox{%
	\parbox{\textwidth}{\textbf{Problem.} Given a set $\ccalX:=\{\bbx_p\}_{p=1}^P$ of graph signal observations supported on the unknown graph $\ccalG(\ccalV,\ccalE, \bbW)$ with $|\ccalV|=N$, the goal is to identify the topology encoded in the entries of a graph-shift operator $\bbS$ that is optimal in some sense. The optimality criterion is usually dictated by the adopted network-dependent model for the signals in $\ccalX$, possibly augmented by priors motivated by physical characteristics of $\bbS$, to effect statistical regularization, or else to favor more interpretable graphs.}
}}
\vspace{0.2cm} 

This is admittedly a very general and somewhat loose formulation, that will be narrowed down in subsequent sections as we elaborate on various criteria stemming from different models binding the (statistical) signal properties to the graph topology.  {Indeed, it is clear that one must assume \emph{some} relation between the signals and the unknown underlying graph, since otherwise the topology inference exercise would be hopeless. This relation will be henceforth given by statistical generative priors (Section~\ref{sec:stat}), and by properties of the signals with respect to the underlying graph such as smoothness (Section~\ref{sec:smooth}) or stationarity (Section \ref{sec:topoid_diffusion}).} The observations in $\ccalX$ can be noisy and incomplete, and accordingly the relationship between $N$, $P$ and the mechanisms of data errors and missingness will all play a role in the graph recovery performance. Mostly the focus will be on inference of undirected and static graphs, an active field for which the algorithms and accompanying theory are nowadays better developed. Section \ref{sec:emerging} will broaden the scope to more challenging directed, dynamic, and multi-graphs.

% % % % % % % % % % % % % % % % % % % % % % % % % % % % % % % % % % % % % % % %
%                         Section III                                         %
% % % % % % % % % % % % % % % % % % % % % % % % % % % % % % % % % % % % % % % %

\section{Statistical methods for network topology inference}
\label{sec:stat}

As presented in the previous section, networks typically encode similarities between signal elements. Thus, a natural starting point towards constructing a graph representation of the data is to associate edge weights with nontrivial correlations or coherence measures between signal profiles at incident nodes. In this vein, informal (but popular) scoring methods rely on ad hoc thresholding of user-defined edgewise score functions. Examples include the Pearson product-moment correlation used to quantify gene-regulatory interactions, the Jaccard coefficient for scientific citation networks, the Gaussian radial basis function to link measurements from a sensor network, or mutual information to capture nonlinear interactions. Often thresholds are manually tuned so that the resulting graph is deemed to accurately capture the relational structure in the data; a choice possibly informed by domain experts.  In other cases, a prescribed number $k$ of the top relations out of each node are retained, leading to the so-called $k$-nearest neighbor graphs that are central to graph smoothing techniques in machine learning.

Such informal approaches fall short when it comes to assessing whether the obtained graph is accurate in some appropriate (often application-dependent) sense. In other words, they lack a framework that facilitates validation. Recognizing this shortcoming, a different paradigm is to cast the graph learning problem as one of selecting the best representative from a family of candidate networks by bringing to bear elements of statistical modeling and inference. The advantage of adopting such a methodology is that one can leverage existing statistical concepts and tools to formally study issues of identifiability, consistency, robustness to measurement error and sampling, as well as those relating to sample and computational complexities. Early statistical approaches to the network topology inference problem are the subject of this section.

% % % % % % % % % % % % % % % % % % % % % % % % % % % % % % % % % % % % % % % %
%                        Subsection III-A                                     %
% % % % % % % % % % % % % % % % % % % % % % % % % % % % % % % % % % % % % % % %

\subsection{Correlation networks}
\label{ssec:correlation}

Arguably the most widely adopted linear measure of similarity between nodal random variables $x_i$ and $x_j$ is the Pearson correlation coefficient defined as
\begin{equation}\label{E:corr_coeff}
\rho_{ij}:=\frac{\textrm{cov}(x_i,x_j)}{\sqrt{\textrm{var}(x_i)\textrm{var}(x_j)}}.
\end{equation}
It can be obtained from entries $\sigma_{ij}:=\textrm{cov}(x_i,x_j)$ in the covariance matrix $\bbSigma:=\E{(\bbx-\bbmu)(\bbx-\bbmu)^T}$ of the random graph signal $\bbx=[x_1,\ldots,x_N]^T$, with mean vector $\bbmu:=\E{\bbx}$. Given this choice, it is natural to define the \emph{correlation network} $\ccalG(\ccalV,\ccalE, \bbW)$ with vertices $\ccalV:=\{1,\ldots,N\}$ and edge set $\ccalE:=\{(i,j)\in\ccalV\times \ccalV:\rho_{ij}\neq 0\}$. There is some latitude on the definition of the weights. To directly capture the correlation strength between $x_i$ and $x_j$ one can set $W_{ij}=|\rho_{ij}|$ or its un-normalized variant $W_{ij}=|\textrm{cov}(x_i,x_j)|$; alternatively the choice $W_{ij}=\ind{\rho_{ij}\neq 0}$ gives an unweighted graph consistent with $\ccalE$. In GSP applications it is often common to refer to correlation network as one with graph-shift operator $\bbS:=\bbSigma$. Regardless of these choices, what is important here is that the definition of $\ccalE$ dictates that the problem of identifying the topology of $\ccalG$ becomes one of inferring the subset of nonzero correlations.  

To that end, given independent realizations $\ccalX:=\{\bbx_p\}_{p=1}^P$ of $\bbx$ one forms empirical correlations $\hat{\rho}_{ij}$ by replacing the ensemble covariances in \eqref{E:corr_coeff}, with the entries $\hat{\sigma}_{ij}$ of the unbiased sample covariance matrix estimate $\hbSigma$. As discussed earlier in this section, one could then manually fix a threshold and assign edges to the corresponding largest values $|\hat{\rho}_{ij}|$. Instead, a more principled approach is to test the hypotheses
\begin{equation}\label{E:corr_tests}
H_0:\rho_{ij}=0\quad \textrm{versus}\quad  H_1:\rho_{ij}\neq 0,
\end{equation}
for each of the ${N \choose 2}=N(N-1)/2$ candidate edges in $\ccalG$, i.e., the number of unordered pairs in $\ccalV \times \ccalV$. While $|\hat{\rho}_{ij}|$ would appear to be the go-to test statistic, a more convenient choice is the Fischer score $z_{ij}:=\tanh^{-1}(\hat{\rho}_{ij})$. The reason is that  under $H_0$ one (approximately) has $z_{ij}\sim \textrm{Normal}(0,\frac{1}{P-3})$; see~\cite[p. 210]{kolaczyk2009book} for further details and the justification based on asymptotic-theory arguments. This simple form of the null distribution facilitates computation of $p$-values, or, the selection of a threshold that guarantees a prescribed significance level (i.e., false alarm probability $P_{FA}$ in the SP parlance) per test. 

However, such individual test control procedures might not be effective for medium to large-sized graphs, since the total number of simultaneous tests to be conducted scales as $O(N^2)$. Leaving aside potential computational challenges, the problem of large-scale hypothesis testing must be addressed~\cite[Ch. 15]{efron2016book}. Otherwise, say for an empty graph with $\ccalE=\emptyset$ a constant false alarm rate $P_{FA}$ per edge will yield on average $O(N^2P_{FA})$ spurious edges, which can be considerable if $N$ is large. A common workaround is to instead focus on controlling the false discovery rate (FDR) defined as
\begin{equation}\label{E:FDR}
\textrm{FDR}:=\E{\frac{R_{f}}{R}\given R>0}\Pr{R>0},
\end{equation}
where $R$ denotes the number of rejections among all $O(N^2)$ edgewise tests conducted, and $R_f$ stands for the number of false rejections (here representing false edge discoveries). Let $p_{(1)}\leq p_{(2)}\leq\ldots\leq p_{(|\ccalV\times\ccalV|)}$ be the ordered $p$-values for all tests. Then a prescribed level $\textrm{FDR}\leq q$ can be guaranteed by following the Benjamini-Hochberg FDR control procedure, which declares edges for all tests $i$ such that $p_{(i)}\leq (2i/N(N-1))q$; see e.g.,~\cite[Ch. 15.2]{efron2016book}. It is worth noting that the FDR guarantee is only valid for independent tests, an assumption that rarely holds in a graph learning setting. Hence results and control levels should be interpreted with care; see also~\cite[p. 212]{kolaczyk2009book} for a discussion on FDR extensions when some level of dependency is present between tests.          

With regards to the scope of correlation networks, apparently they can only capture linear and symmetric pairwise dependencies among vertex-indexed random variables. Most importantly, the measured correlations can be due to latent network effects rather than from strong direct influence among a pair of vertices. For instance, a suspected regulatory interaction among genes $(i,j)$ inferred from their highly correlated micro-array expression-level profiles, could be an artifact due to a third latent gene $k$ that is actually regulating the expression of both $i$ and $j$. If seeking a graph reflective of direct influence among pairwise signal elements, clearly correlation networks may be undesirable. 

Interestingly, one can in principle resolve such a confounding by instead considering \emph{partial} correlations  
\begin{equation}\label{E:par_corr_coeff}
\rho_{ij|\ccalV\setminus ij}:=\frac{\textrm{cov}(x_i,x_j\given \ccalV\setminus ij)}{\sqrt{\textrm{var}(x_i\given \ccalV\setminus ij)\textrm{var}(x_j\given \ccalV\setminus ij)}},
\end{equation}
where $\ccalV\setminus ij$ symbolically denotes the collection of all $N-2$ random variables $\{x_k\}$ after excluding those indexed by nodes $i$ and $j$. A partial correlation network can be defined in analogy to its (unconditional) correlation network counterpart, but with edge set $\ccalE:=\{(i,j)\in\ccalV\times \ccalV:\rho_{ij|\ccalV\setminus ij}\neq 0\}$. Again, the problem of inferring nonzero partial correlations from data 
$\ccalX:=\{\bbx_p\}_{p=1}^P$ can be equivalently cast as one of hypothesis testing. With minor twists, issues of selecting a test statistic and a tractable approximate null distribution, as well as successfully dealing with the multiple-testing problem, can all be addressed by following similar guidelines to those in the Pearson correlation case~\cite[Ch. 7.3.2]{kolaczyk2009book}.

% % % % % % % % % % % % % % % % % % % % % % % % % % % % % % % % % % % % % % % %
%                        Subsection III-B                                     %
% % % % % % % % % % % % % % % % % % % % % % % % % % % % % % % % % % % % % % % %

\subsection{Gaussian graphical models, covariance selection, and graphical lasso}
\label{ssec:glasso}

Suppose now that the graph signal $\bbx=[x_1,\ldots,x_N]^T$ is a Gaussian random vector, meaning that the vertex-indexed random variables are jointly Gaussian. Under such a distributional assumption, $\rho_{ij|\ccalV\setminus ij}=0$ is equivalent to $x_i$ and $x_j$ being conditionally \emph{independent} given all of the other variables in  $\ccalV\setminus ij$. Consequently, the partial correlation network with edges $\ccalE:=\{(i,j)\in\ccalV\times \ccalV:\rho_{ij|\ccalV\setminus ij}\neq 0\}$ specifies conditional independence relations among the entries of $\bbx$, and is known as an undirected Gaussian graphical model or Gaussian Markov random field (GMRF).

A host of opportunities for inference of Gaussian graphical models emerge by recognizing that the partial correlation coefficients can be expressed as
\begin{equation}\label{E:par_corr_precision}
\rho_{ij|\ccalV\setminus ij}=-\frac{\theta_{ij}}{\sqrt{\theta_{ii}\theta_{jj}}},
\end{equation}
where $\theta_{ij}$ is the $(i,j)$-th entry of the precision or concentration matrix $\bbTheta:=\bbSigma^{-1}$, namely the inverse of the covariance matrix $\bbSigma$ of $\bbx$. The upshot of \eqref{E:par_corr_precision} is that it reveals a bijection between the set of nonzero partial correlations (the edges of $\ccalG$) and the sparsity pattern of the precision matrix $\bbTheta$. The graphical model selection problem of identifying the conditional independence relations in $\ccalG$ given i.i.d. realizations $\ccalX:=\{\bbx_p\}_{p=1}^P$ from a multivariate Gaussian distribution, is known as the \emph{covariance selection} problem. 

The term covariance selection was first coined by Dempster back in the early 70s, who explored the role of sparsity in estimating the entries of $\bbTheta$ via a recursive, likelihood-based thresholding procedure on the entries of $\hbTheta:=\hbSigma^{-1}$~\cite{dempster_cov_selec}. Computationally, this classical algorithm does not scale well to contemporary large-scale networks. Moreover, in high-dimensional regimes where $N\gg P$ the method breaks down since the sample covariance matrix $\hbSigma$ is rank deficient. Such a predicament calls for regularization, and next we describe graphical model selection approaches based on $\ell_1$-norm regularized global likelihoods for the Gaussian setting. Neighborhood-based regression methods are the subject of Section \ref{ssec:regression}.

We will henceforth assume zero-mean $\bbx\sim\textrm{Normal}(\mathbf{0},\bbSigma)$, since the focus is on estimating graph structure encoded in the entries of the precision matrix $\bbTheta=\bbSigma^{-1}$. Under this model, the maximum-likelihood estimate (ML) of the precision matrix is given by a strictly concave log-determinant program
\begin{equation}\label{E:precision_ML}
\hbTheta_{ML}=\arg\max_{\bbTheta\succeq\mathbf{0}}\left\{\log\det\bbTheta -\textrm{trace}(\hbSigma \bbTheta)\right\},
\end{equation}
where $\bbTheta\succeq\mathbf{0}$ requires the matrix to be positive semidefinite (PSD) and $\hbSigma:=\frac{1}{P}\sum_{p=1}^P\bbx_p\bbx_p^T$ is the empirical covariance matrix obtained from the data in $\ccalX$. It can be shown that if $\hbSigma$ is singular, the expression in \eqref{E:precision_ML} does not yield the ML estimator, which, in fact, does not exist. This happens e.g., when $N$ is larger than $P$. To overcome this challenge or otherwise to encourage parsimonious (hence more interpretable) graphs, the \emph{graphical lasso} regularizes the ML estimator \eqref{E:precision_ML} with the sparsity-promoting $\ell_1$-norm of $\bbTheta$~\cite{yuanlin2007}, yielding 
\begin{equation}\label{E:glasso}
\hbTheta\in\arg\max_{\bbTheta\succeq\mathbf{0}}\left\{\log\det\bbTheta -\textrm{trace}(\hbSigma \bbTheta)-\lambda\|\bbTheta\|_1\right\}.
\end{equation}
Variants of the model penalize only the off-diagonal entries of $\bbTheta$, or incorporate edge-specific penalty parameters $\lambda_{ij}>0$ to account for structural priors on the graph topology. Estimators of graphs with non-negative edge weights are of particular interest; see  \emph{Learning Gaussian graphical models with Laplacian constraints.}

\begin{mdframed}[hidealllines=true,backgroundcolor=gray!20]
	\textbf{Learning Gaussian graphical models with Laplacian constraints.} There are important differences when estimating precision matrices subject to combinatorial graph Laplacian constraints. The off-diagonal elements $L_{ij}:=-W_{ij}\leq 0$ of a Laplacian must be non-positive, so when $\bbTheta=\bbL$ the resulting GMRF is termed \emph{attractive} {; see also~\cite{slawski2015mmatrices} for estimation of precision matrices under the constraint that all partial correlations are non-negative.} Moreover, the Laplacian matrix is always singular because $\bbL\mathbf{1}=\mathbf{0}$, which yields an \emph{improper} GMRF. A proper GMRF can be obtained via diagonal loading of the sought Laplacian, which motivates the following sparse precision matrix estimation problem~\cite{Lake10discoveringstructure}
	\begin{align}\label{E:GMRF_Laplacian}
	\max_{\bbTheta\succeq\mathbf{0},\gamma\geq 0}&{}\left\{\log\det\bbTheta -\textrm{trace}(\hbSigma \bbTheta)-\lambda\|\bbTheta\|_1\right\}\\
	\textrm{ s. to }&{}\bbTheta = \bbL+\gamma\bbI\nonumber\\
	&{} \bbL\mathbf{1}=\mathbf{0},\: L_{ij}\leq 0, \:i\neq j.\nonumber
	\end{align}
	Given a solution $\{\hbTheta,\hat{\gamma}\}$ of \eqref{E:GMRF_Laplacian}, a combinatorial Laplacian can be recovered as $\hbL:=\hbTheta-\hat{\gamma}\bbI$. There are various probabilistic interpretations of such a diagonal loading, e.g., one where $\gamma^{-1}$ corresponds to the variance of white Gaussian noise modeling isotropic signal fluctuations; see also the related factor analysis model in Section \ref{ssec:smooth_factor_kernel}. 
	
	A general optimization framework for estimating (possibly diagonally-dominant, generalized) Laplacian matrices was presented in~\cite{egilmez2017jstsp}, along with specialized block-coordinate descent algorithms to tackle the resulting graphical model selection problems.  {A customized algorithm based on sequential quadratic approximation was developed in~\cite{mihailo2016topoid} to identify the topology of sparse consensus networks, where the steady-state precision matrix is shown to have Laplacian structure.}  Recently, an efficient soft-thresholding based estimator for a sparse graph Laplacian encoded in the precision matrix of a GMRF was put forth in~\cite{mike_icassp17}. The procedure offers quantifiable performance, in the form of probabilistic recovery error bounds similar to those available for the graphical lasso~\cite{ravikumar2011}.
\end{mdframed}	

Although \eqref{E:glasso} is convex, the objective is non-smooth and has an unbounded constraint set. As shown in~\cite{banerjee2008jlmr} the resulting complexity for off-the-shelf interior point methods adopted in~\cite{yuanlin2007} is $O(N^6)$. In addition, interior point methods require
computing and storing a Hessian matrix of size $O(N^2)$ every iteration. The memory requirements and
complexity are thus prohibitive for even modest-sized graphs, calling for custom-made scalable algorithms that 
are capable of handling larger problems. Such efficient first-order cyclic block-coordinate descent algorithms were developed in~\cite{banerjee2008jlmr} and subsequently refined in~\cite{glasso2008}, which can comfortably tackle sparse problems with thousands of nodes in under a few minutes. In terms of performance guarantees for the recovery of a ground-truth precision matrix $\bbTheta_0$, the graphical lasso estimator \eqref{E:glasso} with $\lambda=2\sqrt{\frac{\log N}{P}}$ satisfies the operator norm bound $\|\hbTheta-\bbTheta_0\|_2\leq \sqrt{\frac{d_{\max}^2\log N}{P}}$ with high probability, where $d_{\max}$ denotes the maximum nodal degree in $\bbTheta_0$~\cite{ravikumar2011}. Support consistency has been also established provided the number of samples scales as $P=\Omega(d_{\max}^2\log N)$; see~\cite{ravikumar2011} for details. 

% % % % % % % % % % % % % % % % % % % % % % % % % % % % % % % % % % % % % % % %
%                        Subsection III-C                                     %
% % % % % % % % % % % % % % % % % % % % % % % % % % % % % % % % % % % % % % % %

\subsection{Graph selection via neighborhood-based sparse linear regression}
\label{ssec:regression}

Another way to estimate the graphical model is to find the set of neighbors $\ccalN_i:=\{j: (i,j)\in\ccalE\}$ of each node $i\in \ccalV$ in the graph, by regressing $x_i$ against all other variables $\bbx_{\setminus i}:=[x_1,\ldots,x_{i-1},x_{i+1},\ldots,x_N]^T\in\reals^{N-1}$. To illustrate this idea, note that in the Gaussian setting where $\bbx\sim\textrm{Normal}(\mathbf{0},\bbTheta^{-1})$, we have that the conditional distribution of $x_i$ given $\bbx_{\setminus i}$ is also Gaussian. The minimum mean-square error (MMSE) predictor of $x_i$ based on $\bbx_{\setminus i}$ is  $\E{x_i\given \bbx_{\setminus i}}=\bbx_{\setminus i}^T\bbbeta^{(i)}$, which is linear in $\bbx_{\setminus i}$ and yields the decomposition
\begin{equation}\label{E:linear_predictor}
x_i=\bbx_{\setminus i}^T\bbbeta^{(i)}+\varepsilon_{\setminus i},
\end{equation}
where $\varepsilon_{\setminus i}$ is the zero-mean Gaussian prediction error, independent of $\bbx_{\setminus i}$ by the orthogonality principle. The dependency between $x_i$ and $\bbx_{\setminus i}$ (what specifies the incident edges to $i\in\ccalV$ in $\ccalG$) is thus entirely captured in the regression coefficients $\bbbeta^{(i)}\in\reals^{N-1}$, which are expressible in terms of the entries of $\bbTheta$ as
\begin{equation}\label{E:linear_predictor_precision}
\beta^{(i)}_j=-\frac{\theta_{ij}}{\theta_{ii}}.
\end{equation}
Importantly, \eqref{E:linear_predictor_precision} together with \eqref{E:par_corr_precision} reveal that a candidate edge $(i,j)$ belongs to $\ccalE:=\{(i,j)\in\ccalV\times \ccalV:\rho_{ij|\ccalV\setminus ij}\neq 0\}$ if and only if $\beta^{(i)}_j\neq 0$ (and also $\beta^{(j)}_i\neq 0$). Compactly, we have $\textrm{supp}(\bbbeta^{(i)}):=\{j:\beta^{(i)}_j\neq 0\}\equiv\ccalN_i$, which suggests casting the problem of Gaussian graphical model selection as one of sparse linear regression using observations $\ccalX:=\{\bbx_p\}_{p=1}^P$. 

The neighborhood-based lasso method in~\cite{meinshausen06} cycles over vertices $i=1,\ldots,N$ and estimates [cf. \eqref{E:linear_predictor}]
\begin{equation}\label{E:linear_predictor_lasso}
\hat{\ccalN}_i=\textrm{supp}(\hbbeta^{(i)}),\quad \textrm{ where }\quad\hbbeta^{(i)}\in\arg\min_{\bbbeta\in\reals^{N-1}}\left\{\sum_{p=1}^P(x_{pi}-\bbx_{p,\setminus i}^T\bbbeta)^2+\lambda\|\bbbeta\|_1\right\}.
\end{equation}
For finite data there is no guarantee that $\hat{\beta}^{(i)}_j\neq 0$ implies $\hat{\beta}^{(j)}_i\neq 0$ and vice versa, so the information in $\hat{\ccalN}_i$ and $\hat{\ccalN}_j$ should be combined to enforce symmetry. To declare an edge $(i,j)\in \ccalE$ the algorithm in~\cite{meinshausen06} requires that either $\hat{\beta}^{(i)}_j$ or $\hat{\beta}^{(j)}_i$ are nonzero (the OR rule), or alternatively consider the AND rule requiring that both coefficients be nonzero. Interestingly, for a judicious choice of $\lambda$ in \eqref{E:linear_predictor_lasso} and under suitable conditions on (possibly) $P\ll N$ as well as the sparsity of the ground-truth precision matrix $\bbTheta_0$, the graph can be consistently identified using either edge selection rule; see~\cite{meinshausen06} for the technical details. 

% % % % % % % % % % % % % % % % % % % % % % % % % % % % % % % % % % % % % % % %
%                        Subsection III-C                                     %
% % % % % % % % % % % % % % % % % % % % % % % % % % % % % % % % % % % % % % % %

 {
\subsection{Comparative summary}}
\label{ssec:comparison_stat}

\begin{figure*}[t]
	\centering
	\includegraphics[width=.7\linewidth]{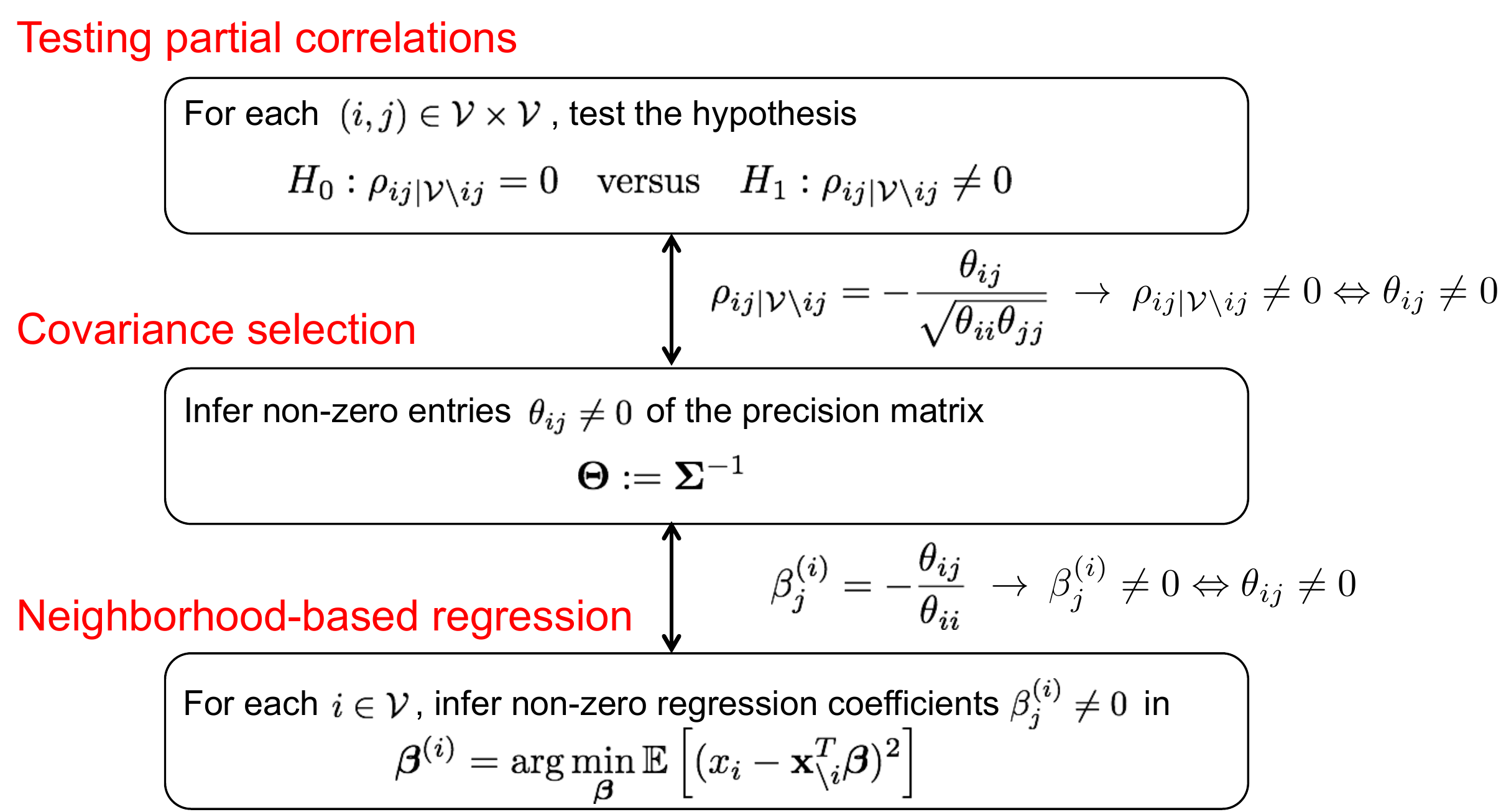}
	
	\caption{Conceptual roadmap for Gaussian graphical model selection with edge set $\ccalE:=\{(i,j)\in\ccalV\times \ccalV:\rho_{ij|\ccalV\setminus ij}\neq 0\}$. The problem of inferring nonzero partial correlations can be cast as one of covariance selection, since there is a bijection between the set of nonzero partial correlations (the edges of $\ccalG$) and the sparsity pattern of the precision matrix $\bbTheta$. Another equivalent approach to estimate the graphical model is to find the set of neighbors $\ccalN_i$ of each node $i\in \ccalV$ in the graph, by regressing $x_i$ against all the other variables in $\bbx_{\setminus i}$. The regression coefficients $\bbbeta^{(i)}$ are expressible in terms of the entries of the precision matrix $\bbTheta$, and it follows that $\textrm{supp}(\bbbeta^{(i)})=\ccalN_i$.}
	\label{F:roadmap_graphical_model_inference}
\end{figure*}

The estimator \eqref{E:linear_predictor_lasso} is  computationally appealing, since all $N$ lasso problems can be solved in parallel. Such a decomposability can be traced to the fact that the neighborhood-based approach relies on conditional likelihoods per vertex  {and does not enforce the PSD constraint $\bbTheta \succeq \mathbf{0}$}, whereas the graphical lasso maximizes a penalized version of the global likelihood $\ccalL(\bbTheta;\ccalX)=\log\det\bbTheta -\textrm{trace}(\hbSigma \bbTheta)$.  {For these reasons, the neighborhood-based lasso method in~\cite{meinshausen06} is computationally faster while the graphical lasso tends to be more (statistically) efficient~\cite{yuanlin2007}}. Another advantage of relying on neighborhood-based conditional likelihoods is that they yield tractable graph-learning approaches even for discrete or mixed graphical models, where computation of global likelihoods is generally infeasible. For binary $\bbx\in\{-1,+1\}^N$, an $\ell_1$-norm penalized logistic regression counterpart of \eqref{E:linear_predictor_lasso} was proposed for Ising model selection in~\cite{ravikumar2010}. To summarize and relate the approaches for Gaussian graphical model selection covered, Figure \ref{F:roadmap_graphical_model_inference} shows a schematic conceptual roadmap of this section. 

 {So far we have shown how to cast the graph topology identification problem as one of statistical inference, where modern topics such as multiple-testing, learning with sparsity, and high-dimensional model selection are prevalent. While fairly mature, the methods of this section may not be as familiar to the broad SP community and provide the needed historical context on the graph learning problem. Next, we shift gears to recent GSP-based topology inference frameworks that postulate the observed signals as either smoothly-varying or stationary with respect to the unknown graph (Sections \ref{sec:smooth} and \ref{sec:topoid_diffusion}, respectively).  To formalize these signal models, in the next section we introduce the required GSP background. In Section \ref{sec:discussion} we come full circle and offer a big picture summary of the new perspectives, benefits, and limitations of the GSP-based approaches relative to the statistical methods of this section.}

%%%%%%%%%%%%%%%   F   I   G   U   R    E   %%%%%%%%%%%%%%%%%%%%%%%%%%%%%%%%%%%%%%%%
%
%%%%%%%%%%%%%%%%%%%%%%%%%%%%%%%%%%%%%%%%%%%%%%%%%%%%%%%%%%%%%%%%%%%

% % % % % % % % % % % % % % % % % % % % % % % % % % % % % % % % % % % % % % %
%                         Section IV                                        %
% % % % % % % % % % % % % % % % % % % % % % % % % % % % % % % % % % % % % % %

\section{Graph signal processing foundations for graph learning advances}
\label{sec:GSP_foundations}

Here we review foundational GSP tools and concepts that have enabled recent topology inference advances, the subject of Sections \ref{sec:smooth} and \ref{sec:topoid_diffusion}. The graph Fourier transform (GFT), graph filter design, implementation and performance analysis, as well as structured signal models induced by graph smoothness or stationarity are all active areas of research on their own right, where substantial progress can be made.

% % % % % % % % % % % % % % % % % % % % % % % % % % % % % % % % % % % % % % % %
%                        Subsection IV-A                                      %
% % % % % % % % % % % % % % % % % % % % % % % % % % % % % % % % % % % % % % % %

\subsection{Graph Fourier transform and signal smoothness}
\label{ssec:GFT}

An instrumental GSP tool is the GFT, which decomposes a graph signal into orthonormal components describing different modes of variation with respect to the graph topology encoded in 
$\bbL$ (or an application-dictated graph-shift operator $\bbS$). The
GFT allows to equivalently represent a graph signal in two different domains -- the vertex domain consisting
of the nodes in $\ccalV$, and the graph frequency domain spanned by the spectral basis of \ccalG.
Therefore, signals can be manipulated in the frequency domain to induce different levels of interactions
between neighbors in the network; see Section \ref{ssec:diffusion} for more on graph filters.

To elaborate on this concept, consider the eigenvector decomposition of the combinatorial graph Laplacian $\bbL$ to define the GFT and the associated notion of graph frequencies. With $\bbLam:=\diag(\lam_1,\ldots,\lam_{N})$ denoting the diagonal matrix of non-negative Laplacian eigenvalues and $\bbV := [\bbv_1,\ldots,\bbv_{N}]$ the orthonormal matrix of eigenvectors, one can always decompose the symmetric graph Laplacian as $\bbL =  \bbV \bbLambda \bbV^T$. 

\begin{definition}[Graph Fourier transform]\normalfont 
The GFT of $\bbx$ with respect to the combinatorial graph Laplacian $\bbL$ is the signal $\tbx=[\tdx_1,\ldots,\tdx_{N}]^T$ defined as $\tbx = \bbV^T\bbx$. The inverse iGFT of $\tbx$ is given by $\bbx = \bbV \tbx$, which is a proper inverse by the orthogonality of $\bbV$.
\end{definition}

\vspace{.1cm}
\noindent The iGFT formula $\bbx = \bbV \tbx=\sum_{k=1}^{N} \tdx_k \bbv_k$ allows one to synthesize $\bbx$ as a sum of orthogonal frequency components $\bbv_k$. The contribution of $\bbv_k$ to the signal $\bbx$ is the GFT coefficient $\tdx_k$. The GFT encodes a notion of signal variability over the graph akin to the notion of frequency in Fourier analysis of temporal signals. To understand this analogy, define the total variation of the graph signal $\bbx$ with respect to the Laplacian $\bbL$ (also known as Dirichlet energy) as the following quadratic form
\begin{align}\label{eqn_TV_general}
	\text{TV}(\bbx) := \bbx ^T \bbL \bbx  = \sum_{i \neq j} W_{ij} (x_i-x_j)^2.
\end{align}
The total variation $\text{TV}(\bbx)$ is a smoothness measure, quantifying how much the signal $\bbx$ changes with respect to the presumption on variability that is encoded by the weights $\bbW$~\cite{EmergingFieldGSP,gsp2018tutorial}. Topology inference algorithms that search for graphs under which the observations are smooth is the subject of Section \ref{sec:smooth}.

Back to the GFT, consider the total variation of the eigenvectors $\bbv_k$, which is given by $ \text{TV}(\bbv_k) = \bbv_k^T \bbL \bbv_k = \lam_k$. It follows that the eigenvalues $0=\lam_1<\lam_2\leq\ldots\leq\lam_{N}$ can be viewed as graph frequencies, indicating how the eigenvectors (i.e., frequency components) vary over the graph $\ccalG$. Accordingly, the GFT and iGFT offer a decomposition of the graph signal $\bbx$ into spectral components that characterize different levels of variability. To the extent that $\ccalG$ is a good representation of the relationship between the components of $\bbx$, the GFT can be used to process $\bbx$ by e.g., exploiting  sparse or low-dimensional representations $\tbx$ in the graph frequency domain.  For instance, Section \ref{ssec:US_economy} illustrates how the disaggregated gross domestic product signal supported on a graph of United States (US) economic sectors can be accurately represented with a handful of GFT coefficients.

So far the discussion has focused on the GFT for symmetric graph Laplacians $\bbL$ associated with undirected graphs. However, the GFT can be defined in more general contexts where the interpretation of components as different modes of variability is not as clean and Parseval's identity may not hold, but its value towards yielding parsimonious spectral representations of network processes remains~\cite{gsp2018tutorial}. Consider instead a (possibly asymmetric) graph-shift operator $\bbS$ which is assumed to be diagonalizable as $\bbS=\bbV \bbLambda \bbV^{-1}$, and redefine the GFT as $\tbx = \bbV^{-1}\bbx$; otherwise one can consider Jordan decompositions~\cite{DSP_freq_analysis}. Allowing for generic graph-shift operators reveals the encompassing nature of the GFT relative to the time-domain discrete Fourier transform (DFT), the multidimensional DFT, and the  {Karhunen-Lo\`{e}ve transform (KLT) [also known as the principal component analysis (PCA) transform in statistics and data analysis]}; see \emph{Encompassing nature of the graph Fourier Transform}. The GFT offers a unifying framework that subsumes all the aforementioned transforms for specific graphs, while it also offers a natural representation to work with signals of increasingly complex structure. 

\begin{mdframed}[hidealllines=true,backgroundcolor=gray!20]
\textbf{Encompassing nature of the graph Fourier transform.} Discrete-time, space, and correlated signals can be reinterpreted as graph signals supported on particular graphs.
\begin{center}
	\vspace*{-0.5cm}

\def \thisplotscale {1.2}

\def \unit {\thisplotscale cm}

\tikzstyle {vertex} = [circle, 
                       draw,
                       minimum width = 0.5*\unit,
                       minimum height = 0.5*\unit,
                       anchor=center,
                       font=\scriptsize]
\tikzstyle {light}  = [opacity = 0.1]
\tikzstyle{bigvertex} = [vertex, 
                        minimum width  = 0.7*\unit,
                        minimum height = 0.7*\unit]

{\footnotesize \begin{tikzpicture}[x = 1*\unit, y = 1*\unit]
    
    %%%%%%%%%%%%%%%%%%%%%%%%%%%%%%%%%%%%%%%%%%%%%%%%%%%%%%%%%%%%%%%%%%%%%%%%%%%%%
    %%% Signal defined on top of the directed cycle graph with six nodes
	% Draw nodes
	\node at (0, 0) (center1) {};
	\path (center1) ++ (  90:1.3) node (1) [fill = blue!20, vertex] {0};
	\path (center1) ++ (  30:1.3) node (2) [fill = blue!20, vertex] {1};
	\path (center1) ++ (- 30:1.3) node (3) [fill = blue!20, vertex] {2};
	\path (center1) ++ (- 90:1.3) node (4) [fill = blue!20, vertex] {3};
	\path (center1) ++ (-150:1.3) node (5) [fill = blue!20, vertex] {4};
	\path (center1) ++ ( 150:1.3) node (6) [fill = blue!20, vertex] {5};
	% Draw signal components 
	\path (1) ++ ( 0.4, 0.35) node [] {$x_0$};
	\path (2) ++ ( 0.4, 0.35) node [] {$x_1$};
	\path (3) ++ ( 0.4, 0.35) node [] {$x_2$};
	\path (4) ++ ( 0.4,-0.35) node [] {$x_3$};
	\path (5) ++ (-0.4, 0.35) node [] {$x_4$};
	\path (6) ++ (-0.4, 0.35) node [] {$x_5$};
	% Draw edges
	\path[-stealth] (1) edge [bend left=20, above] node {} (2);		
	\path[-stealth] (2) edge [bend left=20, above] node {} (3);		
	\path[-stealth] (3) edge [bend left=20, above] node {} (4);		
	\path[-stealth] (4) edge [bend left=20, above] node {} (5);		
	\path[-stealth] (5) edge [bend left=20, above] node {} (6);
	\path[-stealth] (6) edge [bend left=20, above] node {} (1);

    %%%%%%%%%%%%%%%%%%%%%%%%%%%%%%%%%%%%%%%%%%%%%%%%%%%%%%%%%%%%%%%%%%%%%%%%%%%%%
    %%% Signal defined on top of regular grid
	% Draw nodes
    \node at (3.5, 1.21) (center1) {};
	\path (center1) ++ ( 0*1.42,  0*1.21) node (00) [fill = blue!20, vertex] {00};
	\path (center1) ++ ( 1*1.42,  0*1.21) node (01) [fill = blue!20, vertex] {01};
	\path (center1) ++ ( 2*1.42,  0*1.21) node (02) [fill = blue!20, vertex] {02};

	\path (center1) ++ ( 0*1.42, -1*1.21) node (10) [fill = blue!20, vertex] {10};
	\path (center1) ++ ( 1*1.42, -1*1.21) node (11) [fill = blue!20, vertex] {11};
	\path (center1) ++ ( 2*1.42, -1*1.21) node (12) [fill = blue!20, vertex] {12};

	\path (center1) ++ ( 0*1.42, -2*1.21) node (20) [fill = blue!20, vertex] {20};
	\path (center1) ++ ( 1*1.42, -2*1.21) node (21) [fill = blue!20, vertex] {21};
	\path (center1) ++ ( 2*1.42, -2*1.21) node (22) [fill = blue!20, vertex] {22};

%	\path (center1) ++ ( 0*1.0, -1*1.0) node (01) [fill = blue!20, vertex] {10};
		
	% Draw signal components 
	\path (00) ++ (-0.4, 0.35) node [] {$x_{00}$};
	\path (01) ++ (-0.4, 0.35) node [] {$x_{01}$};
	\path (02) ++ (-0.4, 0.35) node [] {$x_{02}$};

	\path (10) ++ (-0.4, 0.35) node [] {$x_{10}$};
	\path (11) ++ (-0.4, 0.35) node [] {$x_{11}$};
	\path (12) ++ (-0.4, 0.35) node [] {$x_{12}$};

	\path (20) ++ (-0.4, 0.35) node [] {$x_{20}$};
	\path (21) ++ (-0.4, 0.35) node [] {$x_{21}$};
	\path (22) ++ (-0.4, 0.35) node [] {$x_{22}$};

	% Draw horizontal edges
    \path[stealth-]         (00) edge ++ (-0.5,0);
	\path[stealth-stealth ] (00) edge (01);
	\path[stealth-stealth ] (01) edge (02);
    \path[stealth-]         (02) edge ++ ( 0.5,0);

    \path[stealth-]         (10) edge ++ (-0.5,0);
	\path[stealth-stealth ] (10) edge (11);
	\path[stealth-stealth ] (11) edge (12);
    \path[stealth-] (12) edge ++ (0.5,0);

    \path[stealth-]         (20) edge ++ (-0.5,0);
	\path[stealth-stealth ] (20) edge (21);
	\path[stealth-stealth ] (21) edge (22);
    \path[stealth-] (22) edge ++ (0.5,0);

	% Draw vertical edges
    \path[stealth-] (00) edge ++ (0, 0.5);
	\path[stealth-] (01) edge ++ (0, 0.5);
	\path[stealth-] (02) edge ++ (0, 0.5);

	\path[stealth-stealth ] (00) edge (10);
	\path[stealth-stealth ] (01) edge (11);
	\path[stealth-stealth ] (02) edge (12); 

	\path[stealth-stealth ] (10) edge (20);
	\path[stealth-stealth ] (11) edge (21);
	\path[stealth-stealth ] (12) edge (22); 

    \path[stealth-] (20) edge ++ (0,-0.5);
	\path[stealth-] (21) edge ++ (0,-0.5);
	\path[stealth-] (22) edge ++ (0,-0.5);
	
    %%%%%%%%%%%%%%%%%%%%%%%%%%%%%%%%%%%%%%%%%%%%%%%%%%%%%%%%%%%%%%%%%%%%%%%%%%%%%
    %%% Signal defined on top of a correlation matrix with five elements
	% Draw nodes
	\node at (9.8, 0) (center1) {};
	\path (center1) ++ ( 135:1.401) node (1) [fill = blue!20, vertex] {0};
	\path (center1) ++ (  45:1.401) node (2) [fill = blue!20, vertex] {1};
	\path (center1) ++ (- 45:1.401) node (3) [fill = blue!20, vertex] {2};
	\path (center1) ++ (-135:1.401) node (4) [fill = blue!20, vertex] {3};
	% Draw signal components 
	\path (1) ++ (-0.4, 0.35) node [] {$x_0$};
	\path (2) ++ ( 0.4, 0.35) node [] {$x_1$};
	\path (3) ++ ( 0.4, 0.35) node [] {$x_2$};
	\path (4) ++ (-0.4, 0.35) node [] {$x_3$};
	% Draw edges
	\path[-stealth] (1) edge [bend left,  above] node {$\sigma_{01}$} (2);
	\path[-stealth] (2) edge [bend left,  right] node {$\sigma_{12}$} (3);
	\path[-stealth] (3) edge [bend left,  below] node {$\sigma_{23}$} (4);
	\path[-stealth] (4) edge [bend left,  left ] node {$\sigma_{30}$} (1);
	\path[-stealth] (1) edge [bend right, above] node {$\sigma_{10}$} (2);
	\path[-stealth] (2) edge [bend right, right] node {$\sigma_{21}$} (3);
	\path[-stealth] (3) edge [bend right, below] node {$\sigma_{32}$} (4);
	\path[-stealth] (4) edge [bend right, left ] node {$\sigma_{03}$} (1);
	% Draw self loops
	\path[-stealth] (1) edge [loop above] node {$\sigma_{00}$} (1);		
	\path[-stealth] (2) edge [loop above] node {$\sigma_{11}$} (2);	
	\path[-stealth] (3) edge [loop below] node {$\sigma_{22}$} (3);	
	\path[-stealth] (4) edge [loop below] node {$\sigma_{33}$} (4);

\end{tikzpicture}} 
\end{center}
A (periodic) discrete-time signal can be viewed as a graph signal supported on the directed cycle, whose circulant adjacency matrix is diagonalized by the DFT basis (left). A spatial signal such as an image can be thought of as a graph signal supported on a regular lattice (middle). A  {zero-mean} correlated signal can be interpreted as a graph signal supported on the covariance graph,   {where $\bbS=\bbSigma=\E{\bbx \bbx^T}$ is diagonalized by the orthogonal basis of principal components} (right). Respective GFTs reduce to the time-domain DFT, the multidimensional DFT, and the  {KLT} or PCA transform.
\end{mdframed}

% % % % % % % % % % % % % % % % % % % % % % % % % % % % % % % % % % % % % % % %
%                        Subsection IV-B										                                     %
% % % % % % % % % % % % % % % % % % % % % % % % % % % % % % % % % % % % % % % %

\subsection{Graph filters as models of network diffusion}
\label{ssec:diffusion}

Here we introduce a fairly general class of linear network diffusion processes on the graph $\ccalG$ with shift operator $\bbS$. Specifically, let $\bby$ be a graph signal supported on $\ccalG$, which is generated from an input graph signal $\bbx$ via linear network dynamics of the form
\begin{align}\label{eqn_diffusion}
	%\textstyle
	 \bby\  =\ \alpha_0 \prod_{l=1}^{\infty} (\bbI-\alpha_l \bbS) \bbx
	\  =\ \sum _{l=0}^{\infty} \beta_l \bbS^l \bbx .
	%   \ :=\  \bbH \bbw.
\end{align}
While $\bbS$ encodes only one-hop interactions, each successive application of the shift in \eqref{eqn_diffusion} diffuses $\bbx$ over $\ccalG$. The product and sum representations in \eqref{eqn_diffusion} are common (and equivalent) models for the generation of linear network processes. Indeed, any process that can be understood as the linear propagation of a seed signal through a static graph can be written in the form in \eqref{eqn_diffusion}, and subsumes heat diffusion, consensus, and the classic DeGroot model of opinion dynamics as special cases.

The diffusion expressions in \eqref{eqn_diffusion} are polynomials in $\bbS$ of possibly infinite degree, yet the Cayley-Hamilton theorem asserts that they are equivalent to polynomials of degree smaller than $N$. This is intimately related to the concept of (linear shift-invariant) graph filter. Specifically, upon defining the vector of coefficients $\bbh:=[h_0,\ldots,h_{L-1}]^T$, a graph filter is defined as
\begin{equation}\label{E:Filter_input_output_time}
	\mathbf{H}:=h_0\bbI+h_1 \mathbf{S}+h_2 \mathbf{S}^2+\ldots+h_{L-1} \mathbf{S}^{L-1}=\sum_{l=0}^{L-1}h_l \mathbf{S}^l.
\end{equation}
Hence, one has that the signal model in \eqref{eqn_diffusion} can be rewritten as
$\bby  = \big(\sum_{l=0}^{L-1}h_l \bbS^l\big)\,\bbx
:= \bbH \bbx$, for some particular $\bbh$ and $L\leq N$. Due to the local structure of $\bbS$, graph filters represent linear transformations that can be implemented in a distributed fashion, e.g., with
$L-1$ successive exchanges of information among neighbors. Since $\bbH$ is a polynomial in $\bbS$, graph filters have the same eigenvectors as the shift. This implies that $\bbH$ and $\bbS$ commute and hence graph filters represent shift-invariant transformations~\cite{SandryMouraSPG_TSP13}. 

Leveraging the spectral decomposition of $\bbS$, graph filters can be represented in the frequency domain.
Specifically, let us use the eigenvectors of $\bbS$ to define the GFT matrix $\bbU:=\bbV^{-1}$, and the eigenvalues $\lambda_i$ of $\bbS$ to define the $N\times L$ Vandermonde matrix  $\bbPsi$, where $\Psi_{ij}:=(\lambda_i)^{j-1}$.
The frequency representation of a filter $\bbh$ is defined as $\tbh:=\bbPsi\bbh$, since the output $\bby\!=\!\bbH\bbx$ of a graph filter in the frequency domain is 
\begin{equation}\label{E:Filter_frequency_domain}
\tby=\diag\big(\bbPsi\bbh\big)\bbU \bbx=\diag\big(\tbh\big)\tbx=\tbh\circ \tbx.
\end{equation}
This identity can be seen as a counterpart of the convolution theorem for temporal signals, where $\tby$ is the elementwise product of $\tbx$ and the filter's frequency response $\tbh$. To establish further connections with the time domain, recall the directed cycle graph with adjacency matrix $\bbW_{dc}$. If $\bbS=\bbW_{dc}$, one finds that: i) $\bby=\bbH\bbx$ can be found as the circular convolution of $\bbh$ and $\bbx$; and ii) both $\bbU$ and $\boldsymbol{\Psi}$ correspond to the DFT matrix. While in the time domain $\bbU=\boldsymbol{\Psi}$, this is not true for general (non-circulant) graphs.

% % % % % % % % % % % % % % % % % % % % % % % % % % % % % % % % % % % % % % % %
%                        Subsection IV-C								                                         %
% % % % % % % % % % % % % % % % % % % % % % % % % % % % % % % % % % % % % % % %

\subsection{Stationary graph processes}
\label{ssec:stationarity}
Having introduced the notions of graph-shift operator, GFT and graph filter, we review here how to use those to characterize a particular class of random graph signals  {(often referred to as graph processes, meaning collections of vertex-indexed random variables)}. In classical SP, stationarity is a fundamental property that facilitates the (spectral) analysis and processing of random signals, by requiring that the data-generation mechanisms (i.e., the joint probability distributions) are invariant to time shifts. Due to the intrinsic irregularity of the graph domain and the associated challenges of defining translation operators, extending the notion of  stationarity to random graph signals is no easy task~\cite{girault2015stationary, EPFL16stationary,marques2017stationary}. 

 {Stationary graph processes  were first defined and analyzed in~\cite{girault2015stationary}. The fundamental problem identified therein is that graph-shift operators do not preserve energy in general and therefore they cannot be isometric. This hurdle is overcome with the definition of an isometric graph shift that preserves the eigenvector space of the Laplacian, but modifies its eigenvalues~\cite{girault2015isometric}. A stationary graph process is then defined as one whose probability distributions are invariant with respect to multiplications with the isometric shift. It is further shown that this definition requires the covariance matrix of the signal to be diagonalized by the eigenvectors of the graph shift, which by construction are also the eigenvectors of the isometric shift. This implies the existence of a graph power spectral density with components given by the covariance eigenvalues. The requirement of having a covariance matrix diagonalizable by the eigenvectors of the Laplacian is itself adopted as a definition in~\cite{EPFL16stationary}, where the requirement is shown to be equivalent to statistical
invariance with respect to the non-isometric translation operator introduced in~\cite{shuman2016translation}.} These ideas are further refined in \cite{marques2017stationary} and extended to general normal (not necessarily Laplacian) graph-shift operators.

Following the approach in~\cite{marques2017stationary}, here we present two (equivalent) definitions of weak stationarity for zero-mean graph signals. We then discuss briefly some of their implications in the context of network topology identification, paving the way for the approaches surveyed in Section \ref{sec:topoid_diffusion}. To that end, we define a standard zero-mean white random graph  {signal} $\bbw$ as one with mean $\E{\bbw}=\bbzero$ and covariance $\bbSigma_{\bbw}:=\E{\bbw\bbw^H} = \bbI$.

\begin{definition}[Weak stationarity - Filtering characterization]\label{D:WeaklyStionaryGraphProcess_0}\normalfont
	Given a normal shift operator $\bbS$, a random  {graph signal} $\bbx$ is weakly stationary with respect to $\bbS$ if it can be written as the response of a linear shift-invariant graph filter $\bbH\!=\!\sum_{l=0}^{N-1} h_l\bbS^l$ to a white input $\bbw$, that is
\begin{equation}\label{E:output_filter}	
\bbx=\sum_{l=0}^{N-1} h_l\bbS^l\bbw=\bbH\bbw.
\end{equation}
\end{definition}

The definition states that stationary graph processes can be written as the output of graph filters when excited with a white input. This generalizes the well-known fact that stationary processes in time can be expressed as the output of linear time-invariant systems driven by white noise. Starting from \eqref{E:output_filter}, the covariance matrix  $\bbSigma_\bbx= \E{\bbx\bbx^H}$ of the  {random vector} $\bbx$ is given by
\begin{equation}\label{E:cov_output_filter}
\bbSigma_\bbx = \E{\bbH\bbw(\bbH\bbw)^H}
= \bbH\bbSigma_{\bbw}\bbH^H
= \bbH\bbH^H,
\end{equation}
which shows that the correlation structure of $\bbx$ is determined by the filter $\bbH$. We can think of Definition~\ref{D:WeaklyStionaryGraphProcess_0} as a constructive definition of stationarity since it describes how a stationary process can be generated. 

Alternatively, one can define stationarity from a descriptive perspective, by imposing requirements on the second-order moment of the random graph  {signal} in the frequency domain.

\begin{definition}[Weak stationarity - Spectral characterization]\label{D:WeaklyStionaryGraphProcess_1}
	%\normalfont
	Given a normal shift operator $\bbS$, a random  {graph signal} $\bbx$ is weakly stationary with respect to $\bbS$ if  $\bbSigma_\bbx$ and $\bbS$ are simultaneously diagonalizable.		
\end{definition}

The second definition characterizes stationarity from a graph frequency perspective by requiring the covariance $\bbSigma_\bbx$ to be diagonalized by the GFT basis $\bbV$. When particularized to time by letting $\bbS=\bbW_{dc}$, Definition~\ref{D:WeaklyStionaryGraphProcess_1} requires $\bbSigma_\bbx$ to be diagonalized by the DFT matrix and, therefore, $\bbSigma_\bbx$ must be circulant.  Definitions \ref{D:WeaklyStionaryGraphProcess_0} and \ref{D:WeaklyStionaryGraphProcess_1} are equivalent in time. For the equivalence to hold also in the graph domain, it is possible to show one only needs $\bbS$ to be normal and with eigenvalues all distinct~\cite{marques2017stationary}. Moreover, it follows from Definition \ref{D:WeaklyStionaryGraphProcess_1} that stationary graph processes are also characterized by a power spectral density. In particular, given a  random  {vector} $\bbx$ that is stationary with respect to $\bbS=\bbV \bbLambda \bbV^H$, the \textit{power spectral density} of such a  {vertex-indexed} process is the vector $\bbp\in\reals_{+}^{N}$ defined as
\begin{equation}\label{E:PSD_stat_graph_process}
	\bbp := \diag\left(\bbV^H \bbSigma_\bbx \bbV\right).
	\end{equation}

Note that since $\bbSigma_\bbx$ is diagonalized by $\bbV$ (see Definition~\ref{D:WeaklyStionaryGraphProcess_1}) the matrix $\bbV^H \bbSigma_\bbx \bbV$ is diagonal, and it follows that the power spectral density in \eqref{E:PSD_stat_graph_process} corresponds to the (non-negative) eigenvalues of the covariance matrix $\bbSigma_\bbx \succcurlyeq \mathbf{0}$. Thus, \eqref{E:PSD_stat_graph_process} is equivalent to $\bbSigma_\bbx = \bbV \diag(\bbp) \bbV^H$. The latter identity shows that stationarity reduces the degrees of freedom of a random graph process --the symmetric matrix $\bbSigma_{\bbx}$ has $N(N+1)/2$ degrees of freedom, while $\bbp$ has only $N$--, thus facilitating its description and estimation. 

In closing, several remarks are in order. First, note that stating that a \emph{graph} process is stationary is an inherently incomplete assertion, since we need to specify which graph we are referring to. Hence, the proposed definitions depend on the graph-shift operator, so that a  {random vector} $\bbx$ can be stationary in $\bbS$ but not in $\bbS' \neq \bbS$. White noise is, on the other hand, an example of a random  {graph signal} that is stationary with respect to \emph{any} graph shift $\bbS$. A second, albeit related, observation is that, by definition, any  {random vector} $\bbx$ is stationary with respect to the shift given by the covariance matrix $\bbS=\bbSigma_{\bbx}$. The same is true for the precision matrix $\bbS=\bbSigma_{\bbx}^{-1}$. These facts will be leveraged in Section \ref{sec:topoid_diffusion}, to draw connections between stationary graph signal-based topology inference approaches and some of the classical statistical methods reviewed in Section \ref{sec:stat}. Third, notice that the stationarity requirement is tantamount to the covariance of the process being a polynomial in the graph-shift operator. Accordingly, under stationarity we ask the mapping from $\bbS$ to the covariance $\bbSigma_{\bbx}$ to be smooth (an analytic function), which is a natural and intuitively pleasing requirement~\cite{AsilomarMultipleShifts17}. 

 {Section \ref{sec:topoid_diffusion} will delve into this last issue when presenting models for topology inference based on graph stationary observations. But before that, in the next section we consider graph learning as an inverse problem subject to a signal smoothness prior.}

%Fourth, and last, it is often the case that one collects observations from a system periodically, giving rise to a sequence $\bbX=[\bbx_1,...,\bbx_T]\in \reals^{N\times T}$ of time-indexed  {random} graph signals. This will be naturally relevant in the context of dynamic network topology identification, or, when seeking to infer (directional) causal relations among a collection of $N$ observable variables. For such measurements, one can state conditions under which the process is stationary both with respect to the graph (each of the $T$ vertex-indexed columns of $\bbX$) as well as with respect to time (each of the $N$ time-indexed rows of $\bbX$); see e.g.,~\cite{perraudin2017towards}. % and \cite{ChapterBookStationarity}.

% % % % % % % % % % % % % % % % % % % % % % % % % % % % % % % % % % % % % % % %
%                        Section V                                     %
% % % % % % % % % % % % % % % % % % % % % % % % % % % % % % % % % % % % % % % %

\section{Learning graphs from observations of smooth signals}
\label{sec:smooth}

In various GSP applications it is desirable to construct a graph on which network data admit certain regularity. Accordingly, in this section we survey a family of topology identification approaches that deal with the following general problem. Given a set $\ccalX:=\{\bbx_p\}_{p=1}^P$ of possibly noisy graph signal observations, the goal is to learn a graph $\ccalG(\ccalV,\ccalE, \bbW)$ with $|\ccalV|=N$ nodes such that the observations in $\ccalX$ are smooth on $\ccalG$. Recall that a graph signal is said to be smooth if the values associated with vertices incident to edges with large weights in the graph tend to be similar. As discussed in Section \ref{ssec:GFT}, the so-defined smoothness of a signal can be quantified by means of a total variation measure given by the Laplacian quadratic form in \eqref{eqn_TV_general}. Such a measure offers a natural criterion to search for the best topology (encoded in the entries of the Laplacian), which endows the signals in $\ccalX$ with the desired smoothness property.

There are several reasons that motivate this graph learning paradigm. First, smooth signals admit low-pass band-limited (i.e., sparse) representations using the GFT basis [cf. the discussion following \eqref{eqn_TV_general}]. From this vantage point, the graph learning problem can be equivalently viewed as one of finding efficient information processing transforms for graph signals. Second, smoothness is a cornerstone property at the heart of several graph-based statistical learning tasks including nearest-neighbor prediction (also known as graph smoothing), denoising, semi-supervised learning, and spectral clustering.  {The success of these methods hinges on the fact that many real-world graph signals are smooth. This should not come as a surprise when graphs are constructed based on similarities between nodal attributes (i.e., signals), or when the network formation process is driven by mechanisms such as homophily or proximity in some latent space. Examples of smooth graph signals include natural images~\cite{Kalofolias2016inference_smoothAISTATS16}, average annual temperatures recorded by meteorological stations~\cite{sundeep_icassp17}, type of practice in a network of lawyer collaborations~\cite[Ch. 8]{kolaczyk2009book}, and product ratings supported over similarity graphs of items or consumers~\cite{weiyu2018matrixcompletion_tsp}, just to name a few.}

% % % % % % % % % % % % % % % % % % % % % % % % % % % % % % % % % % % % % % % %
%                        Subsection V-A                                       %
% % % % % % % % % % % % % % % % % % % % % % % % % % % % % % % % % % % % % % % %

\subsection{Laplacian-based factor analysis model and graph kernel regression}
\label{ssec:smooth_factor_kernel}

A factor analysis-based approach was advocated in~\cite{DongLaplacianLearning} to estimate graph Laplacians, seeking that input
graph signals are smooth over the learned topologies. Specifically, let $\bbL=\bbV\bbLambda\bbV^T$ be the eigendecomposition of the combinatorial Laplacian associated with an unknown, undirected graph $\ccalG(\ccalV,\ccalE,\bbW)$ with $N=|\ccalV|$ vertices. The observed graph signal $\bbx\in\reals^N$ is assumed to have zero mean for simplicity, and adheres to the following graph-dependent factor analysis model
\begin{equation}\label{E:factor_model}
\bbx = \bbV \bbchi +\bbepsilon,
\end{equation}
where the factors are given by the  Laplacian eigenvectors $\bbV$, $\bbchi\in\reals^N$ represents latent variables or factor loadings, and $\bbepsilon\sim \textrm{Normal}(\mathbf{0},\sigma^2\bbI)$ is an isotropic error term. Adopting $\bbV$ as the representation matrix is well motivated, since Laplacian eigenvectors comprise the GFT basis -- a natural choice for synthesizing graph signals as explained in Section \ref{ssec:GFT}. Through this lens the latent variables in \eqref{E:factor_model} can be interpreted as GFT coefficients. Moreover, the Laplacian eigenvectors offer a spectral embedding of the graph vertices, that is often useful for higher-level network analytic tasks such as data visualization, clustering, and community detection.  The representation matrix establishes a first link between the signal model and the graph topology. 
The second one comes from the adopted latent variables' prior distribution $\bbchi\sim\textrm{Normal}(\mathbf{0},\bbLambda^{\dagger})$, where $^{\dagger}$ denotes pseudo-inverse and thus the precision matrix is defined as the eigenvalue matrix $\bbLambda$ of the Laplacian. From the GFT interpretation of \eqref{E:factor_model}, it follows that the prior on $\bbchi$ encourages low-pass bandlimited $\bbx$. Indeed, the mapping $\bbLambda \rightarrow \bbLambda^\dagger$ translates the large eigenvalues of the Laplacian (those associated with high frequencies) to low-power factor loadings in $\bbchi$. On the other hand, small eigenvalues associated with low frequencies are translated to high-power factor loadings -- a manifestation of the model imposing a smoothness prior on $\bbx$.

Given the observed signal $\bbx$, the maximum a posteriori (MAP) estimator of the latent variables is given by ($\sigma^2$ is henceforth assumed known and absorbed into the parameter $\alpha >0$)
\begin{equation}\label{E:h_map}
\hbh_{\textrm{MAP}} = \arg\min_{\bbchi}\left\{\|\bbx-\bbV\bbchi\|^2+\alpha \bbchi^T\bbLambda \bbchi\right\},
\end{equation}
which is of course parameterized by the unknown eigenvectors and eigenvalues of the Laplacian. With $\bby:=\bbV\bbchi$ denoting the predicted graph signal (or error-free representation of $\bbx$), it follows that [cf. \eqref{eqn_TV_general}]
\begin{equation}\label{E:TV_regularization}
\bbchi^T\bbLambda \bbchi=\bby^T\bbV\bbLambda \bbV^T\bby=\bby^T\bbL\bby=\text{TV}(\bby).
\end{equation}
Consequently, one can interpret the MAP estimator \eqref{E:h_map} as a Laplacian-based total variation denoiser of $\bbx$, which effectively imposes a smoothness prior on the recovered signal $\bby=\bbV\bbchi$. One can also view \eqref{E:h_map} as a kernel ridge-regression estimator with (unknown) Laplacian kernel $\bbK:=\bbL^{\dagger}$~\cite[Ch. 8.4.1]{kolaczyk2009book}. 

Building on \eqref{E:h_map} and making the graph topology an explicit variable in the optimization, the idea is to jointly search for the graph Laplacian $\bbL$ and
a denoised representation $\bby = \bbV\bbchi$ of $\bbx$, thus solving 
\begin{equation}\label{E:denoiser}
\min_{\bbL,\bby}\left\{\|\bbx-\bby\|^2+\alpha \bby^T\bbL \bby\right\}.
\end{equation}
The objective function of \eqref{E:denoiser} encourages both: i) data fidelity through a quadratic loss penalizing discrepancies between $\bby$ and the observation $\bbx$; and ii) smoothness on the learned graph via total-variation regularization. Given data in the form of multiple independent observations $\ccalX:=\{\bbx_p\}_{p=1}^P$ that we collect in the matrix $\bbX=[\bbx_1,\ldots,\bbx_P]\in \reals^{N\times P}$, the approach of~\cite{DongLaplacianLearning} is to solve 
\begin{align}\label{E:dong_Laplacian_learning}
\min_{\bbL,\bbY}&{}\left\{\|\bbX-\bbY\|_F^2+\alpha\textrm{trace}\left(\bbY^T\bbL\bbY\right)+\frac{\beta}{2}\|\bbL\|_F^2\right\}\\
\textrm{ s. to } &{} \quad\textrm{trace}(\bbL)=N,\:\bbL\mathbf{1}=\mathbf{0},\: L_{ij}=L_{ji}\leq 0, \:i\neq j,\nonumber
\end{align}
which imposes constraints on $\bbL$ so that it qualifies as a valid combinatorial Laplacian. Notably, $\textrm{trace}(\bbL)=N$ avoids the trivial all-zero solution and essentially fixes the $\ell_1$-norm of $\bbL$. To control the sparsity of the resulting graph, a Frobenius-norm penalty is added to the objective of \eqref{E:dong_Laplacian_learning} to shrink its edge weights. The trade-off between data fidelity, smoothness and sparsity is controlled  via the positive regularization parameters $\alpha$ and $\beta$.

While not jointly convex in $\bbL$ and $\bbY$, \eqref{E:dong_Laplacian_learning} is bi-convex meaning that for fixed $\bbL$ the resulting problem with respect to $\bbY$ is convex, and vice versa. Accordingly, the algorithmic approach of~\cite{DongLaplacianLearning} relies on alternating minimization, a procedure that converges to a stationary point of \eqref{E:dong_Laplacian_learning}. For fixed $\bbY$,  \eqref{E:dong_Laplacian_learning} reduces to a quadratic program (QP) subject to linear constraints, which can be solved via interior point methods. For large graphs, scalable alternatives include the alternating-direction method of multipliers (ADMM), or, primal-dual solvers of the reformulation described in the following section; see \eqref{eq:dong_as_kalofolias}. For fixed $\bbL$, the resulting problem is a matrix-valued counterpart of \eqref{E:denoiser}. The solution is given in closed form as $\bbY=(\bbI+\alpha\bbL)^{-1}\bbX$, which represents a low-pass, graph filter-based smoother of the signals in $\bbX$.

% % % % % % % % % % % % % % % % % % % % % % % % % % % % % % % % % % % % % % % %
%                        Subsection V-B                                       %
% % % % % % % % % % % % % % % % % % % % % % % % % % % % % % % % % % % % % % % %

\subsection{Signal smoothness meets edge sparsity}
\label{ssec:smooth_sparse}

An alternative approach to the problem of learning graphs under a smoothness prior was proposed in~\cite{Kalofolias2016inference_smoothAISTATS16}.  Recall the data matrix $\bbX=[\bbx_1,\ldots,\bbx_P]\in \reals^{N\times P}$, and let $\bar{\bbx}_i^T\in\reals^{1\times P}$ denote its $i$-th row collecting those $P$ measurements at vertex $i$. The key idea in~\cite{Kalofolias2016inference_smoothAISTATS16} is to establish a \emph{link between smoothness and sparsity}, revealed through the identity
\begin{equation}\label{E:smooth_sparse}
\sum_{p=1}^P\textrm{TV}(\bbx_p)=\textrm{tr}(\bbX^T\bbL\bbX)=\frac{1}{2}\|\bbW\circ\bbZ\|_1,
\end{equation}
where the Euclidean-distance matrix $\bbZ\in\reals_{+}^{N\times N}$ has entries $Z_{ij}:=\|\bar{\bbx}_i-\bar{\bbx}_j\|^2$, $i,j\in\ccalV$. The intuition is that when the given distances in $\bbZ$ come from a smooth manifold, the corresponding graph has a sparse edge set, with preference given to those edges $(i,j)$ associated with smaller distances $Z_{ij}$. Identity \eqref{E:smooth_sparse} offers an advantageous way of parameterizing graph learning formulations under smoothness priors, because the space of adjacency matrices  can be described via simpler (meaning entry-wise decoupled) constraints relative to its Laplacian counterpart. It also reveals that once a smoothness penalty is included in the criterion to search for $\ccalG$, adding an extra sparsity-inducing regularization is essentially redundant.

Given these considerations, a general purpose model for learning graphs is advocated in~\cite{Kalofolias2016inference_smoothAISTATS16}, namely 
\begin{align}\label{eq:kalofolias}
\min_{\bbW}&{}\left\{\|\bbW\circ\bbZ\|_1-\alpha\mathbf{1}^T\log(\bbW\mathbf{1})+\frac{\beta}{2}\|\bbW\|_F^2\right\}\\
\textrm{ s. to } &{} \quad\textrm{diag}(\bbW)=\mathbf{0},\: W_{ij}=W_{ji}\geq 0, \:i\neq j,\nonumber
\end{align}
where $\alpha,\beta$ are tunable regularization parameters. Unlike~\cite{DongLaplacianLearning}, the logarithmic barrier on the vector $\bbW\mathbf{1}$ of nodal degrees enforces each vertex to have at least one incident edge.  The Frobenius-norm regularization on the adjacency matrix $\bbW$ controls the graph's edge sparsity pattern by penalizing larger edge weights. Overall, this combination enforces degrees to be positive but does not prevent most individual edge weights from becoming zero. The sparsest graph is obtained when $\beta=0$, and edges form preferentially between nodes with smaller $Z_{ij}$, similar to a $1$-nearest neighbor graph.

The convex optimization problem \eqref{eq:kalofolias} can be solved efficiently with complexity $O(N^2)$ per iteration, by leveraging provably-convergent primal-dual solvers amenable to parallelization. The optimization framework is quite general and it can be used to scale other related state-of-the-art graph learning problems. For instance,  going back to the alternating minimization algorithm in Section \ref{ssec:smooth_factor_kernel}, recall that the computationally-intensive step was to minimize \eqref{E:dong_Laplacian_learning} with respect to $\bbL$, for fixed $\bbY$. Leveraging \eqref{E:smooth_sparse} and noting that $\|\bbL\|_F^2=\|\bbW\mathbf{1}\|^2+\|\bbW\|_F^2$, said problem can be equivalently reformulated as
\begin{align}\label{eq:dong_as_kalofolias}
\min_{\bbW}&{}\left\{\|\bbW\circ\bbZ\|_1-\log(\ind{\|\bbW\|_1=N})+\frac{\beta}{2}\left(\|\bbW\mathbf{1}\|^2+\|\bbW\|_F^2\right)\right\}\\
\textrm{ s. to } &{} \quad\textrm{diag}(\bbW)=\mathbf{0},\: W_{ij}=W_{ji}\geq 0, \:i\neq j.\nonumber
\end{align}
As shown in~\cite[Sec. 5]{Kalofolias2016inference_smoothAISTATS16}, problem \eqref{eq:dong_as_kalofolias} has a favorable structure that can be exploited to develop fast and scalable primal-dual algorithms to bridge the computational gap in~\cite{DongLaplacianLearning}.

% % % % % % % % % % % % % % % % % % % % % % % % % % % % % % % % % % % % % % % %
%                        Subsection V-C                                       %
% % % % % % % % % % % % % % % % % % % % % % % % % % % % % % % % % % % % % % % %

\subsection{Graph learning as an edge subset selection problem}
\label{ssec:edge_selection}

Consider identifying the edge set $\ccalE$ of an undirected, unweighted graph $\ccalG(\ccalV,\ccalE)$ with $|\ccalV|=N$ vertices.
The observations $\bbX=[\bbx_1,\ldots,\bbx_P]\in \reals^{N\times P}$ are assumed to vary smoothly on the sparse graph $\ccalG$ and the actual number of edges $|\ccalE|$ is assumed to be considerably smaller than the maximum possible number of edges $M:={N \choose 2}=N(N-1)/2$. We describe the approach in~\cite{sundeep_icassp17}, whose idea is to cast the graph learning problem as one of \emph{edge subset selection}. As we show in the sequel, it is possible to parametrize the unknown graph topology via a sparse edge selection vector. This way, the model provides an elegant handle to directly control the number of edges in $\ccalG$.   

Let $\bbB:=[\bbb_1,\ldots,\bbb_M]\in\reals^{N\times M}$ be the incidence matrix of the complete graph on $N$ vertices. The rows of the incidence matrix index the vertices of the graph, and the columns its $M=N(N-1)/2$ edges. If edge $m$ connects nodes $i$ and $j$, say, the $m$-th column $\bbb_m:=[b_{m1},\ldots,b_{mN}]^T\in\{-1,0,1\}^N$ is a vector of all zeros except for $b_{mi}=1$ and $b_{mj}=-1$ (the sign choice is inconsequential for undirected graphs). Next, consider the binary edge selection vector  {$\bbomega:=[\omega_1,\ldots,\omega_M]^T\in\{0,1\}^M$, where $\omega_m=1$ if edge $m\in\ccalE$, and $\omega_m=0$} otherwise. In other words, we have  {$\ccalE \equiv \textrm{supp}(\bbomega):=\{m: \omega_m\neq 0\}$}. Using the columns of $\bbB$, one can express the Laplacian of candidate graphs as a function of  {$\bbomega$}, namely
\begin{equation}\label{E:Laplacian_incidence_weights}
 {\bbL(\bbomega)=\sum_{m=1}^M \omega_m \bbb_m \bbb_m^T.}
\end{equation}
For example, an empty graph with $\ccalE=\emptyset$ corresponds to  {$\bbomega=\mathbf{0}$, while the complete graph is recovered by setting $\bbomega=\mathbf{1}$.} Here we are interested in sparse graphs having a prescribed number of edges $K\ll M$, which means  {$\|\bbomega\|_0:=|\textrm{supp}(\bbw)|=K$.}

We can now formally state the problem studied in~\cite{sundeep_icassp17}. Given graph signals $\ccalX:=\{\bbx_p\}_{p=1}^P$, determine an undirected and unweighted graph $\ccalG$ with $K$ edges such that the signals in $\ccalX$ exhibit smooth variations on $\ccalG$. A natural formulation given the model presented so far, is to solve the optimization problem
\begin{equation}\label{E:smoothness_edge_constrained}
 {\min_{\bbomega\in\{0,1\}^M} \textrm{trace}(\bbX^T\bbL(\bbomega)\bbX), \quad \textrm{s. to }\: \|\bbomega\|_0=K. }
\end{equation}
Problem \eqref{E:smoothness_edge_constrained} is a cardinality-constrained Boolean optimization, hence non-convex. Interestingly, the exact solution can be efficiently computed by means of a simple rank ordering procedure. In a nutshell, the solver entails computing edge scores $c_m=\textrm{trace}(\bbX^T(\bbb_m\bbb_m^T)\bbX)$ for all candidate edges, and setting  {$\omega_m=1$} for those $K$ edges having the smallest scores. Computationally, the sorting algorithm costs $O(K\log K)$. 

One can also consider a more pragmatic setting where the observations $\bbx_p=\bby_p+\bbepsilon_p$ are corrupted by Gaussian noise $\bbepsilon_p$, and it is the unobservable noise-free signal $\bby_p$ that varies smoothly on $\ccalG$, for  $p=1,\ldots, P$. In this case, selection of the best $K$-sparse graph can be accomplished by solving
\begin{equation}\label{E:smoothness_edge_constrained_noisy}
 {\min_{\bbY,\bbomega\in\{0,1\}^M} \left\{\|\bbX-\bbY\|_F^2+\alpha\textrm{trace}(\bbY^T\bbL(\bbomega)\bbY)\right\}, \quad \textrm{s. to }\: \|\bbomega\|_0=K, }
\end{equation}
which can be tackled using alternating minimization, or as a semidefinite program obtained via convex relaxation~\cite{sundeep_icassp17}.

% % % % % % % % % % % % % % % % % % % % % % % % % % % % % % % % % % % % % % % %
%                        Subsection III-C                                     %
% % % % % % % % % % % % % % % % % % % % % % % % % % % % % % % % % % % % % % % %

 {
	\subsection{Comparative summary}}
\label{ssec:comparison_smooth}

Relative to the methods in Sections \ref{ssec:smooth_factor_kernel} and \ref{ssec:smooth_sparse}, edge sparsity can be explicitly controlled in \eqref{E:smoothness_edge_constrained} and the graph learning algorithm is simple (at least in the noise-free case). There is also no need to impose Laplacian feasibility constraints as in \eqref{E:dong_Laplacian_learning}, because the topology is encoded via an edge selection vector.  However, \eqref{E:smoothness_edge_constrained} does not encourage connectivity of $\ccalG$, and there is no room for optimizing edge weights.  {The framework in~\cite{Kalofolias2016inference_smoothAISTATS16} is not only  attractive due to is computational efficiency but also due to its generality. In fact, through the choice of $g(\bbW)$ in the advocated inverse problem [cf. \eqref{eq:kalofolias} and \eqref{eq:dong_as_kalofolias}]
\begin{align}\label{eq:kalofolias_general}
\min_{\bbW}&{}\left\{\|\bbW\circ\bbZ\|_1+g(\bbW)\right\}\\
\textrm{ s. to } &{} \quad\textrm{diag}(\bbW)=\mathbf{0},\: W_{ij}=W_{ji}\geq 0, \:i\neq j,\nonumber
\end{align}
one can span a host of approaches to graph inference  from smooth signals. Examples include
the Laplacian-based factor analysis model~\cite{DongLaplacianLearning}  in \eqref{eq:dong_as_kalofolias}, and common graph constructions using the Gaussian kernel
to define edge weights $W_{ij}:=\exp{\left(-\frac{\|\bar{\bbx}_i-\bar{\bbx}_j\|^2}{\sigma^2}\right)}$ are recovered for $g(\bbW)=\sigma^2\sum_{i,j}W_{ij}(\log(W_{ij})-1)$.

Next, we search for graphs under which the observed signals are stationary. This more flexible model 
imposes structural invariants that call for innovative approaches that operate in the graph spectral domain.}

% % % % % % % % % % % % % % % % % % % % % % % % % % % % % % % % % % % % % % %
%                         Section VI                                                                                %
% % % % % % % % % % % % % % % % % % % % % % % % % % % % % % % % % % % % % % %

\section{Identifying the structure of network diffusion processes}
\label{sec:topoid_diffusion}

Here we present a novel variant of the problem of inferring a graph from vertex-indexed signals. 
%In general, it is clear that one must assume \emph{some} relation between the signals and the unknown underlying graph, since otherwise the topology inference exercise would be hopeless.  
 {In previous sections the relation between the signals in $\ccalX$ and the unknown graph $\ccalG$ was given} by statistical generative priors (Section~\ref{sec:stat}), or, by properties of the signals with respect to the underlying graph such as smoothness (Section~\ref{sec:smooth}). Here instead we consider observations of linear diffusion processes in $\ccalG$, such as those introduced in Section \ref{ssec:diffusion}. As we will see, this is a more general model where we require the covariance structure of the observed signals to be \emph{explained} by the unknown network structure. This loose notion of explanatory capabilities of the underlying graph -- when formalized -- has strong ties with the theory of stationary processes on graphs outlined in Section \ref{ssec:stationarity}.  {Stationarity and graph-filtering models can be accurate for real-world signals generated through a physical process akin to network diffusion. Examples include information cascades spreading over social networking platforms~\cite{baingana2014cascadesJSTSP14}, vehicular mobility patterns~\cite{thanou17,shafipour2018topoidnsTSP18}, consensus dynamics~\cite{mihailo2016topoid}, and progression of brain atrophy~\cite{hu2016brain_jstsp}.}

\subsection{Learning graphs from observations of stationary graph processes}
\label{ssec:stationarity_for_topoid}

As in previous sections, we are given a set $\ccalX:=\{\bbx_p\}_{p=1}^P$ of graph signal observations and wish to infer the symmetric shift $\bbS = \bbV \bbLambda \bbV^T$ associated with the unknown underlying graph $\ccalG$. 
We assume that $\bbx_p$ comes from a network diffusion process in $\bbS$. 
Formally, consider a random network process $\bbx = \sum_{l=0}^{L-1} h_l\bbS^l \bbw = \bbH \bbw$ driven by a zero-mean input $\bbw$. We assume for now that $\bbw$ is white, i.e., $\bbSigma_{\bbw}=\E{\bbw\bbw^T} = \bbI$. This assumption will be lifted in Section~\ref{ssec:non-stationary}. As explained in Section~\ref{ssec:diffusion}, the graph filter $\bbH$ represents a global network transformation that can be locally explained by the operator $\bbS$. The goal is to recover the direct relations described by $\bbS$ from the set $\ccalX$ of $P$ independent samples of the random signal $\bbx$. We consider the challenging setup where we have no knowledge of the filter degree $L-1$ or the coefficients $\bbh$, nor we get to observe the specific realizations of the inputs $\{\bbw_p\}_{p=1}^P$. 

The stated problem is severely \emph{underdetermined} and \emph{non-convex}. It is underdetermined because for every observation $\bbx_p$ we have the same number of unknowns in the input $\bbw_p$ on top of the unknown filter coefficients $\bbh$ and the shift $\bbS$, the latter being the quantity of interest. The problem is non-convex because the observations depend on the product of our unknowns and, notably, on the first $L-1$ powers of $\bbS$. To overcome the underdeterminacy we will rely on statistical properties of the input process $\bbw$ as well as on some imposed regularity on the graph to be recovered, such as edge sparsity or least-energy weights. To surmount the non-convexity, we split the overall inference task by first estimating the eigenvectors of $\bbS$ -- that remain unchanged for any power of $\bbS$ -- and then its eigenvalues. This naturally leads to a two-step process whereby we: (i) leverage the observation model to estimate $\bbV$ from the signals $\ccalX$; and (ii) combine $\bbV$ with a priori information about $\ccalG$ and feasibility constraints on $\bbS$ to obtain the optimal eigenvalues $\bbLambda$. We specify these two steps next; see Figure~\ref{F:scheme_topo_id} (a) for a schematic view of the strategy in \cite{segarra2016topoidTSP16}. A similar two-step approach was proposed in~\cite{pasdeloup2016inferenceTSIPN16},  but it relies on a different optimization problem in (ii).

\subsubsection{Step 1 -- Inferring the eigenvectors}\label{sssec:step_1}
From the described model it follows that the covariance matrix $\bbSigma_{\bbx}$ of the signal $\bbx$ is given by
\begin{align}\label{eqn_diagonalize_covariance}
\textstyle \bbSigma_{\bbx}  = \E{\bbH\bbw\big(\bbH\bbw\big)^T}= \bbH^2= \ \bbV\,\big(\sum_{l=0}^{L-1}h_l\bbLam^l\big)^2\,\bbV^T = h_0 \bbI + 2 h_0 h_1 \bbS + (2h_0h_2 + h_1^2) \bbS^2 + \ldots ,
\end{align}
where we have used $\bbSigma_{\bbw}= \bbI$ and the fact that $\bbH = \bbH^T$ since $\bbS$ is assumed to be symmetric.
It is apparent from \eqref{eqn_diagonalize_covariance} that the \textit{eigenvectors} (i.e., the GFT basis) of the shift $\bbS$ and the covariance $\bbSigma_{\bbx}$ are the same. Hence the difference between $\bbSigma_{\bbx}$, which includes indirect relationships between signal elements, and $\bbS$, which contains exclusively direct relationships, is only on their eigenvalues. While the underlying diffusion in $\bbH$ obscures the eigenvalues of $\bbS$ as per the frequency response of the filter, the eigenvectors $\bbV$ remain unaffected in $\bbSigma_{\bbx}$ as templates of the original spectrum. So if we have access to $\bbSigma_{\bbx}$ then $\bbV$ can be obtained by performing an eigendecomposition of the covariance. The attentive reader will realize that obtaining $\bbSigma_{\bbx}$ perfectly from a finite set of signals $\ccalX$ is, in general, infeasible. Hence, in practice we estimate the covariance, e.g. via the sample covariance $\hat{\bbSigma}_\bbx$, leading to a noisy version of the eigenvectors $\hat{\bbV}$. The robustness of this two-step process to the level of noise in $\hat{\bbV}$ is analyzed in Section~\ref{ssec:robust}.

The fact that $\bbS$ and $\bbSigma_{\bbx}$ are simultaneously diagonalizable implies that $\bbx$ is a stationary process on the unknown graph-shift operator $\bbS$ (cf. Definition~\ref{D:WeaklyStionaryGraphProcess_1}). Consequently, one can restate the graph inference problem as one of finding a shift  on which the observed signals are stationary.  Moreover, \eqref{eqn_diagonalize_covariance} reveals that the assumption on the observations being explained by a diffusion process is in fact more general than the statistical counterparts outlined in Section~\ref{sec:stat}; see \emph{Diffusion processes as an overarching model}. Smooth signal models are subsumed as special cases found with diffusion filters having a low-pass  response.

\begin{mdframed}[hidealllines=true,backgroundcolor=gray!20]
\textbf{Diffusion processes as an overarching model.} 
In some settings, such as opinion formation in social networks, it is reasonable to assume the existence of a bona fide diffusion process that shapes the observed signals. For instance, individuals observe the opinions of their neighbors and, as influences or beliefs propagate across the network, they form their own opinion. However, the fact that we are modeling the observations as being \emph{represented} by the output of a diffusion process \emph{does not} require such a diffusion process to be the true generative mechanism. Indeed, this assumption is valid in the absence of \emph{any} generative model since \eqref{eqn_diagonalize_covariance} reveals that it translates into concrete statistical requirements on the observed process. More precisely, assuming a diffusion-based representation is equivalent to assuming that the covariance $\bbSigma_{\bbx}$ of the observed process is an analytic matrix function $\phi$ of the unknown shift $\bbS$. In this sense, correlation networks ($\bbSigma_{\bbx} = \bbS$), covariance selection ($\bbSigma_{\bbx} = \bbS^{-1}$), and symmetric structural equation models driven by white noise ($\bbSigma_{\bbx} = (\bbI - \bbS)^{-2}$) are special cases of this framework.
\begin{center}
\includegraphics[width=0.9\linewidth]{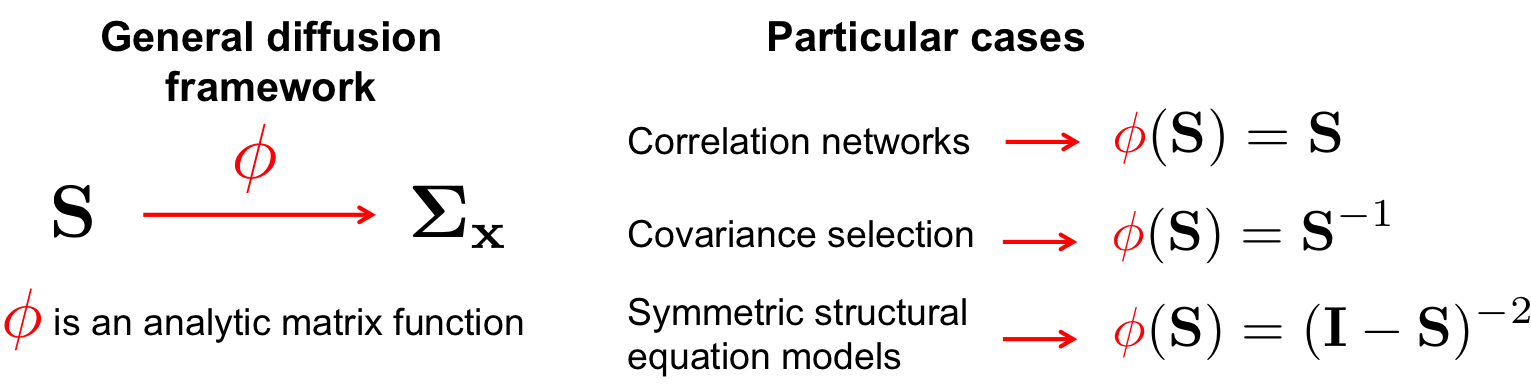}
\end{center}
\end{mdframed}	

\subsubsection{Step 2 -- Inferring the eigenvalues}\label{sssec:step_2}
From the previous discussion it follows that \emph{any} $\bbS$ that shares the eigenvectors with $\bbSigma_{\bbx}$ can \emph{explain} the observations, in the sense that there exist filter coefficients $\bbh$ that generate $\bbx$ through a diffusion process on $\bbS$. 
In fact, the covariance matrix $\bbSigma_{\bbx}$ itself is a graph that can generate $\bbx$ through a diffusion process and so is the precision matrix $\bbSigma_{\bbx}^{-1}$ (of partial correlations under Gaussian assumptions).  
To sort out this ambiguity, which amounts to selecting the eigenvalues of $\bbS$, we assume that the shift of interest is optimal in some sense~\cite{segarra2016topoidTSP16}. Our idea is then to seek for the shift operator $\bbS$ that: (a) is optimal with respect to (often convex) criteria $f (\bbS)$; (b) belongs to a convex set $\ccalS$ that specifies the desired type of shift operator (e.g., the adjacency $\bbW$ or Laplacian $\bbL$); and (c) has the prescribed $\bbV$ as eigenvectors. Formally, one can solve
\begin{equation}\label{E:general_problem}
\min_{\bbS \in \ccalS,\, \bbLambda} \
f (\bbS) ,   \quad
\text{s. to }\: \bbS = \bbV \bbLambda \bbV^T
\end{equation}
which is a convex optimization problem provided $f (\bbS)$ is convex. 

Within the scope of the signal model \eqref{eqn_diffusion}, the formulation \eqref{E:general_problem} entails a general class of network topology inference problems parametrized by the choices of $f(\bbS)$ and $\ccalS$. 
The selection of  $f(\bbS)$ allows to  incorporate physical characteristics of the desired graph into the formulation, while being consistent with the spectral basis $\bbV$. For instance, the matrix (pseudo-)norm $f (\bbS)=\|\bbS\|_0$ which counts the number of nonzero entries in $\bbS$ can be used to minimize the number of edges; while $f ( \bbS)=\|\bbS\|_1$ is a convex proxy for the aforementioned edge cardinality function.  
Alternatively, the Frobenius norm $ f (\bbS)=\|\bbS\|_F$ can be adopted to minimize the energy of the edges in the graph, or $f (\bbS)=\|\bbS\|_{\infty}$ which yields shifts $\bbS$ associated with graphs of uniformly low edge weights. 
This can be meaningful when identifying graphs subject to capacity constraints. Finally, one can minimize $f (\bbS) = -\lambda_2$, where $\lambda_2$ is the second smallest eigenvalue of $\bbS$. If $\bbS$ is further constrained to be a combinatorial Laplacian via $\ccalS=\{\bbS\:|\:\bbS \succeq \mathbf{0}, S_{ij}\leq 0\text{ for }i\neq j, \bbS\mathbf{1}=\mathbf{0}\}$,  then $\bbI-\bbS$ is a shift with fast mixing times.
Alternatively, to impose that $\bbS$ represents the adjacency matrix of an undirected graph with non-negative weights and no self-loops, one can set $\ccalS \!:= \! \{ \bbS \, | \, S_{ij}=S_{ji} \geq 0, \;\,  S_{ii} = 0, \;\, \textstyle\sum_j S_{j1} \! = \! 1 \}$. The first condition in $\ccalS$ encodes the non-negativity of the weights and incorporates that $\ccalG$ is undirected, hence, $\bbS$ must be symmetric. The second condition encodes the absence of self-loops, thus, each diagonal entry of $\bbS$ must be null. Finally, the last condition fixes the scale of the admissible graphs by setting the weighted degree of the first node to $1$, and rules out the trivial solution $\bbS\!=\!\bbzero$.

%%%%%%%%%%%%%%%   F   I   G   U   R    E   %%%%%%%%%%%%%%%%%%%%%%%%%%%%%%%%%%%%%%%%
\begin{figure*}[t]
	\hfill
	\begin{minipage}[b]{.4\linewidth}
		\centering
		\includegraphics[width=\linewidth, trim={0cm, 0cm, 0cm, 1cm}]{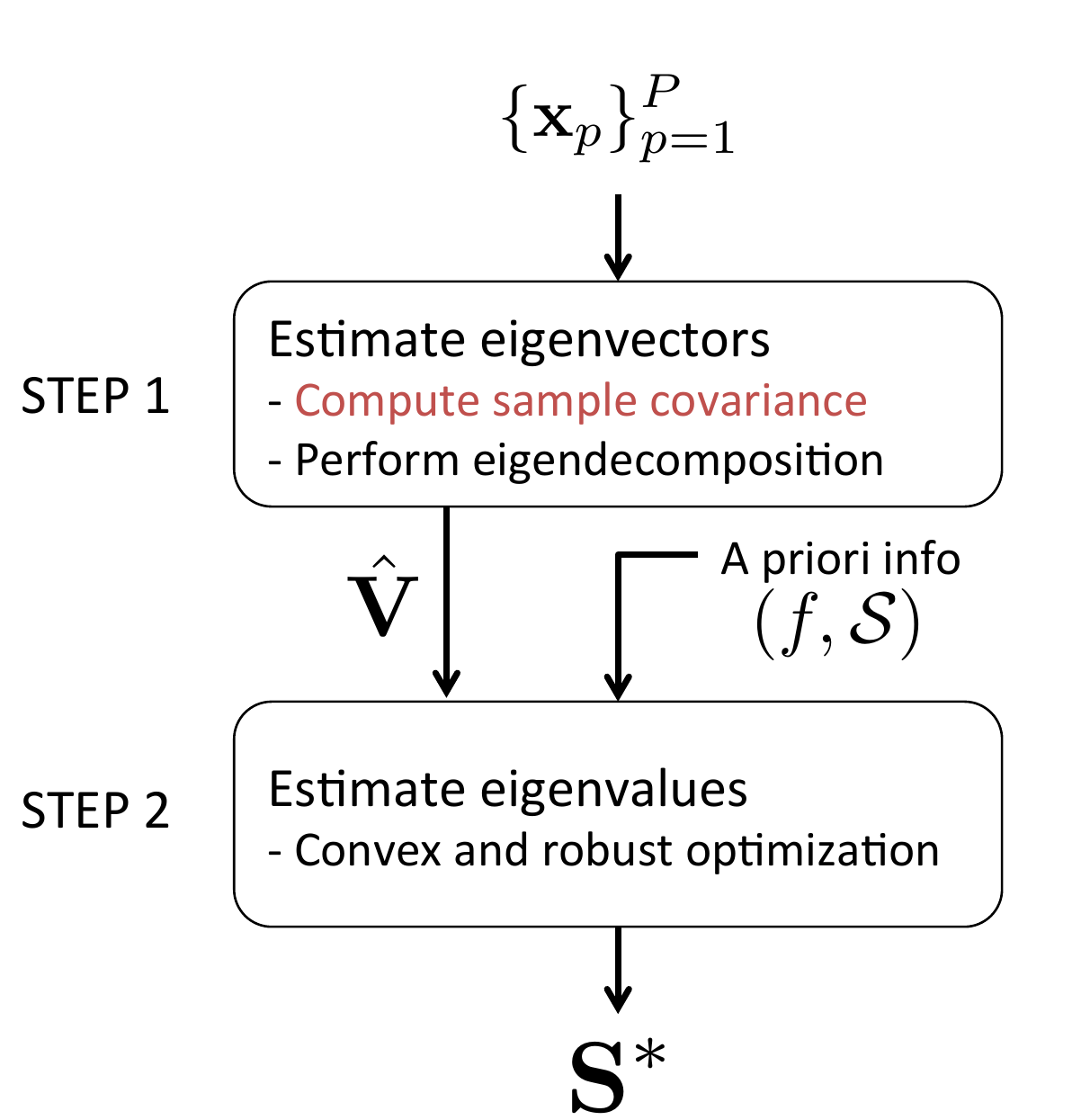}
		\centerline{\hspace{0.7cm}(a)}\medskip
	\end{minipage}
	\hfill
	\begin{minipage}[b]{.4\linewidth}
		\centering
		\includegraphics[width=\linewidth, trim={0cm, 0cm, 0cm, 1cm}]{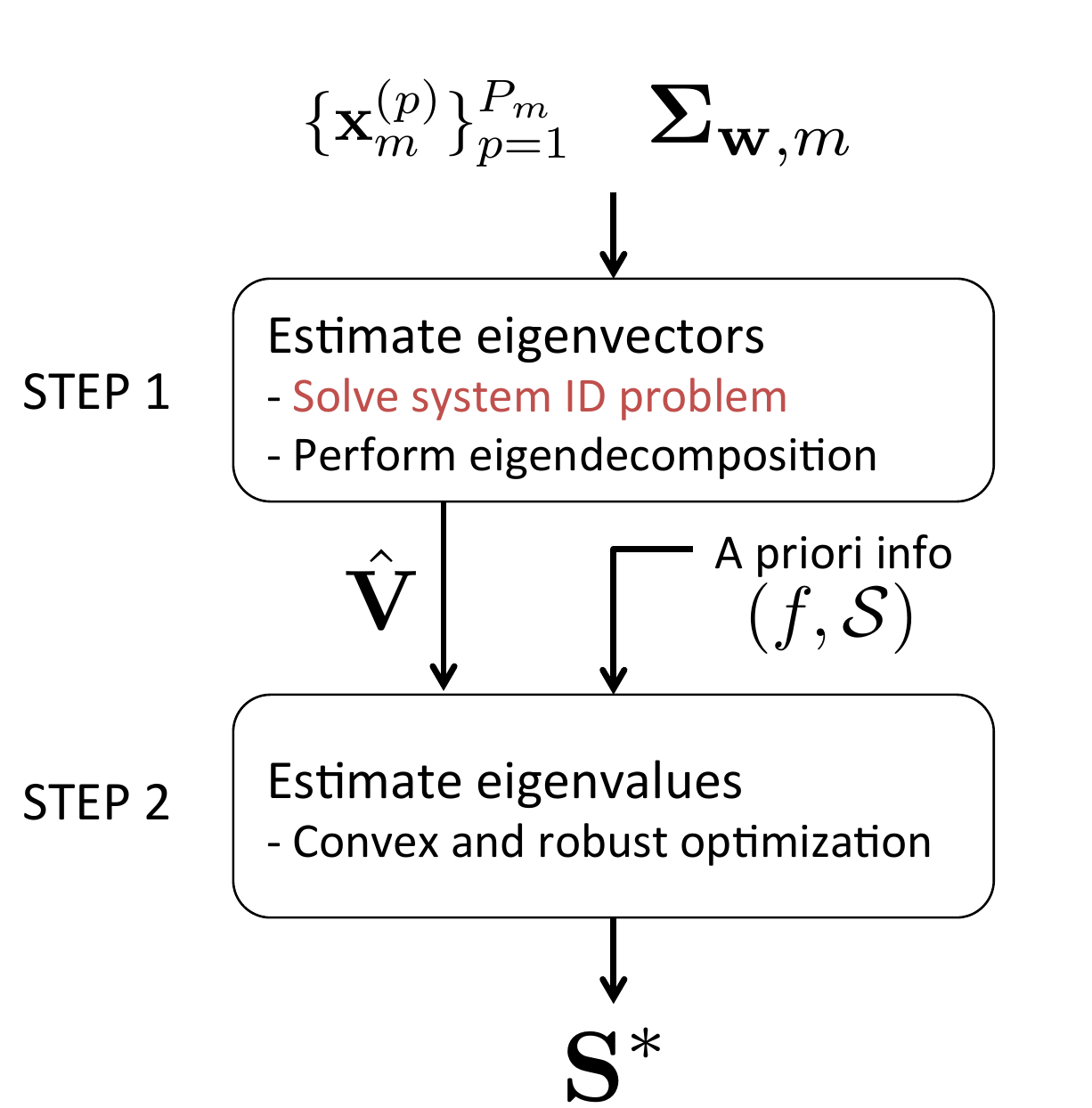}
		\centerline{\hspace{0.7cm}(b)}\medskip
	\end{minipage}
	\hfill
	\vspace*{-0.5cm}
	\caption{Schematic view of the two-step network inference method for (a) stationary and (b) non-stationary diffusion processes. The main differences between both approaches lie in Step 1. For stationary processes, we are given only one set of realizations of a random process, whose covariance is guaranteed to share the eigenvectors with $\bbS$.
	For non-stationary processes, covariance matrices are no longer simultaneously diagonalizable with $\bbS$, thus requiring a more challenging system identification step in order to estimate $\bbH$.
	In both cases the output of Step 1 is an estimate of the eigenvectors $\hat{\bbV}$ of the sought shift. During Step 2, this estimate is combined with a priori information about the shift in an optimization problem to obtain the estimate $\bbS^*$.}
	\label{F:scheme_topo_id}
\end{figure*}
%
%%%%%%%%%%%%%%%%%%%%%%%%%%%%%%%%%%%%%%%%%%%%%%%%%%%%%%%%%%%%%%%%%%%

% % % % % % % % % % % % % % % % % % % % % % % % % % % % % % % % % % % % % % % %
%                        Subsection VI-B                                       %
% % % % % % % % % % % % % % % % % % % % % % % % % % % % % % % % % % % % % % % %

\subsection{Robust network topology inference}
\label{ssec:robust}

The optimization problem formulated in Step 2 assumes perfect knowledge of the eigenvectors $\bbV$, which is only feasible if we have access to the ensemble covariance matrix $\bbSigma_{\bbx}$. In practice, we form the empirical covariance $\hat{\bbSigma}_{\bbx}$ that results in a noisy eigenbasis $\hat{\bbV}$. It is thus prudent to account for the (finite sample) discrepancies between $\hbV$ and the actual eigenvectors of $\bbS$. 
To that end, we modify \eqref{E:general_problem} by relaxing the equality constraint to obtain
\begin{equation}\label{E:general_problem_2}
\bbS^* := \argmin_{\bbS \in \ccalS,\, \bbLambda} \
f (\bbS) ,   \quad
\text{s. to }\: d(\bbS, \hat{\bbV} \bbLambda \hat{\bbV}^T) \leq \epsilon,
\end{equation}
where $d(\cdot, \cdot)$ is a convex matrix distance and $\epsilon$ is a tuning parameter chosen based on a priori information on the noise level.
The form of the distance $d(\cdot, \cdot)$ depends on the particular application. For instance, if $\|\bbS-\hbV\bbLambda\hbV^T\|_F$ is chosen, the focus is more on the similarities across the entries of the compared matrices, while  $\|\bbS-\hbV\bbLambda\hbV^T\|_{M(2)}$ focuses on their spectrum. 

One may ponder how does the noise level in $\hat{\bbV}$ affect the recovery performance of $\bbS$. 
To provide an answer we focus on the particular case of sparse shifts, where we adopt $f (\bbS)=\|\bbS\|_1$ as criterion in \eqref{E:general_problem_2} and $d(\bbS,\hbV\bbLambda\hbV^T)=\|\bbS-\hbV\bbLambda\hbV^T\|_F$ to obtain 
\begin{equation}\label{E:SparseAdj_l1_obj_noisy_matrix_v2}
\bbS_1^* := \argmin_{\bbLambda, \bbS \in \ccalS} \
\|\bbS\|_1 ,   \quad
\text{s. to }\:\|\bbS-\hbV\bbLambda\hbV^T\|_F\leq \epsilon.
\end{equation}
%
%Moreover, further uncertainties can be introduced in the definition of the feasible set $\ccalS$ (see~\cite{segarra2016topoidTSP16} for additional details). 
We denote by $\bbS_0^*$ the sparsest $\bbS$ with the true eigenbasis ${\bbV}$, and we assume that $\epsilon$ is chosen large enough to ensure that $\bbS_0^*$ belongs to the feasibility set of \eqref{E:SparseAdj_l1_obj_noisy_matrix_v2}. It was shown in \cite{segarra2016topoidTSP16} that under two conditions on matrices derived from $\hat{\bbV}$ it can be guaranteed that $\| \bbS_1^* - \bbS_0^* \|_1 < C \epsilon$, where $C$ is a well-defined constant that depends, among other things, on the size of the support of the sparse graph $\bbS_0^*$.
This means that when given noisy versions $\hat{\bbV}$ of the eigenvectors, the recovered shift is guaranteed to be at a distance from the desired shift bounded by the tolerance $\epsilon$ times a constant. This also implies that, for fixed $N$, as the number of observed signals $P$ increases we recover the true shift. 
In particular, the empirical covariance $\hbSigma_\bbx \to\bbSigma_\bbx$ as $P\to\infty$ and, for the cases where $\bbSigma_\bbx$ has no repeated eigenvalues, the noisy eigenvectors $\hbV$ converge to the eigenvectors $\bbV$ of the desired shift $\bbS_0^*$; see, e.g., \cite[Th. 3.3.7]{Ortega90}. Moreover, with better estimates $\hat{\bbV}$ the tolerance $\epsilon$ in \eqref{E:SparseAdj_l1_obj_noisy_matrix_v2} needed to guarantee feasibility can be made smaller, entailing a smaller discrepancy between the recovered $\bbS_1^*$ and the sparsest shift $\bbS_0^*$. 
In the limit when $\hat{\bbV} = \bbV$ and under no additional uncertainties, the tolerance $\epsilon$ can be made zero and solving \eqref{E:SparseAdj_l1_obj_noisy_matrix_v2} guarantees perfect recovery under the two aforementioned conditions. For a comprehensive performance evaluation that includes comparisons with the statistical methods of Section \ref{sec:stat} as well as with graph learning algorithms that rely on smoothness priors (Section \ref{sec:smooth}), the interested reader is referred to~\cite{segarra2016topoidTSP16}. 

An alternative scenario for robustness analysis arises when we have partial access to the eigenbasis $\bbV$ and, as a result, we can only access $K$ out of the $N$ eigenvectors of the unknown shift $\bbS$. This would be the case when, e.g., the given signal ensemble is bandlimited and $\bbV$ is found as the eigenbasis of the low-rank $\bbSigma_\bbx$; when the noise level is high and the eigenvectors associated with low-power components cannot be effectively estimated; or, when $\bbSigma_\bbx$ contains repeated eigenvalues, giving rise to a rotation ambiguity in the definition of the associated eigenvectors. In this latter case, we keep the eigenvectors that can be unambiguously characterized and, for the remaining ones, we include the rotation ambiguity as an additional constraint in the optimization problem.

To state the problem in this setting, assume that the $K$ first eigenvectors $\bbV_K=[\bbv_1,...,\bbv_K]$ are those which are known. For simplicity of exposition, suppose {as well that $\bbV_K$ is estimated} error free. Then, the network topology inference problem with incomplete eigenbasis can be formulated as [cf. \eqref{E:general_problem}]
\begin{equation}\label{E:SparseAdj_l00_onlysomeeig}
\bar{\bbS}^*_1 := \argmin_{\bbS \in \ccalS, \bbS_{\bar{K}}, \bblambda} \
\|\bbS\|_1 ,   \quad
\text{s. to }\:\bbS= \bbS_{\bar{K}} + {\textstyle\sum_{k=1}^{K}} \lambda_k \bbv_k \bbv_k^T, \;\; \bbS_{\bar{K}}\bbV_K=\bb0
\end{equation}
where we already particularized the objective to the $\ell_1$-norm convex relaxation.
The formulation in \eqref{E:SparseAdj_l00_onlysomeeig} constrains $\bbS$ to be diagonalized by the subset of known eigenvectors $\bbV_K$, with its remaining component $\bbS_{\bar{K}}$ being forced to belong to the orthogonal complement of $\text{range}(\bbV_K)$. This implies that the rank of $\bbS_{\bar{K}}$ can be at most $N-K$. An advantage of using only partial information of the eigenbasis as opposed to the whole $\bbV$ is that the set of feasible solutions in \eqref{E:SparseAdj_l00_onlysomeeig} is larger {than that in \eqref{E:general_problem}}. This is particularly important when the desired eigenvectors do not come from a prescribed shift but, rather, one has the freedom to choose $\bbS$ provided it satisfies certain spectral properties (see \cite{segarra2015distributed} for examples in the context of distributed estimation). Performance guarantees can also be derived for \eqref{E:SparseAdj_l00_onlysomeeig}; see~\cite{segarra2016topoidTSP16} for the technical details and formulations to accommodate scenarios where the knowledge of the $K$ templates is imperfect. 

%In closing, notice that scenarios where the knowledge of the $K$ templates is imperfect can be accommodated by combining the formulations in \eqref{E:SparseAdj_l1_obj_noisy_matrix_v2} and \eqref{E:SparseAdj_l00_onlysomeeig}. To that end,  consider the shift $\bbS'$ as a new optimization variable, replace the first constraint in \eqref{E:SparseAdj_l00_onlysomeeig} with $\bbS'=\bbS_{\bar{K}}+{\textstyle \sum_{k=1}^{K}} \lambda_k \hbv_k \hbv_k^T$, and add $d(\bbS,\bbS') \leq \epsilon$ as a new constraint. 

 {Regarding the computational complexity incurred by the two-step network topology inference strategy depicted in Figure \ref{F:scheme_topo_id} (a), there are two major tasks to consider: (i) computing the eigenvectors of the sample covariance which incurs $O(N^3)$ complexity; and (ii) solving iteratively the sparsity minimization problems in \eqref{E:SparseAdj_l1_obj_noisy_matrix_v2} or \eqref{E:SparseAdj_l00_onlysomeeig} to recover the graph-shift operator, which cost $O(N^3)$ per iteration~\cite{segarra2016topoidTSP16}. The cost incurred by the linear-programming based algorithms in~\cite{pasdeloup2016inferenceTSIPN16} is of the same order. Admittedly,  cubic complexity could hinder applicability of these approaches to problems involving high-dimensional signals. To bridge this complexity gap, progress should be made in developing custom-made scalable algorithms that exploit the particular structure of the problems; see the research outlook in Section \ref{sec:conc}.}

% % % % % % % % % % % % % % % % % % % % % % % % % % % % % % % % % % % % % % % %
%                        Subsection VI-C                                       %
% % % % % % % % % % % % % % % % % % % % % % % % % % % % % % % % % % % % % % % %

\subsection{Diffused non-stationary graph signals}
\label{ssec:non-stationary}

We now deal with more general \emph{non-stationary} signals $\bbx$ that adhere to linear diffusion dynamics $\bbx = \sum_{l=0}^{L-1} h_l\bbS^l \bbw = \bbH \bbw$, but where the input covariance $\bbSigma_\bbw = \E{\bbw\bbw^T}$ can be arbitrary. 
In other words, we relax the assumption of $\bbw$ being white, which led to the stationary signal model dealt with so far [cf. Definition \ref{D:WeaklyStionaryGraphProcess_0} and~\eqref{eqn_diagonalize_covariance}]. Such a model is for instance relevant to (geographically) correlated sensor network data, or to models of opinion dynamics, where (even before engaging in discussion) the network agents can be partitioned into communities according to their standing on the subject matter.

For generic (non-identity) $\bbSigma_\bbw$, we face the challenge that the signal covariance [cf. \eqref{E:cov_output_filter}]
\begin{equation}\label{E:covariance_y_ns}
\bbSigma_\bbx=\bbH\, \bbSigma_\bbw\, \bbH^T
\end{equation}
is no longer simultaneously diagonalizable with $\bbS$. This rules out using the eigenvectors of the sample covariance $\hbSigma_\bbx$ as eigenbasis of $\bbS$, as proposed in Step 1 for the stationary case. Still, observe that the eigenvectors of the shift coincide with those of the graph filter $\bbH$ that governs the underlying diffusion dynamics. This motivates adapting Step 1 in Section \ref{ssec:stationarity_for_topoid} when given observations of non-stationary graph processes. In a nutshell, the approach in~\cite{shafipour2018topoidnsTSP18} is to use snapshot observations $\ccalX$ together with additional (statistical) information on the excitation input $\bbw$ to \textit{identify the filter} $\bbH$, with the ultimate goal of estimating its eigenvectors $\bbV$. These estimated eigenvectors $\hat{\bbV}$ are then used as inputs to the shift identification problem \eqref{E:general_problem_2}, exactly as in the robust version of Step 2 in Section \ref{ssec:robust}. Accordingly, focus is henceforth placed on the graph filter (i.e., system) identification task; see Figure~\ref{F:scheme_topo_id} (b).

Identification of the graph filter $\bbH$ from non-stationary signal observations is studied in detail in~\cite{shafipour2018topoidnsTSP18}, for various scenarios that differ on what is known about the input process $\bbw$. Of particular interest is the setting where realizations of the excitation input are challenging to acquire, but information about the \textit{statistical} description of $\bbw$ is still available. 
Concretely, consider $M$ different excitation processes that are zero mean and their covariance $\bbSigma_{\bbw,m}=\E{\bbw_m\bbw_m^T}$ is known for all $m = 1, \ldots, M$. Further suppose that for each input process $\bbw_m$ we have access to a set of independent realizations $\ccalX_m=\{\bbx_m^{(p)}\}_{p=1}^{P_m}$ from the diffused signal $\bbx_m=\bbH\bbw_m$, which are then used to \textit{estimate the output covariance} as $\hbSigma_{\bbx,m}=\frac{1}{P_m}\sum_{p=1}^{P_m}\bbx_m^{(p)}(\bbx_m^{(p)})^T$. 
Since the ensemble covariance is $\bbSigma_{\bbx,m} = \E{\bbx_m \bbx_m^T} = \bbH \bbSigma_{\bbw,m}\bbH^T$ [cf. \eqref{E:covariance_y_ns}], the aim is to identify a filter $\bbH$ such that matrices $\hbSigma_{\bbx,m}$ and $\bbH \bbSigma_{\bbw,m} \bbH^T$ are close in some appropriate sense. 

Assuming for now perfect knowledge of the signal covariances, the above rationale suggests studying the solutions of the {following} system of matrix \textit{quadratic} equations
\begin{equation}\label{E:quadratic_system}
\bbSigma_{\bbx,m}=\bbH \bbSigma_{\bbw,m} \bbH^T, \quad m=1,\ldots,M.
\end{equation}
Given the eigendecomposition of the PSD covariance matrix $\bbSigma_{\bbw,m}=\bbV_{\bbw,m}\bbLambda_{\bbw,m}\bbV_{\bbw,m}^T$, its {principal square root} is given by $\bbSigma_{\bbw,m}^{1/2}=\bbV_{\bbw,m}\bbLambda_{\bbw,m}^{1/2}\bbV_{\bbw,m}^T$.
With this notation in place, let us introduce the matrix $\bbSigma_{\bbw\bbx\bbw,m}:=\bbSigma_{\bbw,m}^{1/2} \bbSigma_{\bbx,m} \bbSigma_{\bbw,m}^{1/2}$. We now study the set of solutions of \eqref{E:quadratic_system} for two different settings where we: i) assume that $\bbH$ is PSD, and ii) do not make any assumption on $\bbH$ other than symmetry.

\subsubsection{Positive semidefinite graph filters}
PSD graph filters arise, for example, with heat diffusion processes of the form $\bbx=(\sum_{l=0}^\infty \beta^l \bbL^l) \bbw$, $\beta>0$, where the graph Laplacian  $\bbL$ is PSD and the filter coefficients $h_l=\beta^l$ are all positive. In this setting, if $\bbSigma_{\bbw,m}$ is nonsingular then the filter $\bbH$ can be recovered via \cite{shafipour2018topoidnsTSP18}
\begin{equation}\label{E:filt_estimate_psd}
\bbH = \bbSigma_{\bbw,m}^{-1/2} \bbSigma_{\bbw\bbx\bbw,m}^{1/2} \bbSigma_{\bbw,m}^{-1/2}.
\end{equation} 
The solution in \eqref{E:filt_estimate_psd} reveals that the assumption $\bbH\succeq\mathbf{0}$ gives rise to a strong identifiability result. 
Indeed, if $\{\bbSigma_{\bbx,m}\}_{m=1}^M$ are known perfectly, the graph filter is identifiable even for $M=1$. 

However, in pragmatic settings where only empirical covariances are available, the observation of multiple ($M>1$) diffusion processes improves the performance of the system identification task. Given empirical covariances $\{\hbSigma_{\bbx,m}\}_{m=1}^M$  respectively estimated with enough samples $P_m$ to ensure that they are full rank, define $\hbSigma_{\bbw\bbx\bbw,m}:=\bbSigma_{\bbw,m}^{1/2} \hbSigma_{\bbx,m} \bbSigma_{\bbw,m}^{1/2}$ for each $m$. Motivated by \eqref{E:filt_estimate_psd}, one can estimate the graph filter by solving the constrained linear  least-squares (LS) problem~\cite{shafipour2018topoidnsTSP18}
\begin{equation}\label{E:opt_filter_PSD_opt}
\hbH=\mathop{\mathrm{argmin}}_{\bbH\succeq\mathbf{0}}\sum_{m=1}^M\left\|\hbSigma_{\bbw\bbx\bbw,m}^{1/2}-\bbSigma_{\bbw,m}^{1/2}\bbH\bbSigma_{\bbw,m}^{1/2}\right\|_F^2.
\end{equation}
Whenever the number of samples $P_m$ -- and accordingly the accuracy of the empirical covariances $\hbSigma_{\bbx,m}$ -- differs significantly across diffusion processes $m=1,\ldots, M$, it may be prudent to introduce non-uniform coefficients to downweigh those residuals in \eqref{E:opt_filter_PSD_opt} associated with inaccurate covariance estimates.  

\subsubsection{General symmetric graph filters}

Consider now a more general setting whereby $\bbH$ is only assumed to be symmetric, and denote by $\bbV_{\bbw\bbx\bbw,m}$ the unitary matrix containing the eigenvectors of $\bbSigma_{\bbw\bbx\bbw,m}$.  While for PSD graph filters the solution to \eqref{E:quadratic_system} is unique and given by \eqref{E:filt_estimate_psd}, when $\bbH$ is symmetric any matrix obtained by changing the sign of one (or more) of the eigenvalues of $\bbSigma_{\bbw\bbx\bbw,m}^{1/2}$ is also a feasible solution. Leveraging this and provided that the input covariance matrix $\bbSigma_{\bbw,m}$ is nonsingular, it follows that all symmetric solutions of $\bbSigma_{\bbx,m}=\bbH\bbSigma_{\bbw,m}\bbH^T$ are described by the set
\begin{equation}\label{E:solution_set_symmetric_single}
\ccalH_{m}^{\text{sym}}=\left\{\bbH\;|\:\bbH\!=\!\bbSigma_{\bbw,m}^{-1/2}\bbSigma_{\bbw\bbx\bbw,m}^{1/2}\bbV_{\bbw\bbx\bbw,m}\diag(\bbb_m)\bbV_{\bbw\bbx\bbw,m}^T\bbSigma_{\bbw,m}^{-1/2}\;\;\text{and}\;\;\bbb_m\!\in\!\{-1,1\}^N\right\}.
\end{equation}
Inspection of $\ccalH_{m}^{\text{sym}}$ confirms that in the absence of the PSD assumption, the problem for $M=1$ is non-identifiable. Indeed, for each $m$ there are $2^N$ possible solutions to the quadratic equation \eqref{E:covariance_y_ns}, which are parameterized by the binary vector $\bbb_m$. If $\bbH\succeq\mathbf{0}$ the solution is unique and corresponds to $\bbb_m=\mathbf{1}$, consistent with \eqref{E:filt_estimate_psd}.
For $M>1$, the set of feasible solutions to the system of equations \eqref{E:quadratic_system} is naturally given by $\ccalH_{1:M}^{\text{sym}}=\bigcap_{m=1}^M\ccalH_{m}^{\text{sym}}$.

If only empirical covariances $\{\hbSigma_{\bbx,m}\}_{m=1}^M$ are available, \eqref{E:solution_set_symmetric_single} can be leveraged to define the  matrices $\hbA_{m} := (\bbSigma_{\bbw,m}^{-1/2}\hbV_{\bbw\bbx\bbw,m}) \odot ( \bbSigma_{\bbw,m}^{-1/2}\hbSigma_{\bbw\bbx\bbw,m}^{1/2}\hbV_{{\bbw\bbx\bbw},m})$ and  solve the binary-constrained LS problem
\begin{align}\label{E:general_problem_symmetric_shifts} 
\underset{ \{\bbb_m\}_{m=1}^M}{\text{min}} \; \sum_{m,m'}\left\| \hbA_{m}\bbb_{m} - \hbA_{m'}\bbb_{m'}\right\|^2
\qquad \text{s. to }\;
\bbb_m\in\{-1,1\}^N,\;m=1,\ldots,M.
\end{align}
Both terms within the $\ell_2$-norm in \eqref{E:general_problem_symmetric_shifts} should equal $\normalfont\textrm{vec}(\bbH)$ in a noiseless setting. Thus, we are minimizing the residuals across the $M$ processes considered. While the objective in \eqref{E:general_problem_symmetric_shifts} is convex in the $\{\bbb_m\}_{m=1}^M$, the binary constraints render the optimization non-convex and particularly hard. Interestingly, this problem can be tackled using a convexification technique known as semidefinite relaxation~\cite{luo2010semidefinite}. More precisely, \eqref{E:general_problem_symmetric_shifts} can be recast as a Boolean quadratic program and then equivalently expressed as a semidefinite program subject to a rank constraint.
Dropping this latter constraint, one arrives at a convex relaxation with provable approximation bounds; see \cite{shafipour2018topoidnsTSP18} for full algorithmic,  {complexity}, and performance details.

% % % % % % % % % % % % % % % % % % % % % % % % % % % % % % % % % % % % % % % %
%                        Subsection VI-D                                       %
% % % % % % % % % % % % % % % % % % % % % % % % % % % % % % % % % % % % % % % %

\subsection{Learning heat diffusion graphs}
\label{ssec:heat_diffusion}

A different graph topology identification method was put forth in~\cite{thanou17}, which postulates that the observed signals consist of superimposed heat diffusion processes on the unknown graph.
%That is, the measured signals are observations at different time instants of a few processes that start at different nodes and diffuse with time.
Mathematically, the observed graph signals are modeled as a linear combination of a few (sparse) components
from a dictionary consisting of heat diffusion filters with different heat rates. The graph learning task is then formulated as a regularized inverse problem where both the graph -- hence, the filters -- and the sparse combination coefficients are unknown.

Similar to Section~\ref{sec:smooth}, let us define the matrix $\bbX = [\bbx_1, \ldots, \bbx_P]$ collecting the $P$ observed graph signals, as well as the vector $\bbtau = [\tau_1, \ldots, \tau_S]^T$ of heat rates corresponding to each of the $S$ diffusion filters $\bbH_s=e^{\tau_s \bbL}=\sum_{l=0}^\infty \frac{(\tau_s\bbL)^l}{l!}$. With those notations in place, the inference problem is formulated as
\begin{align}\label{E:thanou_heat_kernel}
\min_{\bbL,\bbR, \bbtau}&{}\left\{\Big|\!\Big| \bbX- \left[e^{\tau_1 \bbL}, e^{\tau_2 \bbL}, \ldots, e^{\tau_S \bbL} \right] \bbR \Big|\!\Big|_F^2 +\alpha \sum_{p=1}^P \| \bbr_p \|_1 + \beta \| \bbL \|^2_F\right\}\\
\textrm{ s. to } &{} \quad\textrm{trace}(\bbL)=N,\quad \bbL\mathbf{1}=\mathbf{0}, \quad L_{ij}=L_{ji}\leq 0, \:i\neq j, \quad \tau_i \geq 0,\nonumber
\end{align}
where $\bbr_p\in\reals^{NS}$, which corresponds to the $p$-th column of $\bbR$, collects the (sparse) coefficients that combine the columns of the dictionary to approximate the graph signal $\bbx_p$. The objective function in \eqref{E:thanou_heat_kernel} has three components. The first term seeks to explain the observations with a dictionary model, where the atoms of the dictionary are the potential outputs of heat diffusion processes centered at every possible node and for several candidate heat diffusion rates $\tau_s$. The model postulates that every observation $\bbx_p$ can be synthesized as a few diffusion processes, thus, the coefficients associated with the dictionary should be sparse. Accordingly, the second term in the objective function imposes sparsity on the columns of $\bbR$. Finally, the last term regularizes the unknown Laplacian $\bbL$. The constraints in \eqref{E:thanou_heat_kernel} basically ensure that $\bbL$ is a well-defined Laplacian and that heat diffusion rates are non-negative; see~\cite{thanou17} for more details.

The optimization problem in \eqref{E:thanou_heat_kernel} is non-convex, thus potentially having multiple local minima and hindering its solution. 
Moreover, solving \eqref{E:thanou_heat_kernel} only with respect to $\bbL$ is challenging due to the matrix exponential, rendering the problem
non-convex even for fixed $\bbtau$ and $\bbR$. This discourages traditional alternating minimization techniques.
To overcome this difficulty, the approach in~\cite{thanou17}  is to apply a proximal alternating linearized
minimization algorithm, which can be interpreted as alternating the steps of a proximal forward-backward scheme. 
The basis of the algorithm is alternating minimization between $\bbL$, $\bbR$, and $\bbtau$, but in each step the non-convex fitting term is linearized with a first-order function at the solution obtained from the previous iteration.
In turn, each step becomes the proximal regularization of the non-convex function, which can be solved in polynomial time.  {The computational cost of the aforementioned graph learning algorithm is $O(N^3)$ per iteration, stemming from the computation of matrix exponentials, gradients, and required Lipschitz constants. Savings can be effected by relying on truncated (low-degree) polynomial approximations of the heat diffusion filters $\bbH_s$; see~\cite[Sec. IV-C]{thanou17}.}

% % % % % % % % % % % % % % % % % % % % % % % % % % % % % % % % % % % % % % % %
%                        Subsection III-C                                     %
% % % % % % % % % % % % % % % % % % % % % % % % % % % % % % % % % % % % % % % %

 {
	\subsection{Comparative summary}}
\label{ssec:comparison_diffusion}

Inspection of \eqref{E:thanou_heat_kernel} reveals the main differences between the method in~\cite{thanou17} and the ones outlined in Sections~\ref{ssec:robust} and~\ref{ssec:non-stationary}. Namely, \eqref{E:thanou_heat_kernel} assumes a specific filter type (heat diffusion) parametrized by a single scalar (the diffusion rate). Moreover, the inputs to these filters are required to be sparse.
On the other hand, in the previous methods the filters were arbitrary -- thus, not necessarily modeling heat diffusion -- while the available information on the inputs was statistical (white or known covariance) instead of structural (like sparsity). 
In this respect, when there are strong reasons to believe that the true diffusion model is (close to) a heat diffusion, then the more model-specific approach in~\cite{thanou17} would be preferable. Otherwise, a more data-driven approach like the one explained in Sections~\ref{ssec:robust} and~\ref{ssec:non-stationary} can attain better estimation performance {, possibly at the price of a larger sample size}. This trade-off is nicely conveyed through the numerical tests reported in~\cite{thanou17}.

% % % % % % % % % % % % % % % % % % % % % % % % % % % % % % % % % % % % % % % %
%                         Section VII                                           %
% % % % % % % % % % % % % % % % % % % % % % % % % % % % % % % % % % % % % % % %

 {
	\section{Further insights on choosing a suitable graph learning method}
	\label{sec:discussion}

\begin{table}[t]
		 {
		\tiny
		\flushleft
		\caption{Comparison of surveyed undirected network topology inference algorithms}
		\label{tab:tablon}
		\begin{tabular}{p{29mm} c p{18mm} c l p{60mm} c} \hline
			Method & Equation & Observed signals & Target  & Complexity & Salient characteristics  \\ \hline\hline
			Correlation network\newline \cite[Ch. 7.3.1]{kolaczyk2009book}  &  \eqref{E:corr_coeff} & i.i.d. & $\bbSigma$ &  $O(N^2 P)$ &  
			$\checkmark$ Flexible signal model, intuitive notion of pairwise interaction \newline 
			%$\xmark$ Limited to linear and symmetric interactions \newline
			%$\checkmark$ Ad-hoc thresholding yields simple implementation  \newline 
			$\xmark$ Misses latent effects, limited to linear and symmetric interactions %\newline 
			%$\xmark$  Statistical and computational issues of large-scale hypothesis testing  
			\\\hline
			Partial correlation network\newline \cite[Ch. 7.3.2]{kolaczyk2009book} &  \eqref{E:par_corr_coeff}  & i.i.d & $\bbSigma^{-1}$ & $O(N^3)$  & 
			$\checkmark$ Flexible signal model, controlling for latent effects  \newline 
			%$\xmark$ Limited to linear and symmetric interactions \newline
			$\xmark$ Statistical and computational issues of large-scale hypothesis testing    \\\hline
			%Covariance selection~\cite{dempster_cov_selec} & \eqref{E:par_corr_precision}  &  jointly Gaussian  &  &  - Exploits \! Gaussian \! model &  - Only low dimensional setting   \\\hline
			Graphical lasso \newline \cite{yuanlin2007, banerjee2008jlmr,glasso2008}   & \eqref{E:glasso}  & jointly Gaussian & $\bbSigma^{-1}$ & $^1O(N^3)$ &  
			$\checkmark$ Sparse regularization to handle high dimensional setting ($N\gg P$) \newline 
			$\checkmark$ Efficient first-order algorithms, statistical support consistency \newline
			$\xmark$ Gaussianity may be restrictive, intractable for discrete models    \\ \hline
			Laplacian-constrained GMRF \newline \cite{Lake10discoveringstructure, egilmez2017jstsp}   &  \eqref{E:GMRF_Laplacian} & jointly Gaussian & $\bbSigma^{-1}=\bbL$ & $^1O(N^3)$ & 
			$\checkmark$ Incorporates Laplacian and other structural constraints  \newline
			$\checkmark$ Non-negativity of edge weights can aid interpretability  \newline 
			$\xmark$ Attractive and improper GMRF can be restrictive   \\\hline
			Neighborhood-based regression~\cite{meinshausen06}    & \eqref{E:linear_predictor_lasso}  & jointly Gaussian\newline discrete distributions & $\bbSigma^{-1}$ & $^2{O(N^2P)}$ &  
			$\checkmark$ Scalable via per-node parallelization, statistical support consistency   \newline 
			$\checkmark$ Tractable even for discrete or mixed graphical models \newline 
			$\xmark$ Symmetry and positive-definiteness is not naturally enforced    \\ \hline
			Laplacian-based factor analysis \newline\cite{DongLaplacianLearning}    & \eqref{E:dong_Laplacian_learning}  & smooth & $\bbL$ & $^3O(N^2)$ & $\checkmark$ Natural graph-based factor analysis model (akin to iGFT synthesis)  \newline 
			$\xmark$ Bi-convex criterion lacking global optimality guarantees   \\\hline
			Smoothness-based graph \newline learning~\cite{Kalofolias2016inference_smoothAISTATS16}    & \eqref{eq:kalofolias}, \eqref{eq:kalofolias_general}  & smooth  & $\bbW$ & $^3O(N^2)$  & 
			$\checkmark$ General graph learning framework under smoothness prior \newline 
			$\checkmark$ Efficient, scalable primal-dual solver \newline 
			$\xmark$ No explicit generative model for the observations  \\ \hline
			Edge subset selection \newline \cite{sundeep_icassp17}   & \eqref{E:smoothness_edge_constrained}, \eqref{E:smoothness_edge_constrained_noisy}  & smooth & $\bbL$ & $^4O(|\ccalE| \log |\ccalE|)$ &  
			$\checkmark$ Explicit handle on edge sparsity  \newline 
			$\xmark$ No control on graph connectivity or edge weights    \\\hline
			Spectral templates \newline \cite{segarra2016topoidTSP16,pasdeloup2016inferenceTSIPN16,shafipour2018topoidnsTSP18}    & \eqref{E:SparseAdj_l1_obj_noisy_matrix_v2}, \eqref{E:SparseAdj_l00_onlysomeeig}  &  graph stationary \newline network diffusion & $\bbS$ & $^3O(N^3)$ & $\checkmark$ Flexible model, data covariance as analytic function of the shift  \newline 
			$\checkmark$ Robust formulations to accommodate imperfections \newline  
			$\xmark$ Limited sample size can hinder covariance eigenvector estimates \\\hline
			%Spectral templates (extended) \newline \cite{shafipour2018topoidnsTSP18}    & \eqref{E:opt_filter_PSD_opt}, \eqref{E:general_problem_symmetric_shifts}   & graph diffusion & $\bbS$  & $^3O(N^3)$ & $\checkmark$ Flexible signal model \newline $\checkmark$ Data-driven inference of underlying model \newline  $\xmark$ Computationally costly   \\\hline
			Heat diffusion graphs \newline \cite{thanou17}    &  \eqref{E:thanou_heat_kernel} & heat diffusions  & $\bbL$ & $^3O(N^3)$ & 
			$\checkmark$ Dictionary model of superimposed heat diffusion processes \newline 
			$\checkmark$ Can capture localized properties of the data \newline 
			$\xmark$ Non-convex criterion lacking global optimality guarantees \\\hline\hline
		\end{tabular}
		$^1$This is the complexity for dense graphs. In practice, the computation time markedly decreases for sparse graphs.\\
		$^2$Complexity \emph{per node}.\\
		$^3$Complexity \emph{per step} of an iterative algorithm. \\
		$^4$This complexity is attained for the noiseless case. In the presence of noise, a convex optimization problem with $N^2$ variables must be solved. \\
	}		
\vspace{-0.5cm}
\end{table}

Having presented several methods for (undirected) network topology inference that we summarize in Table \ref{tab:tablon}, it is prudent to reflect for a moment on a few general questions. First, what is the most suitable algorithm for a given network-analytic task? On a related note, what are the key considerations in making such decision? Second, what are the new perspectives, benefits, and limitations of the GSP-based approaches in Sections \ref{sec:smooth} and \ref{sec:topoid_diffusion}, relative to e.g., the statistical methods for graphical model selection in Section \ref{sec:stat}? While there are certainly no definite answers to at least some of these questions, here we shed some light based on our experience with graph learning problems. To this end, the qualitative comparison that follows will delve into three central characteristics of the approaches, namely: (i) signal models and their relationships; (ii) computational and sample complexities; and (iii) relevance to applications.

\subsection{Graph signal models and their relationships}
\label{ssec:models}

A general principle to unveil (network) structure from data is to adopt a parametric model, thus modeling has been a recurrent theme in our presentation of topology inference algorithms; see the third column in Table \ref{tab:tablon}. Correlation networks advocate an intuitive notion of similarity between signal elements, where the interactions are modeled by the covariance matrix $\bbSigma$. They are widely adopted especially when implemented using simple ad hoc thresholding rules on the $\hat{\rho}_{ij}$ to define edges. If seeking a graph reflective of \emph{direct} pairwise influence among signal elements, then partial-correlation networks represent a more sensible alternative. For both these network models, recall that their scope is limited to linear and symmetric dependencies. Formal network inference in this context requires conducting multiple hypothesis tests (one per vertex pair), so e.g., FDR control procedures should be implemented. A word of caution is due here since classical multiple-testing theory considers independent tests -- an assumption that can be grossly violated for network data~\cite[Ch. 7]{kolaczyk2009book}. 

Gaussian models are ubiquitous in machine learning and statistical analysis of real-valued network data, because of their widespread applicability and analytical tractability. Most recent advances to GMRF model selection have explored ways of incorporating Laplacian or otherwise graph topological constraints in the precision matrix estimation task~\cite{pavez2018tsp,egilmez2017jstsp,mike_icassp17,mihailo2016topoid}. These  approaches are well suited to settings when prior information dictates that e.g., feasible graphs should have a tree structure or edge weights should be positive given the physics of the problem. Interestingly, one can motivate the signal modeling framework in Section \ref{sec:smooth} through the lens of Gaussian graphical models. To this end, it suffices to notice that smooth signals have a higher likelihood under GMRF models with a Laplacian-constrained precision matrix. Going back to Section \ref{sec:stat}, it is thus not surprising to see that the ML estimator \eqref{E:precision_ML} minimizes  $\textrm{trace}(\hbSigma \bbTheta)\propto \sum_{p=1}^P\bbx_p^T\bbTheta\bbx_p$, a term that represents a smoothness penalty for graphical models with $\bbTheta=\bbL$ [cf. \eqref{eqn_TV_general}]. This connection notwithstanding, the optimization problems \eqref{eq:kalofolias} and \eqref{eq:kalofolias_general} are motivated by smooth signal priors but there is no explicit generative model for the observations, unlike the graphical models in Section \ref{sec:stat} which have a clear interpretation. 

Stationarity models on the other hand are suitable when second-order statistical invariance to graph shifts is a suspected property of the data. Moreover, Definition \ref{D:WeaklyStionaryGraphProcess_0} offers a clear generative mechanism in terms of network diffusion~\cite{segarra2016topoidTSP16,pasdeloup2016inferenceTSIPN16}, well suited to model observations from cascading or epidemic processes. This perspective also suggests that  smoothness models can be recovered via diffusion filters with low-pass frequency response.  Equivalently, stationarity implies a graph-dependent model $\bbSigma=\phi(\bbS)$ for the data covariance, where $\phi$ is some analytic function of the graph-shift operator. Through this lens, one can recover correlation networks and covariance selection as special cases when $\phi$ is the identity or inverse operator, respectively~\cite{segarra2016topoidTSP16}. A similar model $\bbTheta=\phi(\bbL)$ was recently advocated in~\cite{egilmez2018tsipn} for graph Laplacian learning within the GMRF framework. Notice how all these insightful links and rich interpretations are facilitated through the fresh perspective GSP brings to the topology inference problem.

\subsection{Computational and sample complexities}
\label{ssec:complexities}

In the current era of information, computation and data have arguably emerged as the key resources over which to optimize performance of signal and information processing algorithms. We encountered this computational versus statistical trade-off in Section \ref{sec:stat}: neigborhood-based regression algorithms for GMRF model selection are faster, while graphical lasso is statistically more efficient. Both of these schemes along with the methods from Section \ref{ssec:smooth_sparse} come with efficient solvers that scale relatively well to large-size problems; see the fifth column of Table \ref{tab:tablon} for a summary of the incurred computational complexities.

Methods in Section \ref{sec:topoid_diffusion} that require computing eigenvectors from the empirical covariance matrix are likely to fail when there are few data samples, unless some form of regularization is introduced in the process. This is to be contrasted with the $P\ll N$ regime under which sparse GMRFs can be successfully identified. In this direction, fundamental statistical questions on the sample complexity of GSP-based topology inference algorithms are yet to be addressed. Returning to the approach in Section \ref{ssec:robust}, an analytical (even approximate) characterization of recovery performance as a function of $P=|\ccalX|$ remains elusive. Such an analysis has to jointly account for the signal model \eqref{eqn_diffusion} (plus possibly a model for $\bbS$), the imperfections in estimating $\hbV$ from the empirical covariance, and how these errors affect the result of the optimization \eqref{E:general_problem_2}. The challenges could be compounded for the approaches in~\cite{DongLaplacianLearning,thanou17,sardellitti2016transform_globalsip}, which rely on non-convex criteria lacking global optimality guarantees. Methods based on graphical models can be analyzed in theory under the model assumptions, for instance~\cite{meinshausen06,ravikumar2011,mike_icassp17} show statistical consistency. Consequently, for given problem size and some prior knowledge on graph sparsity (possibly informed by physical constraints or interpretability considerations), existing  sample-complexity bounds can inform the amount of data required to attain a prescribed performance goal.

\subsection{Relevance to applications}
\label{ssec:applications}

It is ultimately the applications and the characteristics of the data involved that largely dictate what is a suitable graph learning algorithm for the information processing task at hand. For instance, graph-filtering based models of network diffusion have been adopted to unveil urban mobility patterns in New York City from Uber pick-up data~\cite{thanou17,shafipour2018topoidnsTSP18}. A sparse graph explaining the (presumed smooth) temperature observations collected across weather stations in the French region of Brittany was obtained in~\cite{sundeep_icassp17}. 

The graph frequency decomposition of neuroimaging data shows promise for analyzing brain signals and connectivity~\cite{weiyu2016brain_jstsp}; see also the numerical test in Section~\ref{ssec:brain}. For supervised classification of brain states (in response to different visual stimuli), GFT-based dimensionality reduction of functional magnetic resonance imaging (fMRI) data has been shown to outperform state-of-the art reduction techniques relying on PCA or independent component analysis (ICA)~\cite{menoret2017neuro_globalsip}; see also~\cite{rui2017brain_icassp} for related approaches dealing with electroencephalogram (EEG) data. Results in~\cite{menoret2017neuro_globalsip} indicate that the smooth signal prior along with the graph learning approach in~\cite{Kalofolias2016inference_smoothAISTATS16} yield the best performance for the aforementioned classification task. This is a valuable insight, since most software for constructing functional connectivity network relies on the correlation methods of Section \ref{ssec:correlation}~\cite{sporns2011}. 

Graph frequency analyses require a description of the underlying network, which suggests learning graphs that yield orthonormal transforms over which signals admit parsimonious (i.e., bandlimited) representations. Such a design principle was recently advocated in~\cite{sardellitti2016transform_globalsip}, through the following two-step procedure: i) learn the GFT basis and the sparse signal representation jointly from the observed signals; and ii) infer the graph weighted Laplacian, and then the graph topology, from the estimated Laplacian eigenvectors (much alike~\cite{segarra2016topoidTSP16,pasdeloup2016inferenceTSIPN16}).  This signal representation perspective to graph learning is also implicit to the factor analysis model \eqref{E:factor_model} that is central to the method in~\cite{DongLaplacianLearning}. These ideas have been successfully applied to recovering brain functional connectivity networks associated to epilepsy~\cite{sardellitti2016transform_globalsip}, and to learn climate graphs from evapotranspiration data recorded by meteorological stations in California~\cite{DongLaplacianLearning}. 

GSP tools are also envisioned to have major impact to image, point cloud, and video processing applications~\cite{gsp2018tutorial}. 
Though a digital image contains pixels that reside on a regular two-dimensional lattice, if one can design an
appropriate underlying graph connecting pixels with weights that reflect the image structure, then one can 
interpret the image as a graph signal and apply GSP tools for processing and  analysis of the signal in graph spectral 
domain~\cite{cheung2018imageproc_pieee}. For image restoration tasks
such as denoising and deblurring, a major challenge is how to design appropriate signal priors to regularize otherwise
ill-posed inverse problems. Learning graph Laplacians that endow the signal representations with desired sparsity or 
smoothness properties is thus well motivated and an active area of research. 
	
Increasingly, applications call for learning graph representations of dynamic, multi-aspect data, possibly accounting for nonlinear and directional (causal) effects among nodal signals. While a thorough treatment is beyond the scope of this paper, for completeness we offer a brief account in the next section. For a comprehensive survey of these emerging topics, the reader is referred to~\cite{ggtopoid2018piee}. 
}	
	
% % % % % % % % % % % % % % % % % % % % % % % % % % % % % % % % % % % % % % %
%                         Section VIII                                        %
% % % % % % % % % % % % % % % % % % % % % % % % % % % % % % % % % % % % % % %

\section{Emerging topic areas}
\label{sec:emerging}

So far the focus has been on learning static and undirected graphs from data. In this section we first consider identification of digraphs given nodal time series, which is intimately related to the problem of causal inference.  We then cross the boundary of linear time-invariant network models and outline recent advances for tracking topologies of dynamic graphs, as well as mechanisms to account for nonlinear pairwise interactions among vertex processes.

% % % % % % % % % % % % % % % % % % % % % % % % % % % % % % % % % % % % % % % %
%                        Subsection VIII-A                                      %
% % % % % % % % % % % % % % % % % % % % % % % % % % % % % % % % % % % % % % % %

\subsection{Directed graphs and causality}
\label{ssec:digraphs_causal}

Undirected graphs, like correlation networks, can inform proximity between nodal signals but cannot inform causality.
Here we will lift the assumption that graph-shift operators are symmetric and consider estimation of directed graphs with the intent of inferring causality from snapshot observations.

To that end, structural equation modeling encapsulates a family of statistical methods that
model causal relationships between interacting variables in a complex system. This is pursued through estimation
of linear relationships among  endogenous as well as exogenous traits, and structural equation models (SEMs) have been extensively
adopted in economics, psychometrics, social sciences, and genetics, among others; see e.g.,~\cite{kaplan_book}. The appeal of SEMs can be attributed to simplicity and the inherent ability to capture edge directionality in graphs, represented through a (generally) asymmetric adjacency matrix $\bbW\in\reals^{N\times N}$ whose entry $w_{ij}$ is nonzero only if a directed edge connects nodes $i$ and $j$ (pointing from $j$ to $i$). 

SEMs postulate a linear time-invariant network model of the form
\begin{equation}\label{E:SEM}
x_{it}=\sum_{j=1, j\neq i}^N w_{ij}x_{jt}+\omega_{ii}u_{it}+\epsilon_{it}, \: i\in \ccalV \:\: \Rightarrow \bbx_t=\bbW\bbx_t+\mathbf{\Omega}\bbu_t+\bbepsilon_t,
\end{equation}
where $\bbx_t=[x_{1t},\ldots,x_{Nt}]^T$ represents a graph signal of endogenous variables at discrete time $t$ and $\bbu_t=[u_{1t},\ldots,u_{Nt}]^T$ is a vector of exogenous influences. The term $\bbW\bbx_t$ in \eqref{E:SEM} models network effects, implying $x_{it}$ is a linear combination of the instantaneous values $x_{jt}$ of node $i$'s in-neighbors $j\in\ccalN_i$. The signal $x_{it}$ also depends on $u_{it}$, where weight $\omega_{ii}$ captures the level of influence of external sources and we defined $\mathbf{\Omega}:=\textrm{diag}(\omega_{11},\ldots,\omega_{NN})$. Vector $\bbepsilon_t$ accounts for measurement errors and unmodeled dynamics. 

Depending on the context, $\bbx_t$ can be thought of as an output signal while $\bbu_t$ corresponds to the excitation or control input. In the absence of noise and letting $\mathbf{\Omega}=\bbI$ for simplicity, \eqref{E:SEM} becomes $\bbx_t=(\bbI-\bbW)^{-1}\bbu_t$, where $\bbH:=(\bbI-\bbW)^{-1}$ is a polynomial graph filter in the graph-shift operator $\bbS=\bbW$. If $\bbW$ is further assumed to be symmetric, one recovers a particular instance of the signal model adopted in Section \ref{ssec:non-stationary}; see also~\cite{topoid_directed_dsw18}.

Given snapshot observations $\ccalX:=\{\bbx_t,\bbu_t\}_{t=1}^T$, SEM parameters $\bbW$ and $\bbomega:=[\omega_{11},\ldots,\omega_{NN}]^T$ are typically estimated via penalized LS, for instance by solving~\cite{BazerqueGeneNetworks}
\begin{align}\label{E:SEM_estimation}
\min_{\bbW,\bbomega}&{}\left\{\sum_{t=1}^T\|\bbx_t-\bbW\bbx_t+\mathbf{\Omega}\bbu_t\|^2+\alpha\|\bbW\|_1\right\}\\
\textrm{ s. to } &{} \quad\mathbf{\Omega}=\textrm{diag}(\bbomega),\quad W_{ii}= 0,\: i=1,\ldots,N\nonumber
\end{align}
where the $\ell_1$-norm penalty promotes sparsity in the adjacency matrix. Both edge sparsity as well as the endogenous inputs play a critical role towards guaranteeing the SEM parameters \eqref{E:SEM} are uniquely identifiable; see also~\cite{ggtopoid2018piee}.

While SEMs only capture contemporaneous relationships among the nodal variables (i.e., SEMs are memoryless), the class of sparse vector autoregressive models (SVARMs) account for linear time-lagged (causal) influences instead; see e.g.,~\cite{varm_group_sparse,timeseries2005book}. Specifically, for given model order $L$ and unknown sparse evolution matrices $\{\bbW^{(l)}\}_{l=1}^L$, SVARMs postulate a multivariate linear dynamical model of the form
\begin{equation}\label{E:SVARM}
\bbx_t=\sum_{l=1}^L\bbW^{(l)}\bbx_{t-l}+\bbepsilon_t.
\end{equation}
Similar to the neighborhood-based regression approaches we encountered in Section \ref{ssec:regression}, here a directed edge from vertex $j$ to $i$ is present in $\ccalG$ if at least one of $\{w_{ij}^{(l)}\}_{l=1}^L$ is nonzero (the OR rule). The other common alternative relies on the AND rule, which requires $w_{ij}^{(l)}\neq 0$ for all $l=1,\ldots,L$ to have $(i,j)\in \ccalE$. The AND rule is often explicitly imposed as a constraint during estimation of SVARM parameters, through the requirement that all matrices $\bbW^{(l)}$ have a common support. This can be achieved for instance via a group lasso penalty, that promotes sparsity over edgewise coefficients $\bbw_{ij}:=[w_{ij}^{(1)},\ldots,w_{ij}^{(L)}]^T$ jointly~\cite{varm_group_sparse}. The sparsity assumption is often well-justified due to physical considerations or for the sake of interpretability, but here it is also critical to reliably estimate the graph from limited and noisy data.

The benefits of SEMs and SVARMs can be leveraged jointly through so-termed \emph{structural} VARMs, which augment the right-hand-side of \eqref{E:SVARM} with a term $\bbW^{(0)}\bbx_t$ to also capture instantaneous relationships among variables as in \eqref{E:SEM}; see also~\cite{ggtopoid2018piee}. In~\cite{willet_autoregressive}, the inference of the autoregressive parameters and associated network structure is studied within a generalized SVARM framework that includes discrete Poisson and Bernoulli autoregressive processes. SVARMs are also central to popular digraph topology identification approaches based on the principle of Granger causality~\cite{granger}. Said principle is based on the concept of precedence and predictability,
where node $j$'s time series is said to ``Granger-cause'' the time series at node $i$ if knowledge of $\{x_{j,t-l}\}_{l=1}^L$
improves the prediction of $x_{it}$ compared to using only $\{x_{i,t-l}\}_{l=1}^L$. Such form of causal dependence defines the status of a candidate edge from $j$ to $i$, and it can be assessed via judicious hypothesis testing~\cite{timeseries2005book}. 

Recently, a notion different from Granger's was advocated to associate a graph with causal network effects among vertex time series, effectively blending VARMs with graph filter-based dynamical models. The so-termed causal graph process (CGP) introduced in~\cite{mei_tsp2017} has the form
\begin{align}\label{E:causal_graph_process}
\bbx_t={}&{} \sum_{l=1}^L\sum_{i=0}^l h_{li}\bbS^i\bbx_{t-l}+\bbepsilon_t\nonumber\\
={}&{}(h_{10}\bbI +h_{11}\bbS)\bbx_{t-1}+\ldots+(h_{L0}\bbI+\ldots+h_{LL}\bbS^L)\bbx_{t-L}+\bbepsilon_t
\end{align}
where $\bbS$ is the (possibly asymmetric) graph-shift operator encoding the unknown graph topology. The CGP model corresponds to a generalized VARM [cf. \eqref{E:SVARM}], with coefficients given by graph filters $\bbH_l(\bbS,\bar{\bbh}):=\sum_{i=0}^l h_{li}\bbS^i$,and  where  $\bar{\bbh}:=[h_{10}, h_{11},\ldots,h_{li},\ldots h_{LL}]^T$. This way, the model can possibly account for multi-hop nodal influences per time step. Unlike SVARMs, matrices $\bbH_l(\bbS,\bar{\bbh})$ need not be sparse for larger values of $l$, even if $\bbS$ is itself sparse. 

Given data $\ccalX:=\{\bbx_t\}_{t=1}^T$ and a prescribed value of $L$, the approach to estimating $\bbS$ entails solving the non-convex optimization problem
\begin{equation}\label{E:mei_graph_estimation}
\min_{\bbS,\bar{\bbh}}\left\{\sum_{t=L+1}^{T}\Big\|\bbx_t-\sum_{l=1}^L\bbH_l(\bbS,\bar{\bbh})\bbx_{t-l}\Big\|^2+\alpha\|\bbS\|_1+\beta\|\bar{\bbh}\|_1\right\}.
\end{equation}
Similar to sparse SEMs in \eqref{E:SEM_estimation} and SVARMs, the estimator encourages sparse graph topologies. Moreover, the $\ell_1$-norm regularization on the filter coefficients $\bar{\bbh}$ effectively implements a form of model-order selection. A divide-and-conquer heuristic is advocated in~\cite{mei_tsp2017} to tackle the challenging problem \eqref{E:mei_graph_estimation}, whereby one: (i) identifies the filters $\bbH_l:=\bbH_l(\bbS,\bar{\bbh})$ so that $\bbx_t \approx \sum_{l=1}^L\sum_{i=0}^l \bbH_l\bbx_{t-l}$, exploiting that $\bbH_l$ and $\bbH_{l'}$ commute for all $l,l'$; (ii) recovers a sparse $\bbS$ using the estimates $\{\hbH_l\}$ and leveraging the shift-invariant property of graph filters; and (iii) estimates $\bar{\bbh}$ given $\{\hbH_l,\hbS\}$ via the lasso. For full algorithmic details and an accompanying convergence analysis under some technical assumptions, please see~\cite{mei_tsp2017}.

% % % % % % % % % % % % % % % % % % % % % % % % % % % % % % % % % % % % % % % %
%                        Subsection VIII-B                                      %
% % % % % % % % % % % % % % % % % % % % % % % % % % % % % % % % % % % % % % % %

\subsection{Dynamic networks and multi-layer graphs}
\label{ssec:dynamic_multi}

As data become more complex and heterogeneous, possibly generated in a streaming fashion by non-stationary sources, it is becoming increasingly common to rely on models comprising \textit{multiple} related networks describing the interactions between various entities. While dynamic graphs with time-varying topologies naturally fall within this general class of models, the multiple graphs of interest need not be indexed by time, but possibly instead by different subjects, demographic variables, or sensing modalities. This is for instance the case in neuroscience, where observations for different patients are available and the objective is to estimate their functional brain networks; and in computational genomics where the goal is to identify pairwise interactions between genes when measurements for different tissues of the same patient are available. In order to unveil hidden structures, detect anomalies, and interpret the temporal dynamics of such data, it is essential to understand the relationships between the different entities and how these relationships evolve over time or across different modalities. Joint identification of multiple graph-shift operators can be useful even when interest is only in one of the networks, since joint formulations exploit additional sources of information and, hence, they are likely to result in more accurate topology estimates. Although noticeably less than its single-network counterpart, joint inference of multiple graphs has attracted attention, especially for the case of GMRFs and in the context of dynamic (time-varying) topologies~\cite{ggtopoid2018piee,Kalofolias2017inference_dynamic,baingana2014cascadesJSTSP14,shen2017tensors,time_varying_graphical_lasso}. 
All of the aforementioned works consider that the multiple graphs $\ccalG_t(\ccalV,\ccalE_t,\bbW_t)$, $t=1,\ldots,T$, share a common vertex set while being allowed to have different edge sets and weights, a structure oftentimes referred to as a multi-layer graph. Given the previous motivation, here we extend several of the problem formulations in the previous sections to accommodate (dynamic) multi-layer graphs.

To state the joint network topology inference problem in its various renditions, consider a scenario with $T$ different graphs $\ccalG_t(\ccalV,\ccalE_t,\bbW_t)$ defined over a common set $\ccalV$ of nodes, with $|\ccalV|=N$. This implies the existence of $T$ different graph-shift operators $\{\bbS_t\}_{t=1}^T$ that we want to recover, all represented by $N\times N$ matrices, whose sparsity pattern and nonzero values may be different across $t$. Suppose that for each graph we have access to a set of graph signals $\ccalX_t:=\{\bbx_t^{(p)}\}_{p=1}^{P_t}$ collecting information associated with the nodes. Equivalently, it will be convenient to represent $\ccalX_t$ through the matrix $\bbX_t:=[\bbx_t^{(1)},\ldots,\bbx_t^{(P_t)}]\in\reals^{N\times P_t}$ containing the $P_t$ signals associated with graph $\ccalG_t$. A popular approach to joint inference of multi-layer networks assumes that graphs $\ccalG_t$ and $\ccalG_{t-1}$ are \emph{similar} in some application-dependent sense, which we encode as some matrix distance $r(\bbS_t,\bbS_{t-1})$ being small. For instance, this could be well motivated for identification of a sequence of slowly time-varying graphs. In this context, a general formulation of the multi-layer graph learning problem entails solving
\begin{equation}\label{E:tv_estimator_general}
\{\bbS^{*}_t\}_{t=1}^T :=\arg\min_{\{\bbS_t\in \ccalS\}_{t=1}^T }\left\{\sum_{t=1}^T\Phi_t(\bbS_t,\bbX_t)+\alpha \sum_{t=2}^Tr(\bbS_t,\bbS_{t-1})\right\},
\end{equation}
where $\Phi_t$ is an often convex objective function stemming from the adopted topology identification criterion, and $\alpha>0$ is a regularization parameter. In~\cite{time_varying_graphical_lasso}, the so-termed time-varying graphical lasso estimator was proposed to identify a collection of GMRFs encoded in the precision matrices $\bbS_t=\bbTheta_t:=\bbSigma_t^{-1}$.  Therein,  $\Phi_t(\bbS_t,\bbX_t)=-\log\det\bbTheta_t +\textrm{trace}(\hbSigma_t \bbTheta_t)+\lambda\|\bbTheta_t\|_1$ corresponds to the penalized global likelihood for $\bbx_t\sim\textrm{Normal}(\mathbf{0},\bbSigma_t)$ we encountered in Section \ref{ssec:glasso}. They also propose a comprehensive list of distance functions $r$ to encode different network evolutionary patterns including smooth and abrupt topology transitions; see~\cite{time_varying_graphical_lasso} for additional details. From an algorithmic standpoint, an alternating-direction method of multipliers (ADMM) solver is adopted to tackle \eqref{E:tv_estimator_general} efficiently.

A general estimator for learning slowly time-varying graphs over which signals in $\ccalX_t$ exhibit smooth variations was put forth in~\cite{Kalofolias2017inference_dynamic}. Let $\bbS_t=\bbW_t$ denote the adjacency matrix of $\ccalG_t$ and recall the Euclidean-distance matrix $\bbZ_t$ defined in Section \ref{ssec:smooth_sparse}. The idea in~\cite{Kalofolias2017inference_dynamic} is to set $\Phi_t(\bbW_t,\bbX_t)=\|\bbW_t\circ\bbZ_t\|_1-\alpha\mathbf{1}^T\log(\bbW_t\mathbf{1})+\frac{\beta}{2}\|\bbW_t\|_F^2$ in \eqref{E:tv_estimator_general} along with $r(\bbS_t,\bbS_{t-1})=\|\bbS_t-\bbS_{t-1}\|_F^2$, and rely on the primal-dual algorithmic framework of~\cite{Kalofolias2016inference_smoothAISTATS16} to tackle the resulting separable optimization problem. Through a different choice of $\Phi_t(\bbW_t,\bbX_t)$, the framework therein can also accommodate a time-varying counterpart of the model in~\cite{DongLaplacianLearning}. Network topology inference approaches that rely on observations of stationary signals have been extended to the multi-layer graph setting as well. As done for $T=1$, we assume that $\ccalX_t:=\{\bbx_t^{(p)}\}_{p=1}^{P_t}$ are independent realizations of a random process $\bbx_t$ whose structure is represented by $\bbS_t$ [cf. the model of network diffusion in~\eqref{eqn_diffusion}]. Mimicking the development in Section~\ref{sssec:step_2}, given estimates $\{\hbV_t\}_{t=1}^T$ of the eigenvectors of each of the sought graph-shift operators (e.g., obtained from the sample covariances of the different sets of observations $\{\ccalX_t\}_{t=1}^T$), recovery of the shifts boils down to selecting the optimal eigenvalues $\{\bbLambda_t\}_{t=1}^T$. To this end, \eqref{E:general_problem} can be adapted to obtain~\cite{AsilomarMultipleShifts17}
\begin{align}\label{E:general_problem_multiple}
	\{\bbS^{*}_t\}_{t=1}^T := & \argmin_{\{\bbLambda_t,\, \bbS_t \in \ccalS \}_{t=1}^T } \left\{
	\sum_{t=1}^T  f(\bbS_t) +  \alpha\sum_{t>t'} r(\bbS_t,\bbS_{t'}) \right\}\\
	& \text{s. to } \,\,\, d(\bbS_t,\hbV_t \bbLambda_t \hbV^T_t)\leq \epsilon, \,\, t=1,\ldots,T. \nonumber
\end{align}
As discussed in Section~\ref{sssec:step_2}, the optimality criterion $f$ and the constraint set $\ccalS$ can be selected to promote or enforce desirable properties in the sought graphs $\ccalG_t$. Specific to formulation \eqref{E:general_problem_multiple}, choosing the distance function $r$ as $\|\bbS_t-\bbS_{t'}\|_1$ would promote the pair of shifts to have the same sparsity pattern and weights, whereas the choice $\|\bbS_t-\bbS_{t'}\|_{F}^2$ as in~\cite{Kalofolias2017inference_dynamic} would promote similar weights regardless of the sparsity pattern. Check~\cite{AsilomarMultipleShifts17} for further details and results including provable guarantees of the associated algorithms. Last but not least, for a recent work dealing with inference of both directed and multi-layer graphs from observations of diffused graph signals, we refer the readers to~\cite{topoid_arma1_ssp18}. 

On a related note, network inference from temporal traces of infection events has recently
emerged as an active area of research, which is relevant to epidemic processes, propagation of viral news events between blogs, 
or acquisition of new buying habits by consumer groups. It has been observed in these settings that information often spreads in
cascades by following implicit links between nodes in a time-varying graph $\ccalG_t$. Reasoning that infection times depend on both topological (endogenous)
and external (exogenous) influences, a \emph{dynamic} SEM-based scheme was proposed
in~\cite{baingana2014cascadesJSTSP14} for cascade modeling. With $c=1,\ldots,C$ indexing cascades, similar to \eqref{E:SEM} we model 
topological influences as linear
combinations of infection times $x_{it}^{(c)}$ of other nodes in the network, whose weights
correspond to entries in a time-varying asymmetric adjacency matrix $\bbW_t$. External
influences $u_{i}^{(c)}$ such as those due to on-site reporting in news propagation contexts
are useful for model identifiability, and they are taken to be time invariant for simplicity. 
It is assumed that the networks $\ccalG_t$ vary slowly with time, facilitating adaptive SEM parameter estimation by minimizing a
sparsity-promoting exponentially-weighted LS criterion~\cite{baingana2014cascadesJSTSP14}
\begin{align}\label{E:dynamic_SEM_estimation}
\min_{\bbW_t,\bbomega_t}&{}\left\{\sum_{c=1}^C\sum_{t=1}^T\gamma^{t-T}\|\bbx_t^{(c)}-\bbW_t\bbx_t^{(c)}+\mathbf{\Omega}_t\bbu^{(c)}\|^2+\alpha_t\|\bbW\|_1\right\}\\
\textrm{ s. to } &{} \quad\mathbf{\Omega_t}=\textrm{diag}(\bbomega_t),\quad W_{ii,t}= 0,\: i=1,\ldots,N,\: t=1,\ldots,T\nonumber
\end{align}
where $\gamma\in (0, 1]$ is the forgetting factor that forms estimates using
all measurements acquired until time $T$. Whenever $\gamma < 1$, past data are
exponentially discarded thus enabling tracking of dynamic network topologies; see also Section \ref{ssec:cascades} for a numerical test with real data from blog posts in 2011. Related likelihood-based approaches have been proposed to identify traces of network diffusion~\cite{gomez}, as well as tensor-based topology identification for dynamic SEMs that can account for (abruptly) switching topological states representing the layers of the graph~\cite{shen2017tensors}.

% % % % % % % % % % % % % % % % % % % % % % % % % % % % % % % % % % % % % % % %
%                        Subsection VIII-C                                      %
% % % % % % % % % % % % % % % % % % % % % % % % % % % % % % % % % % % % % % % %

\subsection{Nonlinear models of interaction}
\label{ssec:nonlinear}

Network models such as SEMs or SVARMs are linear, and the same is true for most measures of pairwise similarity we encountered; notably those based on Pearson or partial correlations. However, in complex systems such as the brain there is ample evidence that dynamics involve nonlinear interactions between neural regions, and accordingly linear models fall short in capturing such dependencies.  

Measures capable of summarizing nonlinear association, such as mutual information, might be used depending on the suspected dependencies in the data. Building on the relationship between linear prediction and partial correlations, one can by analogy construct nonlinear measures of interaction among nodal time series by relying on nonlinear (e.g., kernel-based) predictors instead~\cite{Karanikolas_icassp16}. Subsequent hypothesis testing can be performed to decide between presence or absence of edges in the graph. Special care should be exercised  when selecting a test statistic, since the challenges faced in determining a (tractable, even approximate) null distribution may be compounded. Issues of multiple-testing should be accounted for as well, similar to those discussed for (partial) correlation networks. Such an approach cannot infer directionality and so results in an undirected graph of nonlinear correlations. Kernelized counterparts of structural VARMs (and subsumed SEMs) have been proposed in~\cite{shen2016kernelsTSP16} to identify the topology of digraphs, while explicitly accounting for nonlinearities. For a comprehensive treatment of the problem of learning graphs from data involving nonlinear dependencies, the interested reader is referred to the recent survey~\cite{ggtopoid2018piee}, which also touches upon prediction of (nonlinear and dynamic) processes supported on graphs.

% % % % % % % % % % % % % % % % % % % % % % % % % % % % % % % % % % % % % % % %
%                         Section IX                                         %
% % % % % % % % % % % % % % % % % % % % % % % % % % % % % % % % % % % % % % % %

\section{Applications}
\label{sec:applications}

This section presents numerical tests conducted on real data to demonstrate the effectiveness of selected graph learning methods, ranging from  ad hoc thresholding-based network constructions all the way to algorithms for identification of directed, time-varying graphs. Through test cases we show impact to diverse application domains including the economy, computational biology, neuroscience, and online social media.

% % % % % % % % % % % % % % % % % % % % % % % % % % % % % % % % % % % % % % % %
%                        Subsection IX-A                                      %
% % % % % % % % % % % % % % % % % % % % % % % % % % % % % % % % % % % % % % % %

\subsection{Efficient representation of signals supported on a network of US economic sectors}
\label{ssec:US_economy}

%%%%%%%%%%%%%%%   F   I   G   U   R    E   %%%%%%%%%%%%%%%%%%%%%%%%%%%%%%%%%%%%%%%%
\begin{figure*}[t]
	\begin{minipage}[b]{.48\linewidth}
		\centering
		\includegraphics[width=\linewidth]{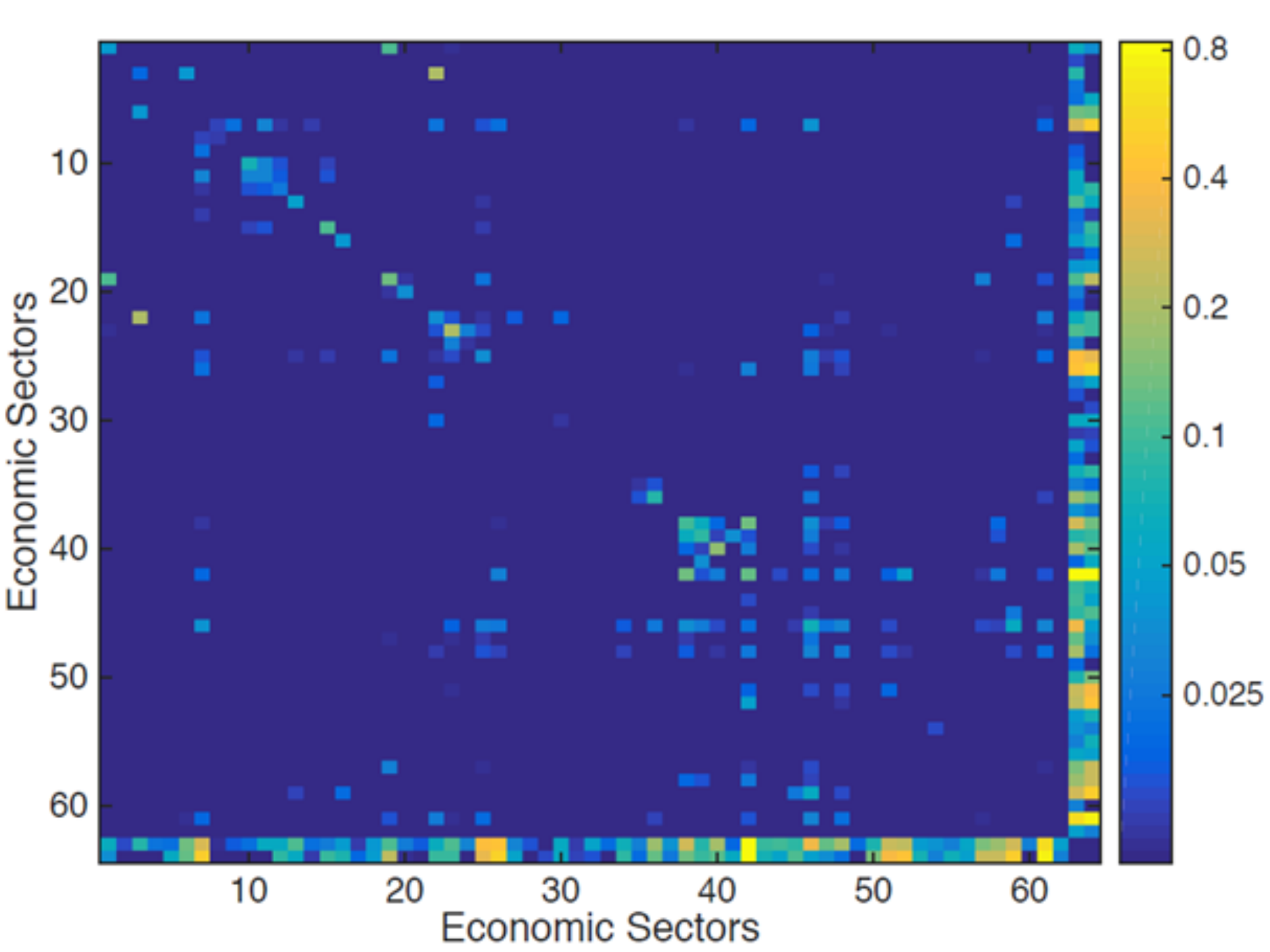}
		\centerline{(a)}\medskip
	\end{minipage}
	\hfill
	\begin{minipage}[b]{.48\linewidth}
		\centering
		\includegraphics[width=\linewidth]{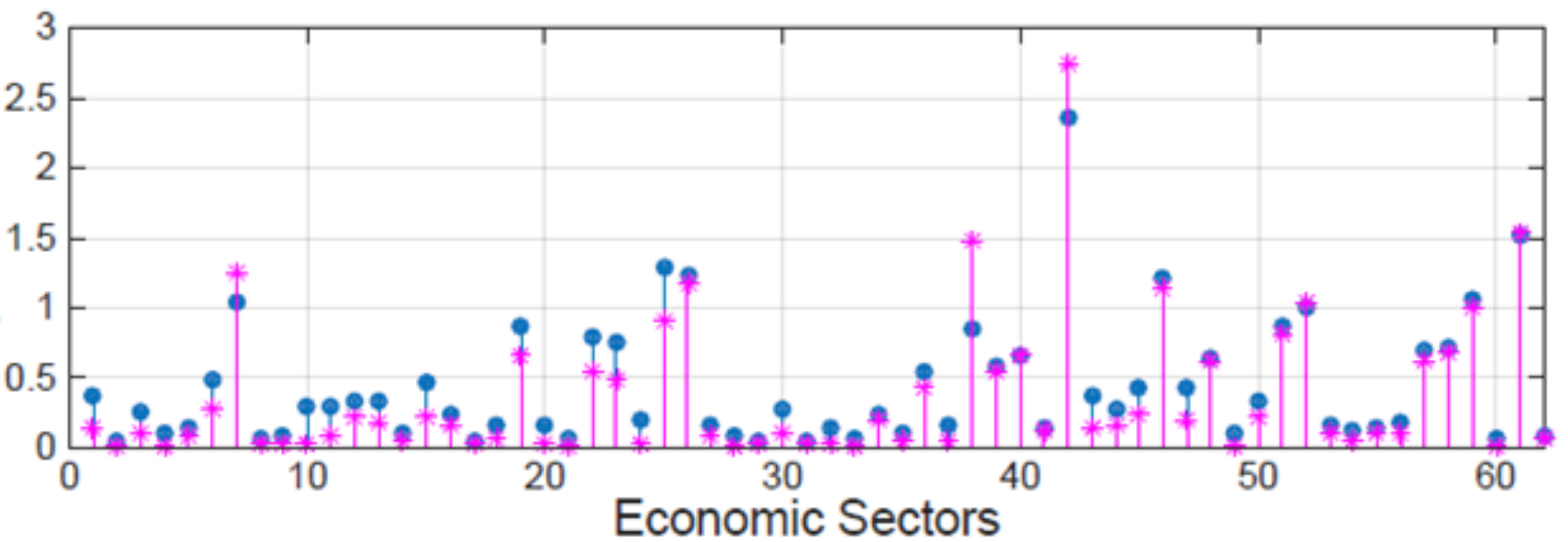}
		\vspace{0.4cm}
		\includegraphics[width=\linewidth]{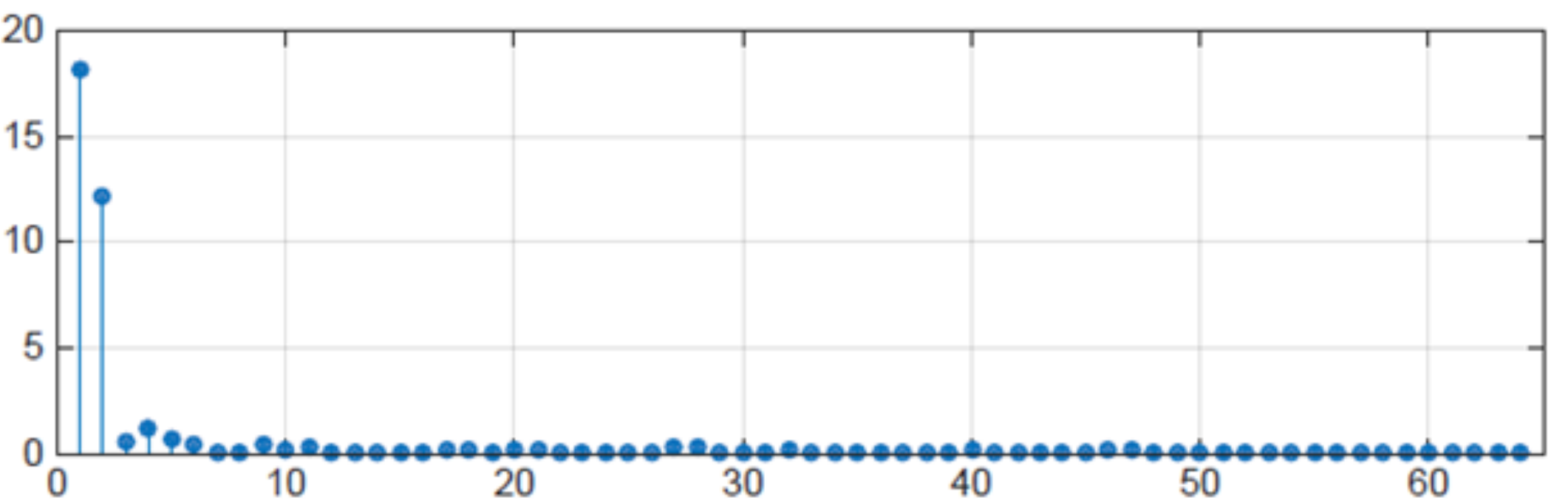}
		\centerline{(b)}\medskip
	\end{minipage}
	\hfill
	\vspace*{-0.5cm}
	\caption{(a) Heat map of the graph-shift operator $\bbS$ of the US economic network.
		It is sparse across the real economic sectors (from sector 1 to 62) while
		the synthetic sectors AV and FU are highly connected.  (b) Disaggregated GDP signal $\bbx$ (blue) and its reconstruction $\hbx$ (magenta) when keeping only the first 4 frequency components (top). Frequency representation of the graph signal $\bbx$ in the basis of eigenvectors of the graph-shift $\bbS$ (bottom). The signal is approximately bandlimited~\cite{antonio2016sampling_tsp}.} 
	\label{F:network_US_economy}
\end{figure*}
%
%%%%%%%%%%%%%%%%%%%%%%%%%%%%%%%%%%%%%%%%%%%%%%%%%%%%%%%%%%%%%%%%%%%

The Bureau of Economic Analysis of the US Department of Commerce publishes a yearly table of inputs and outputs
organized by economic sectors. More precisely, we have a set of 62 industrial sectors as defined by the North American
Industry Classification System and a similarity function $\bbW: \ccalV\times \ccalV \to \reals_{+}$, where $W_{ij}$ represents how much of the production of sector $i$, expressed in trillions of US dollars per year, was used as an input of sector $j$ on average during years 2008-2010. Moreover, for each sector we are given two economic markers: the added value (AV) generated and the
level of production destined to the market of final users (FU). Thus, we define a graph $\ccalG$ on the set of $N = 64$ nodes comprising
the original 62 sectors plus the two synthetic ones (AV and FU) and an associated symmetric graph-shift operator $\barbS$ defined as
$\bar{S}_{ij} := (W_{ij} + W_{ji})/2$. We then threshold $\barbS$ by setting to 0 all the values lower than 0.01 to obtain the sparser, symmetric shift operator $\bbS=\bbV\bbLambda\bbV^T$; see Figure \ref{F:network_US_economy} (a).

While this is an intriguing network that helps to understand the interactions among sectors and reveals how these sectors are clustered, arguably signals defined on such graph are equally interesting and the subject of economic studies. For instance, consider the disaggregated gross domestic product (GDP) signal $\bbx\in\reals_{+}^{64}$ on $\ccalG$, where $x_i$ represents the total production
(in trillion of dollars) of sector $i$ (including AV and FU) during year 2011. As shown in the bottom plot of Figure \ref{F:network_US_economy} (b), signal $\bbx$ is approximately bandlimited in $\bbS$ since most of the elements of $\tbx=\bbV^T\bbx$ are close to zero. In particular, the reconstructed signal $\hbx=\sum_{k=1}^4\tdx_k \bbv_k$ obtained by just keeping the first four GFT coefficients attains a relative reconstruction error of $3.5\times10^{-3}$; see the top plot of Figure \ref{F:network_US_economy} (b) which shows the original GDP signal superimposed to $\hbx$. To present a reasonable scale for illustration, sectors AV and FU are not included in Figure \ref{F:network_US_economy} (b), since the signal takes out-of-scale values for these sectors. All in all, this example shows that -- while heuristic -- the adopted graph construction approach still yields a useful graph to sparsely represent the disaggregated GDP signal. In a way, this serves as validation of $\ccalG$ and it also highlights the value of the GFT decomposition.

% % % % % % % % % % % % % % % % % % % % % % % % % % % % % % % % % % % % % % % %
%                        Subsection IX-B                                      %
% % % % % % % % % % % % % % % % % % % % % % % % % % % % % % % % % % % % % % % %

\subsection{Identifying protein structure via network deconvolution}
\label{ssec:deconvolution}

%%%%%%%%%%%%%%%   F   I   G   U   R    E   %%%%%%%%%%%%%%%%%%%%%%%%%%%%%%%%%%%%%%%%
%
\begin{figure*}[t]
	\centering
	\input{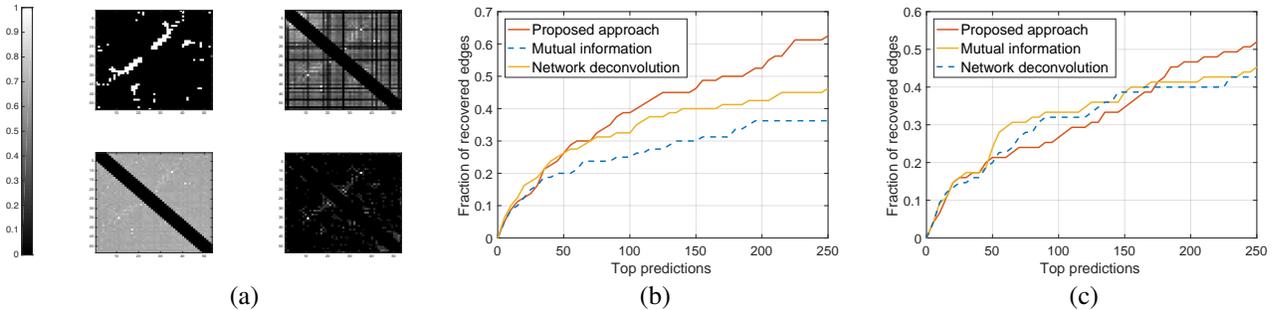}	
	\vspace*{-1cm}
	\caption{(a) Real and inferred contact networks between amino-acid residues for protein BPT1 BOVIN. Ground truth contact network (top left), mutual information of the co-variation of amino-acid residues (top right), contact network inferred by network deconvolution~\cite{FeiziNetworkDeconvolution} (bottom left), contact network inferred by the approach in~\cite{segarra2016topoidTSP16} (bottom right). (b) Fraction of the real contact edges between amino-acids recovered for each method as a function of the number of edges considered. (c) Counterpart of (b) for protein YES HUMAN~\cite{segarra2016topoidTSP16}.}
	\label{F:network_sparsification}
\end{figure*}
%
%%%%%%%%%%%%%%%%%%%%%%%%%%%%%%%%%%%%%%%%%%%%%%%%%%%%%%%%%%%%%%%%%%%

The network deconvolution problem is to identify a sparse adjacency matrix $\bbS=\bbW$ that encodes \emph{direct} dependencies, when given a symmetric adjacency $\bbT$ of \emph{indirect} relationships. The problem broadens the scope of e.g., channel deconvolution to networks and can be tackled by adopting the model~\cite{FeiziNetworkDeconvolution}
\begin{equation}\label{eq_deconv}
\bbT = \bbS \, (\bbI - \bbS)^{-1}=\sum_{l=1}^\infty\bbS^l.
\end{equation} 
This solution assumes a diffusion as in \eqref{eqn_diffusion} but for the particular case of a single-pole, single-zero graph filter, with very specific filter coefficients. This way, the indirect dependencies observed in $\bbT$ arise due to the higher order terms $\bbS^2+\bbS^3+\ldots$ superimposed to the fundamental interactions in $\bbS$. In this numerical test case, we  adopt a more general approach in assuming that $\bbT$ can be written as a polynomial in $\bbS$, but being agnostic to the form of the filter~\cite{segarra2016topoidTSP16}. This naturally leads back to formulation \eqref{E:general_problem}, with $\bbV$ given by the eigenvectors of $\bbT$ and $f(\bbS)$ chosen as an edge sparsity-promoting criterion. 
Different from the problem dealt with in Section \ref{sec:topoid_diffusion}, note that matrix $\bbT$ is not necessarily a covariance matrix.

Consider identifying the structural properties of proteins from a mutual information graph of the co-variation between the constitutional amino-acids~\cite{FeiziNetworkDeconvolution,segarra2016topoidTSP16}. Pictorially, for a particular protein we want to recover the structural graph in the top left of Figure~\ref{F:network_sparsification} (a) when given the graph of mutual information in the top right corner.  The graph recovered by network deconvolution~\cite{FeiziNetworkDeconvolution} is illustrated in the bottom left corner of Figure~\ref{F:network_sparsification} (a), whereas the one recovered using the approach in \eqref{E:general_problem} (with the sparsity-promoting $f(\bbS)=\|\bbS\|_1$) is depicted in the bottom right corner. The comparison of the recovered graphs demonstrates that using a general filter model translates to a sparser graph that captures more accurately the desired structure. To quantify this latter assertion, Figure~\ref{F:network_sparsification} (b) depicts the fraction of the real contact edges recovered for each method as a function of the number of edges considered. For example, if for a given method the 100 edges with largest weight in the recovered graph contain $40\%$ of the edges in the ground truth graph, we say that the 100 top edge predictions achieve a fraction of recovered edges equal to $0.4$. From Figure~\ref{F:network_sparsification} (b) it follows that the method in Section \ref{ssec:stationarity_for_topoid} outperforms network deconvolution and the raw mutual information data. For instance, when considering the top $200$ edges, the mutual information and the network deconvolution methods recover $36\%$ and $43\%$ of the desired edges, respectively, while the solution of  \eqref{E:general_problem} achieves a recovery of $53\%$.  Figure~\ref{F:network_sparsification} (c) shows the results for a different protein network and similar trends can be appreciated.

% % % % % % % % % % % % % % % % % % % % % % % % % % % % % % % % % % % % % % % %
%                        Subsection IX-C                                      %
% % % % % % % % % % % % % % % % % % % % % % % % % % % % % % % % % % % % % % % %

\subsection{Graph frequency analysis of brain signals during learning}
\label{ssec:brain}

%%%%%%%%%%%%%%%   F   I   G   U   R    E   %%%%%%%%%%%%%%%%%%%%%%%%%%%%%%%%%%%%%%%%
\begin{figure*}[t]
	\begin{minipage}[b]{.31\linewidth}
	\centering
	\includegraphics[width=\linewidth]{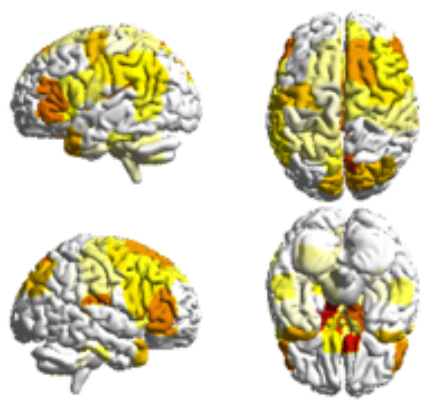}
	  \centerline{(a)}\medskip
\end{minipage}
\hfill
\begin{minipage}[b]{.31\linewidth}
	\centering
	\includegraphics[width=\linewidth]{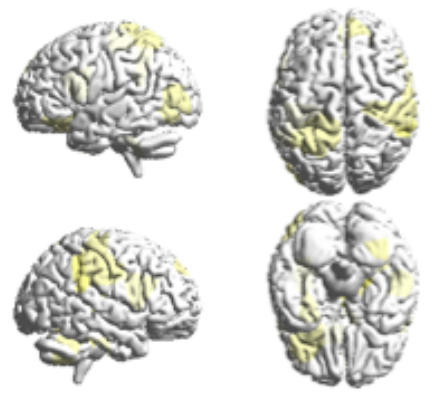}
		  \centerline{(b)}\medskip
\end{minipage}
\hfill
\begin{minipage}[b]{.31\linewidth}
	\centering
	\includegraphics[width=\linewidth]{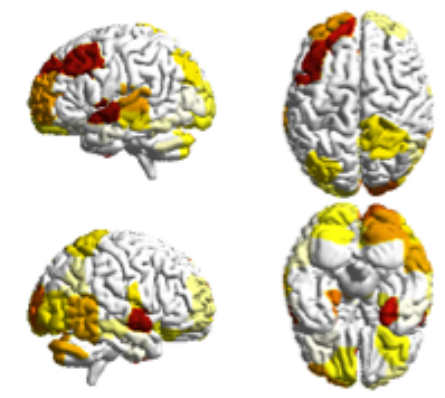}
		  \centerline{(c)}\medskip
\end{minipage}
\vspace*{-0.5cm}
	\caption{Brain activity patterns for a visual-motor learning task decomposed in (a) low, (b) medium, and (c) high frequency components of the GFT relative to the brain connectivity network. The signals are dominated by low and high frequency components with minimal contribution from medium frequency components~\cite{weiyu2016brain_jstsp}.}
	\label{F:network_brain}
\end{figure*}
%
%%%%%%%%%%%%%%%%%%%%%%%%%%%%%%%%%%%%%%%%%%%%%%%%%%%%%%%%%%%%%%%%%%%

Graph frequency analyses have been recently applied to study brain activity signals under the setup of an experiment in which subjects learned a simple motor task~\cite{weiyu2016brain_jstsp}. Participants responded to sequentially presented stimuli with their dominant hand. Sequences were presented using a horizontal array of 5 square stimuli with the responses mapped from left to right such that the thumb corresponded to the leftmost stimulus. The next square in the sequence was highlighted immediately following each correct key press; the sequence was paused awaiting the depression of the appropriate key if an incorrect key was pressed. All participants completed the task inside a magnetic resonance imaging (MRI) scanner.

To construct brain networks, the whole brain is parcellated into a set of $N = 112$ regions that correspond to the set $\ccalV$ of 112 cortical and subcortical structures specified in the Harvard-Oxford atlas. For each individual functional (f)MRI dataset, the regional mean blood-oxygen-level dependent (BOLD) time series is estimated by averaging voxel time series in each of the $N$ regions. Pearson correlations between the activities of all possible pairs of regions are then evaluated and tested (at $5\%$ significance level) to construct $N\times N$ functional connectivity matrices $\bbW$. The graph-shift operator is defined as the associated Laplacian $\bbS := \bbL= \bbV \bbLambda \bbV^T$.
Regarding brain signals, we normalize the regional mean BOLD observations $\hbx_t$ at any time $t$ and consider $\bbx_t := \hbx_t / \| \hbx_t\|_2$, such that the total energy of activities at all structures is consistent at different $t$ to avoid extreme spikes due to head motion or drift artifacts in fMRI; see~\cite{weiyu2016brain_jstsp} for details.

Signal $\bbx$ is not highly bandlimited in $\bbS$, however, recall from Section \ref{ssec:GFT}, when the graph-shift operator $\bbS=\bbL$, the eigenvalue $\lambda_k$ associated with a given eigenvector $\bbv_k$ expresses a level of spatial variation with respect to the brain network. Following this direction, we analyze the decomposed signals of $\bbx$ associated with different levels of spatial variation. Figure \ref{F:network_brain} (a) presents the distribution of the decomposed signal $\bbx^{(\text{L})} :=\sum_{k=1}^{40}\tdx_k \bbv_k$ corresponding to smooth spatial variations. Figure \ref{F:network_brain} (b) displays the decomposed signals $\bbx^{(\text{M})} :=\sum_{k=41}^{72}\tdx_k \bbv_k$ associated with moderate spatial variation; Figure \ref{F:network_brain} (c) represents $\bbx^{(\text{H})} :=\sum_{k=73}^{112}\tdx_k \bbv_k$ corresponding to the fast spatial variation. 

A deep analysis yields several interesting observations. First, decomposed signals of a specific level of variation, e.g. $\bbx^{(\text{L})}$, are highly similar with respect to different sets of participants~\cite{weiyu2016brain_jstsp}. This reflects the fact that frequency decomposition is formed by applying graph filters with different pass bands upon signals and therefore should express some consistent aspects of brain signals. Second, because of the signal normalization at every sample point and for all subjects,  $\bbx^{(\text{L})}, \bbx^{(\text{M})}$ and $\bbx^{(\text{H})}$ would be similarly distributed across the brain if bandpass graph filtering segments brain signals into three equivalent pieces. However, it is observed (see Figure \ref{F:network_brain}) that many brain regions possess magnitudes higher than a threshold in $\bbx^{(\text{L})}$ and $\bbx^{(\text{H})}$, while not many brain regions exceed the thresholding with respect to $\bbx^{(\text{M})}$. Besides, brain regions with high magnitude values in $\bbx^{(\text{L})}$ and $\bbx^{(\text{H})}$ are highly alike to the visual and sensorimotor cortices, whose associations with motor learning task have been well recognized; see~\cite{weiyu2016brain_jstsp} and references therein. It has long been understood that the brain is a complex system combining some degree of disorganized behavior with some degree of regularity and that the complexity of a system is high when order and disorder coexist~\cite{sporns2011}. The low-pass signal $\bbx^{(\text{L})}$ varies smoothly across the brain network and therefore can be regarded as regularity (order), whereas $\bbx^{(\text{H})}$ fluctuates vibrantly and consequently can be considered as randomness (disorder). This observation can be leveraged in designing preprocessing steps to extract brain signals that are more informative and pertinent with learning, and to utilize the GFT to further analyze the association between different level of spatial variability with learning success.

% % % % % % % % % % % % % % % % % % % % % % % % % % % % % % % % % % % % % % % %
%                        Subsection IX-D                                      %
% % % % % % % % % % % % % % % % % % % % % % % % % % % % % % % % % % % % % % % %

\subsection{Tracking the propagation of information cascades}
\label{ssec:cascades}

Here we test the dynamic-SEM estimator \eqref{E:dynamic_SEM_estimation} for unveiling
sparse time-varying topologies, given real information cascade data obtained by
monitoring blog posts and news articles on the web between March 2011 and
February 2012 (45 weeks). Popular textual phrases (a.k.a. memes) due
to globally-popular topics during this period were identified and the times $x_{it}^{(c)}$
when they were mentioned on the websites were recorded. In order to illustrate tracking of dynamic propagation graphs, we extracted cascade traces $\bbx_t^{(c)}$  related to the keyword ``Kim Jong-un,'' the current leader of North Korea whose popularity rose after the death of his father (and predecessor in power) Kim Jong-il. Only significant cascades that propagated to at least 7
websites were retained. This resulted in a dataset with $N = 360$ websites, $C = 466$ cascades,
and $T = 45$ weeks. The exogenous inputs typically capture prior knowledge about susceptibility of nodes to contagions. In the web context dealt with here,
$u_{i}^{(c)}$ could be aptly set to the average search engine ranking of website $i$ on
keywords pertaining to $c$. In the absence of such real data for the observation
interval, the signal values $u_{i}^{(c)}$ were uniformly sampled over the interval $[0, 0.01]$; see~\cite{baingana2014cascadesJSTSP14}.

\begin{figure*}[t]
	\begin{minipage}[b]{.30\textwidth}
		\centering
		\includegraphics[width=\linewidth]{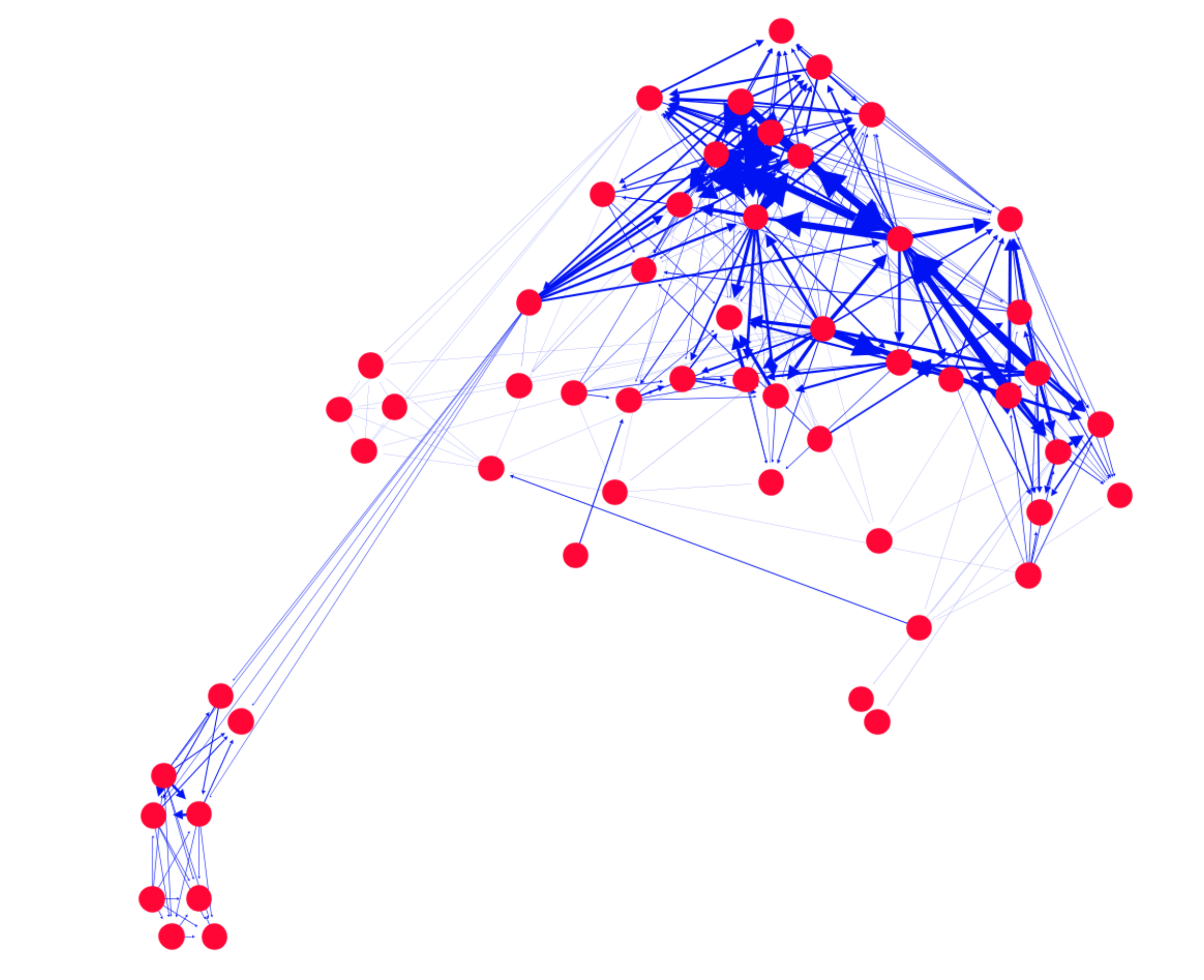}
		\centerline{(a)} %\medskip
	\end{minipage}
	%
	%\hfill
	\begin{minipage}[b]{.30\textwidth}
		\centering
		\includegraphics[width=\linewidth]{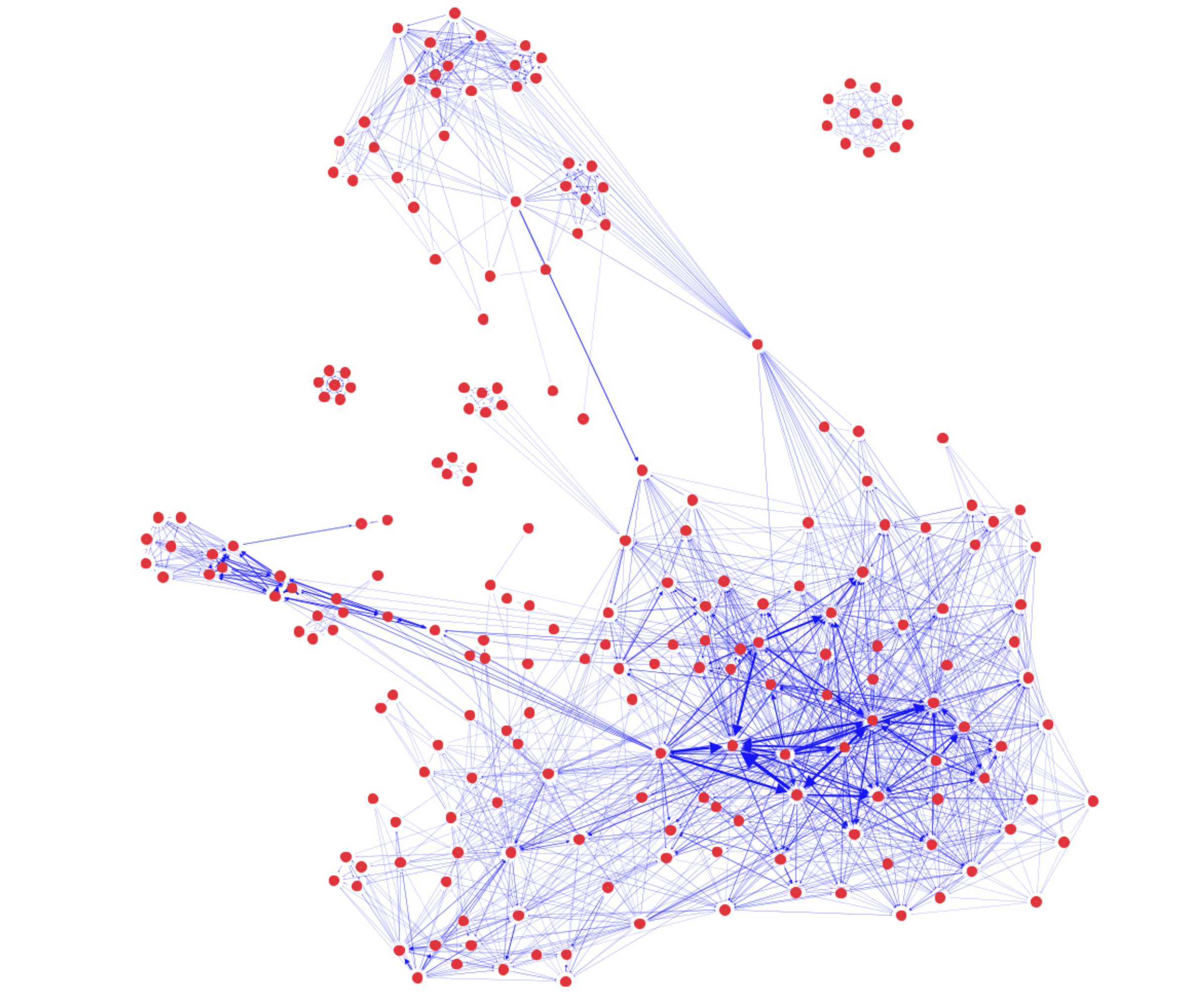}
		\centerline{(b)} %\medskip
	\end{minipage}
	%
	%\hfill
	\begin{minipage}[b]{0.39\textwidth}
		\centering
		\includegraphics[width=\linewidth]{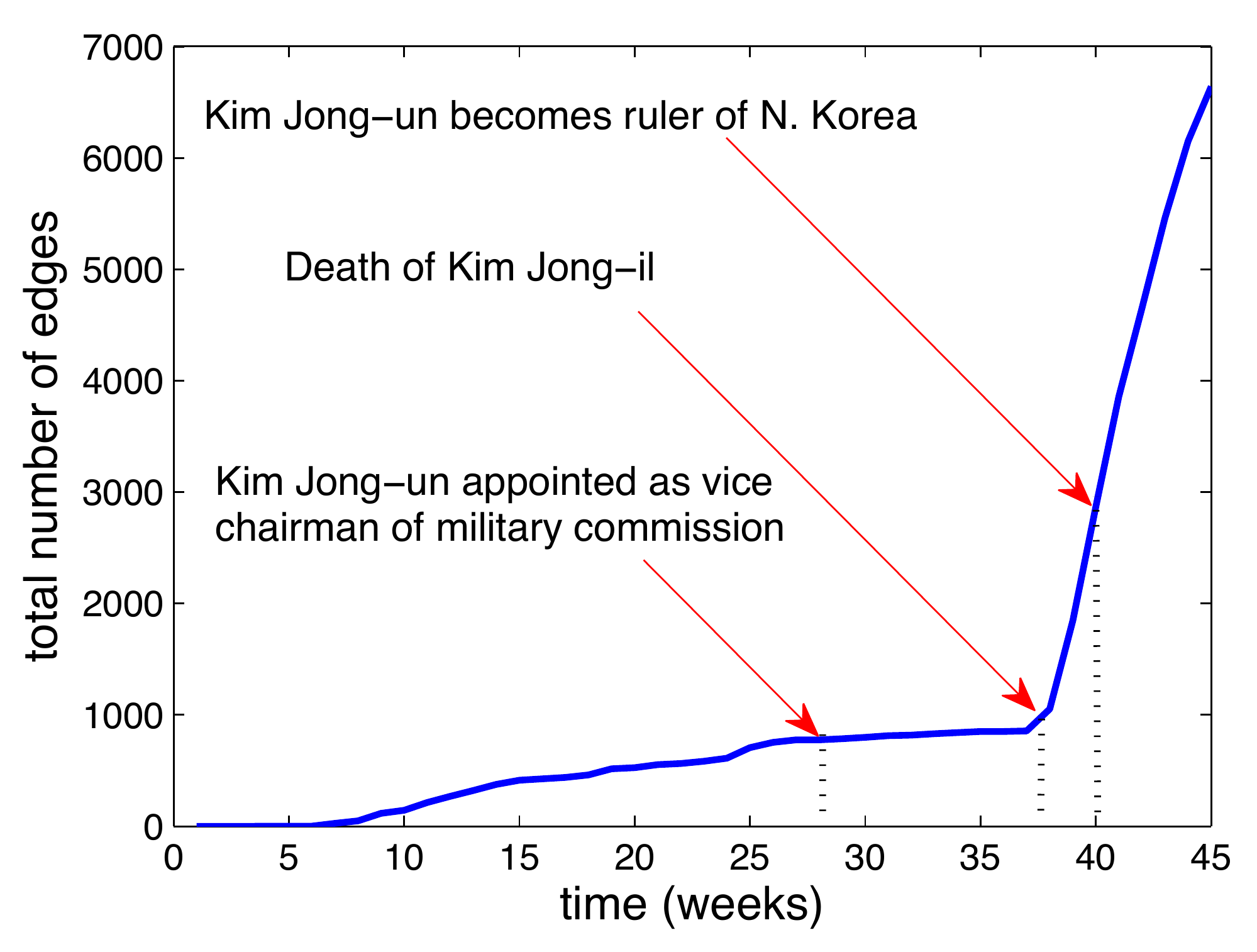}
		\centerline{(c)} %\medskip
	\end{minipage}
	
	\caption{Visualization of the estimated networks obtained by tracking 
		information cascades related to the topic ``Kim Jong-un'' at times (a) $t=10$ weeks and (b) $t=40$ weeks. The total number of inferred edges per week is shown in (c). The abrupt increase
		in edge density can be explained by three key events:  i) Kim Jong-un was appointed as the vice chairman
		of the North Korean military commission ($t=28$); ii) Kim Jong-il died ($t=38$); 
		and iii) Kim Jong-un became the ruler of North Korea ($t=40$)~\cite{baingana2014cascadesJSTSP14}.}
	\label{fig:kimjongun}
\end{figure*}

The algorithm in~\cite{baingana2014cascadesJSTSP14} was run on the dataset and Figures \ref{fig:kimjongun} (a) and (b) 
show visualizations of the inferred network at $t = 10$ and $t = 40$ weeks. Speculation about the
possible successor of the dying North Korean ruler, Kim Jong-il, rose until his
death on December 17, 2011 (week 38). He was succeeded by Kim Jong-un on
December 30, 2011 (week 40). The visualizations show an increasing number
of edges over the 45 weeks, illustrating the growing interest of international
news websites and blogs in the new ruler, about whom little was known in
the first 10 weeks. Unfortunately, the observation horizon does not go beyond
$T = 45$ weeks. A longer span of data would have been useful to investigate
the rate at which global news coverage on the topic eventually subsided.
Figure \ref{fig:kimjongun} (c) depicts the time evolution of the total number of
edges in the inferred dynamic network. Of particular interest are the weeks
during which: i) Kim Jong-un was appointed as the vice chairman of the
North Korean military commission; ii) Kim Jong-il died; and iii) Kim Jong-un
became the ruler of North Korea. These events were the topics of many
online news articles and political blogs, an observation that is reinforced by
the experimental results shown in the plot.

% % % % % % % % % % % % % % % % % % % % % % % % % % % % % % % % % % % % % % % %
%                         Section X                                        %
% % % % % % % % % % % % % % % % % % % % % % % % % % % % % % % % % % % % % % % %

\section{Concluding remarks and research outlook}
\label{sec:conc}

Contending that GSP provides novel insights and relevant tools to address network topology inference problems, this paper outlines recent approaches that use information available from graph signals to learn the underlying network structure in a variety of settings. Aligned with current trends in data-driven scientific inquiry into complex networks, the overarching aim is to shift from: (i) \emph{descriptive} accounts to \emph{inferential} GSP techniques that can explain as well as predict network behavior; and from (ii) \emph{ad hoc} graph constructions to \emph{rigorous} formulations rooted in well-defined models and principles. Accordingly, this tutorial stretches in a comprehensive and unifying manner all the way from (nowadays rather mature) statistical approaches including correlation analyses and graphical model selection, to recent GSP advances facilitated by spectral representations of network diffusion processes. A diverse
gamut of network inference challenges and application domains was selectively covered, based on importance and relevance to SP expertise, as well as on our own experience and contributions to the area. Admittedly, some important topics have been overlooked including tomographic network topology identification~\cite[Ch. 7.4]{kolaczyk2009book}, and inference of directed acyclic graphs (a particular favorable class of directed graphical models also known as Bayes networks)~\cite{chikeringDAGs}.

A wide variety of potential research avenues naturally follow from the developments here presented.
In terms of formal performance guarantees, the GSP-based methods in Sections~\ref{sec:smooth} and \ref{sec:topoid_diffusion} are less understood than their statistical counterparts (Section \ref{sec:stat}), posing a clear opportunity for improvement.
In particular, one might investigate if smoothness alone is sufficient to provide topology recovery guarantees,
or if there is a fundamental requirement for edge sparsity that cannot be forgone.
In practice, it sometimes holds that signals are discrete (such as pre-specified rating levels) or belong to a finite alphabet (such as node labels or classes). However, diffusion-based techniques for these types of signals have not been explored in sufficient depth. 
Moreover, graph topology inference is oftentimes an intermediate step in a larger processing or learning task. 
In this direction, one can think of bi-level topology inference formulations where the graph is designed explicitly taking into account the ultimate task, e.g., searching for a graph that induces a classifier that is maximally discriminative for some training data.
In addition, in line with the diffusion-based methods in Section~\ref{sec:topoid_diffusion}, it would be interesting to explore innovative generative models of nonlinear interactions for network data such as those given by median or other nonlinear graph filters.

In terms of computational complexity, there is room for improving scalability of some of the algorithms described via parallelization and decentralized implementations.
Moreover, adaptive algorithms that can track the (possibly) time-varying structure of the network, and achieve both memory and computational savings by processing the signals on-the-fly are naturally desirable, but so far largely unexplored.

Lastly, one can explore the links between network deconvolution -- as described in Section \ref{ssec:deconvolution} -- and graph sparsification approaches. The latter consists on approximating a given graph via a sparser one while preserving the associated Laplacian-based total-variation measure.
From a GSP perspective, this implies approximately preserving the smoothness of signals supported on those graphs. 
Clearly, this approximation notion is not too different from network deconvolution, where the premise is that the sparse graph approximate should preserve the eigenbasis (hence the GFT) of the original network.
Formalizing this intuition could enable further cross-pollination between established graph-theoretical problems and the exciting family of GSP approaches for network topology inference.

% % % % % % % % % % % % % % % % % % % % % % % % % % % % % % % % % % % % % % % %
%                         References                                          %
% % % % % % % % % % % % % % % % % % % % % % % % % % % % % % % % % % % % % % % %

\bibliographystyle{IEEEtranS}
\bibliography{biblio}

\end{document}